\documentclass[pra,aps,amssymb,amsmath,amstext,amsfont,noshowkeys,groupedaddress,12pt,nofootinbib]{revtex4}

\usepackage{accents}
\usepackage{enumerate}

\newcommand\zero[1]{\accentset{(0)}{#1}}

\newcommand\starr[1]{\accentset{(*)}{#1}}

\newcommand\one[1]{\accentset{(1)}{#1}}
\newcommand\two[1]{\accentset{(2)}{#1}}
\newcommand\eye[1]{\accentset{(i)}{#1}}

 \numberwithin{equation}{section}

\begin{document}

\title{Orbit Space Curvature as a Source of Mass in Quantum Gauge Theory}
\author{Vincent Moncrief}
\affiliation{Department of Physics and Department of Mathematics, \\ Yale University, P.O. Box 208120, New Haven, CT 06520, USA. \\ E-mail address: vincent.moncrief@yale.edu}
\author{Antonella Marini}
\affiliation{Department of Mathematics, \\ Yeshiva University, 500 West 185th Street, New York, NY 10033, USA. \\ and \\ Department of Mathematics, \\ University of L'Aquila, Via Vetoio, 67010 L'Aquila, AQ ITALY. \\ E-mail address: marini@yu.edu}
\author{Rachel Maitra}
\affiliation{Department of Applied Mathematics, \\ Wentworth Institute of Technology, 550 Huntington Avenue, Boston, MA 02115-5998, USA. \\ E-mail address: maitrar@wit.edu}
\date{\today}

\begin{abstract}
It has long been realized that the natural `orbit space' for non-abelian Yang-Mills dynamics (i.e., the reduced configuration space of gauge equivalence classes of spatial connections) is a \textit{positively curved} (infinite dimensional) Riemannian manifold. Expanding upon this result I.M.~Singer was led to propose that strict positivity of the corresponding Ricci tensor (computable from the rigorously defined curvature tensor through a suitable zeta function regularization procedure) could play a fundamental role in establishing that the associated Schr{\"o}dinger operator admits a \textit{spectral gap}. His argument was based on representing the (suitably regularized) kinetic term in the Schr{\"o}dinger operator as a Laplace-Beltrami operator on this positively curved orbit space. In this article we revisit Singer's proposal and show how, when the contribution of the Yang-Mills (magnetic) potential energy is taken into account, the role of the original orbit space Ricci tensor is instead played by a certain `Bakry-Emery Ricci tensor' computable from the ground state wave functional of the quantum theory. We next review the authors' ongoing \textit{Euclidean-signature-semi-classical} program for deriving asymptotic expansions for such wave functionals and discuss how, by keeping the dynamical nonlinearities and non-abelian gauge invariances fully intact at each level of the analysis, our approach surpasses that of conventional perturbation theory for the generation of such approximate wave functionals.

Though our main focus is on Yang-Mills theory we derive the corresponding orbit space curvature for scalar electrodynamics and prove that, whereas the Maxwell factor remains flat, the interaction naturally induces positive curvature in the (charged) scalar factor of the resulting orbit space. This has led us to the conjecture that such orbit space curvature effects could furnish a source of mass for ordinary Klein-Gordon type fields provided the latter are (minimally) coupled to gauge fields, even in the abelian case.

Finally we ask whether such an orbit space curvature mechanism could even play a role in the generation of an effective cosmological constant in quantum gravity theory. While we have, so far, no conclusive argument in this direction, we discuss the surprisingly promising extent to which our Euclidean-signature semi-classical program is applicable to the Wheeler-DeWitt equation of canonically quantized Einstein gravity.

\end{abstract}
\maketitle

\section{Introduction}
\label{sec:introduction}
A fundamental question in quantum gauge theory is whether the Schr{\"o}dinger operator for certain non-abelian Yang-Mills fields admits a spectral gap. Such a gap, if it exists, could represent the energy difference between the actual vacuum state and that of the lowest energy `glueball' states and confirm the expectation that massless gluons cannot propagate freely as photons do but must instead exhibit a form of `color confinement'. It seems to be well understood that this question lies beyond the scope of conventional perturbation theory and will require a more global analytical treatment for its ultimate resolution.

Many years ago I.M.~Singer proposed an elegant, geometrical approach to this fundamental problem based on the fact that the classical, reduced configuration space for Yang-Mills dynamics --- namely the `orbit space' of spatial connections modulo gauge transformations --- has a naturally induced, \textit{curved} Riemannian metric with everywhere non-negative sectional curvature \cite{Singer:1981}. The classical Hamiltonian for the reduced dynamics --- a real-valued functional defined on the cotangent bundle of this orbit space --- consists of a `kinetic' term induced from the spatial integral of the square of the vectorial electric component of the full, spacetime Yang-Mills curvature tensor and a `potential' term induced from the spatial integral of the square of its complementary, vectorial magnetic component. The non-vanishing curvature of the Riemannian metric defined by the kinetic term arises from the implementation of the Gauss-law constraint during the process of reduction to the quotient, orbit space and was independently computed by several investigators \cite{Singer:1981,Babelon:1981,Vergeles:1983}. The classical reduced dynamics is thus that for a system point (namely a gauge equivalence class of spatial connections) moving on a positively curved, infinite dimensional manifold under the influence of a (non-negative) potential energy.

Upon canonical quantization the Schr{\"o}dinger operator for this (pure Yang-Mills) dynamical system will thus include a kinetic term that, formally at least, encompasses the (negative\footnote{We here adopt the usual physicists' sign convention for the definition of a Laplacian.}) Laplace-Beltrami operator for an infinite dimensional, curved Riemannian manifold --- namely the orbit space alluded to above. Whereas the (covariant) Hessian of sufficiently smooth (wave) functionals can be rigorously defined in such infinite dimensional contexts, its associated trace need not make sense without some suitable regularization since the Hessian will not, in general, be trace class. Singer, in particular, proposed an elegant zeta function regularization scheme to define the needed Laplacian \cite{Singer:1981}.

A classical result in Riemannian geometry due to A.~Lichnerowicz \cite{Lichnerowicz:1958} shows that the Laplace operator for a complete, connected (finite-dimensional) Riemannian manifold necessarily exhibits a spectral gap provided that the Ricci tensor of this manifold is bounded, positively, away from zero\footnote{It follows from the Bonnet-Myers theorem that such a manifold is necessarily compact \cite{Myers:1941}.}. Such a result however cannot be expected to extend, in any straightforward way at least, to the infinite dimensional manifolds arising in quantum Yang-Mills theory. First of all, as Singer pointed out, their Ricci tensors, which would result from taking traces of corresponding (rigorously computable) curvature tensors, are not in general well-defined --- the curvature tensors in question not being trace class --- and would require a suitable regularization for their meaningful formulation. Again Singer proposed zeta function regularization as an elegant means of accomplishing this. Some such regularization, however, is actually a desirable feature of the quantum procedure, at least in 4 spacetime dimensions, since it allows the introduction of a length scale into the quantum formalism. In the absence of such a scale no hypothetical spectral energy gap could even be expressed in terms of the naturally occurring parameters of the theory (Planck's constant, the speed of light and the Yang-Mills coupling constant).

Another difficulty with attempting to extend the Lichnerowicz argument to the infinite dimensional setting of interest here is that, thanks to the Bonnet-Myers theorem, one knows that a complete, finite dimensional Riemannian manifold with positive Ricci curvature bounded away from zero is necessarily compact \cite{Myers:1941}. For a connected such manifold the lowest eigenvalue of its associated (negative) Laplacian always vanishes and corresponds to a globally constant eigenfunction. That such an eigenfunction is nevertheless always normalizable follows from the manifold's compactness. The spectral gap referred to in Lichnerowicz's theorem is thus simply the lowest non-vanishing eigenvalue of the manifold's (negative) Laplacian which, in view of compactness, necessarily has a discrete spectrum.

Generalizations of Lichnerowicz's theorem have been established under less stringent conditions on the Ricci tensors provided that the manifolds under study have finite diameters \cite{Ling:2006,Yang:1999}. L.~Andersson has proven that Riemannian Hilbert manifolds have finite diameters whenever their full sectional curvatures are positively bounded away from zero \cite{Andersson:1986} but this result does not apply to the orbit space sectional curvatures of interest here since these latter admit (infinite dimensional) families of 2-planes on which they actually vanish. In any case the diameters of these Yang-Mills orbit spaces are known to be infinite \cite{orbit-misc01}.

The true, normalizable ground state wave functional must necessarily reflect the presence of the potential energy term in the Schr{\"o}dinger operator. In Section~\ref{sec:spectral-gap-estimates} we show how to modify the original Lichnerowicz argument (in a finite dimensional setting) to allow for the occurrence of such a potential energy term and show that a corresponding gap estimate follows therefrom provided that a suitably defined `Bakry-Emery Ricci tensor' is bounded positively away from zero. This Bakry-Emery Ricci tensor differs from the actual Ricci tensor by a term in the (covariant) Hessian of the logarithm of the true ground state wave function. Its positivity could hold on a flat or even negatively curved space and thus its applicability is not limited to manifolds of finite diameter.

Furthermore the natural integration measure arising in this (generalized Lichnerowicz) analysis includes the squared modulus of the ground state wave function itself so that the total space, even it it has infinite diameter, now has finite measure simply by virtue of the normalizability of the vacuum state. This should prove to be especially significant for any potential extensions to infinite dimensional problems wherein formal Lebesgue measures no longer make sense but for which normalizable vacuum state wave functionals are nevertheless expected to exist.

In Section~\ref{sec:euclidean} we discuss an ongoing program, under development by the authors, to derive asymptotic expansions for the wave functionals of certain interacting quantum field theories including, in particular Yang-Mills fields \cite{Moncrief:2012,Marini:2016,Maitra:inprep}. Our `Euclidean signature semi-classical' analysis extends the applicability of certain elegant, microlocal methods to the case of bosonic field theories of renormalizable type. It has the significant advantage over conventional, Rayleigh-Schr{\"o}dinger perturbation theory of keeping the non-linearities and (if present) \textit{non-abelian} gauge invariances of an interacting system fully intact at every level of the analysis. Our expectation is that it should yield an asymptotic expansion for the needed, fully gauge invariant, logarithm of the ground state wave functional that is far superior to any attainable by conventional perturbation methods. The latter, by requiring an expansion in the Yang-Mills coupling constant, disturb both the nonlinear structure and the closely associated (non-abelian) gauge invariance of the Yang-Mills dynamical system at the outset and attempt to reinstate those vital features only gradually, order-by-order in the expansion.

Though our main focus is on the Yang-Mills system we show in Section~\ref{sec:orbit-space-curvature-scalar-electrodynamics} how (non-vanishing) orbit space curvature also arises naturally through the (minimal) coupling of a Maxwell field to a charged scalar field. In this case curvature arises only for the scalar factor of the (product) orbit space and not for the Maxwell factor which remains flat. We are thus led to conjecture that orbit space curvature could even serve as an independent source of mass for matter fields themselves provided that they are (minimally) coupled to (abelian or non-abelian) gauge fields.

Let \({}^{(4)}\!V := (\mathbb{R}^4,\eta)\), where
\begin{equation}\label{eq:301}
\eta = \eta_{\mu\nu}\; dx^\mu \otimes dx^\nu = -c^2\; dt \otimes dt + \sum_{i=1}^3 dx^i \otimes dx^i,
\end{equation}
designate Minkowski space expressed in a standard (Lorentz frame) coordinate system \(\lbrace x^\mu\rbrace = \lbrace ct, x^i\rbrace\) and consider the Yang-Mills action functional (for a compact gauge group \textit{G}) defined over domains \(\Omega\) of the form \(\Omega = I \times \mathbb{R}^3\) where \(I = [t_0,t_1]\). Variation of this action with respect to the time component of the spacetime Yang-Mills connection field yields the so-called Gauss-law constraint equation which, for each fixed \(t \in I\), may be viewed as an elliptic equation on \(\mathbb{R}^3\) for this time component --- a Lie-algebra valued function. If, with suitable boundary conditions imposed, one solves this constraint and substitutes the solution back into the action, the resulting reduced kinetic term (a quadratic form in the `velocity' of the spatial connection) is found to be degenerate along gauge orbit directions but smooth, gauge invariant and positive definite in the transversal directions \cite{Singer:1981,Babelon:1981,Vergeles:1983}. It thus follows that this kinetic term defines a smooth, Riemannian metric on the natural `orbit space' of spatial connections modulo gauge transformations. This orbit space is (at least almost everywhere) itself a smooth, infinite dimensional manifold and provides the geometrically natural (reduced) configuration space for (classical) Yang-Mills dynamics.

A corresponding smooth \textit{potential energy} functional is induced on this orbit space by the integral over \(\mathbb{R}^3\) (at fixed \textit{t}) of the square of the curvature of the spatial connection field --- the `magnetic' component of the curvature of the full spacetime connection field. A Legendre transformation leads in turn to the Hamiltonian functional for the classical dynamics which takes the `standard' form of a sum of (curved space) kinetic and potential energies.

The sectional curvature of this reduced configuration space was independently computed in \cite{Singer:1981,Babelon:1981,Vergeles:1983} and shown to be everywhere non-negative but almost everywhere non-vanishing whenever the gauge group \textit{G} is non-abelian. Though Singer discussed the need for a suitable regularization scheme to make sense of the formally (positively) divergent Ricci tensor of the orbit space metric, the actual form of such a regulated Ricci tensor seems still to be unknown. It would be most interesting if a suitably defined Ricci tensor could be shown to be bounded, positively away from zero on this orbit space, especially inasmuch as we think it quite unlikely that the Bakry-Emery `enhancement' of this tensor would nullify its (hypothetical) positivity properties but perhaps, more likely, complement them\footnote{This would be true for example if the relevant logarithm were (almost everwhere) convex.}. Furthermore, as we shall amplify near the end of Section~\ref{sec:spectral-gap-estimates}, it seems quite plausible that strict positivity of the Bakry-Emery Ricci tensor, though sufficient for the implication of a spectral gap, is not absolutely necessary for this conclusion to hold.

In view of the promising character of these orbit-space-curvature ideas for Minkowski space gauge theories we have felt encouraged to ask whether such ideas could also be relevant to the problem of quantum gravity. Here the natural question would seem to be whether such (orbit-space-curvature) effects could be shown, by themselves, to induce a non-vanishing cosmological constant. Since research in this direction has only just begun we do not, by any means, have convincing arguments for this conclusion. We do, however, have considerable evidence for the applicability of our \textit{Euclidean-signature-semi-classical} technology to the Wheeler-DeWitt equation for (canonically quantized) Einstein gravity. As a first step in this direction we shall review, in Section~\ref{sec:euclidean-cosmology}, how the scope of the aforementioned microlocal methods can be extended to apply to the (partial differential) Wheeler-DeWitt equation for spatially homogeneous, Bianchi IX (or, 'Mixmaster') cosmological models. The key issue addressed therein is how globally smooth `eigenfunctions' for the Wheeler-DeWitt operator can be constructed at all by these methods when the corresponding eigenvalues (for both `ground' and `excited' states) are required to \textit{vanish identically} to all orders in Planck's constant. We also discuss therein how certain (ultra long-wavelength) `graviton excitation numbers' emerge naturally from this (generalized microlocal) analysis in spite of the absence of any (non-vanishing) eigenvalues.

But the Mixmaster Wheeler-DeWitt equation is a quantum mechanical one whereas that for full (canonically quantized) Einstein gravity is a field theoretic, functional differential system. Can the Euclidean-signature-semi-classical technology be nevertheless further generalized to be applicable thereto? In Section~\ref{sec:euclidean-signature-wheeler-dewitt} we shall sketch how such a program could indeed be carried out and draw attention to several remarkably attractive features of such an approach including, in particular, how it apparently avoids some of the serious conceptual and mathematical complications that obstructed progress on the, somewhat similar-in-spirit, \textit{Euclidean path integral} approach to quantum gravity.

\section{Spectral Gap Estimates}
\label{sec:spectral-gap-estimates}
A lower bound for the spectral gap of the Laplacian of a complete Riemannian manifold having strictly positive Ricci curvature was derived in a classic work by Lichnerowicz \cite{Lichnerowicz:1958}. In view of the Bonnet-Myers theorem however such a manifold must be compact and, in particular, have its diameter bounded from above in terms of the assumed, positive lower bound on the Ricci tensor \cite{Myers:1941}. For compact manifolds the spectrum must of course be discrete, and thus exhibit a gap, but, in the absence of positive Ricci curvature, further geometrical information about the manifold would be needed to bound the actual gap. A flat torus, for example can have an arbitrarily large diameter and a corresponding, arbitrarily small gap.

For Schr\"{o}dinger operators on the other hand, wherein the Laplacian is supplemented with a potential energy term, one can modify Lichnerowicz's argument so that the role of the Ricci tensor in the spectral gap estimate is now played by the so-called \textit{Bakry-Emery Ricci tensor} which includes, indirectly, information about the potential energy function. For pure geometry problems, which need have no Schr\"{o}dinger interpretation, the relevant Bakry-Emery tensor often arises from the study of so-called \textit{metric measure spaces} wherein the natural Riemannian volume element is multiplied by a smooth positive function \cite{Lott:2003,Wei:2009}.

In the Schr\"{o}dinger context in particular, however, manifold compactness may no longer be needed since, in the revised argument, only positivity of the Bakry-Emery Ricci tensor is required to bound the spectral gap from below and, depending upon the nature of the potential energy involved, this condition may well hold in the presence of vanishing or even negative ordinary Ricci curvature. In an infinite dimensional, field theoretic setting on the other hand further possibilities may also arise in that positive Ricci curvature, which typically requires a suitable regularization to even be defined, need no longer imply manifold compactness. Setting such complications momentarily aside though, we sketch below the derivation of the relevant `Bochner identity' for a conventional, kinetic-plus-potential Schr\"{o}dinger operator defined over a (smooth, connected, complete and orientable) Riemannian \textit{n}-manifold \(\lbrace M,g\rbrace\).

Let \(\Delta_g\) designate the covariant Laplacian (i.e., Laplace-Beltrami operator) given, in local coordinates for \(\lbrace M,g\rbrace\) by
\begin{equation}\label{eq:201}
\Delta_g := \frac{1}{\mu_g} \;\partial_i (\mu_g g^{ij}\partial_j)
\end{equation}
where \(\mu_g := \sqrt{\det{g}}\), the natural volume element for the given manifold. If \(V:M \rightarrow \mathbb{R}\) is a smooth function we define a corresponding Schr\"{o}dinger (Hamiltonian) operator \(\hat{H}\), for a `particle' with mass \(m > 0\), by
\begin{equation}\label{eq:202}
\hat{H} := -\frac{\hbar^2}{2m}\; \Delta_g + V
\end{equation}
(with \(\hbar := h/2\pi\) the reduced Planck constant) and assume that \(\lbrace M,g\rbrace\) and \textit{V} have been chosen so that \(\hat{H}\) is well-defined and self-adjoint on a suitable domain in \(L^2 (M,g)\).

We also assume that the time-independent Schr\"{o}dinger equation,
\begin{equation}\label{eq:203}
\hat{H}\Psi = E\Psi ,
\end{equation}
admits a (square integrable) ground state wave function,
\begin{equation}\label{eq:204}
\zero{\Psi} = \zero{N} e^{-\mathcal{S}/\hbar},
\end{equation}
with corresponding eigenvalue \(E = \zero{E} \in \mathbb{R}\), where \(\mathcal{S}:M \rightarrow \mathbb{R}\) is a smooth function and \(\zero{N} \in \mathbb{C}\) a normalization constant (unique up to phase) chosen so that
\begin{equation}\label{eq:205}
\int_M \mu_g \zero{\Psi}^{\,\dagger} \zero{\Psi} d^nx = |\zero{N}|^2 \int_M \mu_g e^{-2\mathcal{S}/\hbar} d^nx = 1.
\end{equation}

Normalized excited state wave functions, orthogonal to the ground state, are expressible in the form
\begin{equation}\label{eq:206}
\starr{\Psi} = \starr{\varphi} e^{-\mathcal{S}/\hbar},
\end{equation}
with \(\starr{\varphi}:M \rightarrow \mathbb{C}\), and satisfy
\begin{equation}\label{eq:207}
\begin{split}
\langle\starr{\Psi}|\starr{\Psi}\rangle &:= \int_M \mu_g \starr{\Psi}^{\,\dagger} \starr{\Psi} d^nx\\
 &= \int_M \mu_g \starr{\varphi}^{\,\dagger} \starr{\varphi} e^{-2\mathcal{S}/\hbar} d^nx = 1
\end{split}
\end{equation}
and
\begin{equation}\label{eq:208}
\langle\starr{\Psi}|\zero{\Psi}\rangle := \int_M \mu_g \starr{\varphi}^{\,\dagger} \cdot \zero{N} e^{-2\mathcal{S}/\hbar} d^nx = 0
\end{equation}
where \(\starr{\Psi}^{\,\dagger} = \starr{\varphi}^{\,\dagger} e^{-\mathcal{S}/\hbar}\) is the complex conjugate of \(\starr{\Psi}\).

Noting that
\begin{equation}\label{eq:209}
\begin{split}
(\hat{H} - \zero{E})\starr{\Psi} &= (\hat{H} - \zero{E}) (\starr{\varphi} e^{-\mathcal{S}/\hbar})\\
 &= \frac{-\hbar^2}{2m} \left[\Delta_g \starr{\varphi} - \frac{2}{\hbar}\; \mathcal{S}^{\,|k} \starr{\varphi}_{|k}\right] e^{-\mathcal{S}/\hbar}
\end{split}
\end{equation}
where \(\mathcal{S}^{\,|k} \starr{\varphi}_{|k} := g^{k\ell} (\nabla_k\mathcal{S})(\nabla_\ell \starr{\varphi})\), with \(|k = \nabla_k\) designating covariant differentiation with respect to \textit{g}, we see that if \(\starr{\Psi}\) is an actual eigenstate of \(\hat{H}\), with eigenvalue \(E = \starr{E} \in \mathbb{R}\), then
\begin{equation}\label{eq:210}
(\hat{H} - \zero{E}) \starr{\Psi} = (\starr{E} - \zero{E}) \starr{\Psi}
\end{equation}
or, equivalently
\begin{equation}\label{eq:211}
\frac{-\hbar^2}{2m} \left[\Delta_g \starr{\varphi} - \frac{2}{\hbar}\; \mathcal{S}^{\,|k} \starr{\varphi}_{|k}\right] = (\starr{E} - \zero{E}) \starr{\varphi}.
\end{equation}

The operator
\begin{equation}\label{eq:212}
\hat{\mathcal{H}} := -\frac{\hbar^2}{2m}\; \left[\Delta_g - \frac{2}{\hbar}\; \mathcal{S}^{\,|k} \nabla_k\right],
\end{equation}
which is self-adjoint with respect to the measure \(\mu_g e^{-2\mathcal{S}/\hbar} d^nx\) on \textit{M}, encompasses the so-called Bakry-Emery or Witten Laplacian (on functions) and its lowest nontrivial eigenvalue (in the case of a discrete spectrum) defines the spectral gap, \(\one{E} - \zero{E}\) of principal interest herein.

From equations (\ref{eq:209}--\ref{eq:211}) one finds that
\begin{equation}\label{eq:213}
\begin{split}
(\starr{E} - \zero{E}) \int_M \starr{\Psi}^{\,\dagger} \starr{\Psi} \mu_g d^nx &= (\starr{E} - \zero{E}) \int_M \starr{\varphi}^{\,\dagger} \starr{\varphi} e^{-2\mathcal{S}/\hbar} \mu_g d^nx\\
 &= \frac{\hbar^2}{2m} \int_M \mu_g e^{-2\mathcal{S}/\hbar} \starr{\varphi}^{\,\dagger}_{|k} \starr{\varphi}^{\,|k} d^nx - \frac{\hbar^2}{2m} \int_M \mu_g \left(\starr{\varphi}^{\,\dagger} \starr{\varphi}^{\,|k} e^{-2\mathcal{S}/\hbar}\right)_{|k} d^nx\\
 &= \frac{\hbar^2}{2m} \int_M \mu_g e^{-2\mathcal{S}/\hbar} \starr{\varphi}^{\,\dagger}_{|k} \starr{\varphi}^{\,|k} d^nx
\end{split}
\end{equation}
where the vanishing of the integral of the divergence follows from the (assumed) self-adjoincy of \(\hat{H} - \zero{E}\). In view of its assumed orthogonality to the ground state \(\starr{\varphi}\) cannot be constant and thus (\ref{eq:213}) immediately implies that \((\starr{E} - \zero{E}) > 0\) (in this case of a discrete spectrum). To put a quantitative lower bound on this gap however requires a further argument.

To this end define, for any smooth function \(\tilde{\varphi}:M \rightarrow \mathbb{C}\), the quantity
\begin{equation}\label{eq:214}
\tilde{\mathcal{Q}}_{\tilde{\varphi}} := g^{ij} (\nabla_i\tilde{\varphi}^{\,\dagger}) (\nabla_j\tilde{\varphi}) e^{-2\mathcal{S}/\hbar}
\end{equation}
and apply the covariant Laplacian thereto. The result can be expressed as
\begin{equation}\label{eq:215}
\begin{split}
\Delta_g \tilde{Q}_{\tilde{\varphi}} &= \nabla^k \nabla_k (g^{ij} \tilde{\varphi}^{\,\dagger}_{|i} \tilde{\varphi}_{|j} e^{-2\mathcal{S}/\hbar})\\
 &= -2 \left(\frac{2m}{\hbar^2}\right)^2 \left[ (\hat{H} - \zero{E}) (\tilde{\varphi}^{\,\dagger} e^{-\mathcal{S}/\hbar})\right] \left[ (\hat{H} - \zero{E}) (\tilde{\varphi} e^{-\mathcal{S}/\hbar})\right]\\
 &\hphantom{= }\> + 2\tilde{\varphi}^{\,\dagger}_{|ij} \tilde{\varphi}^{\,|ij} e^{-2\mathcal{S}/\hbar} + 2 \mathcal{R}_{ij} \tilde{\varphi}^{\dagger|i} \tilde{\varphi}^{\,|j} e^{-2\mathcal{S}/\hbar} + \frac{4}{\hbar} \mathcal{S}^{\,|ij} \tilde{\varphi}^{\,\dagger}_{|i} \tilde{\varphi}_{|j} e^{-2\mathcal{S}/\hbar}\\
 &\hphantom{= }\> + \left\lbrace\tilde{\varphi}^{|j} e^{-2\mathcal{S}/\hbar} \left(\tilde{\varphi}^{\dagger|k}_{|k} - \frac{2}{\hbar} \mathcal{S}^{\,|k} \tilde{\varphi}^{\,\dagger}_{|k}\right) + \tilde{\varphi}^{\dagger |j} e^{-2\mathcal{S}/\hbar} \left(\tilde{\varphi}^{\,|k}_{|k} - \frac{2}{\hbar} \mathcal{S}^{\,|k} \tilde{\varphi}_{|k}\right) - \frac{2}{\hbar} \mathcal{S}^{\,|j} \tilde{\varphi}^{\dagger|k} \tilde{\varphi}_{|k} e^{-2\mathcal{S}/\hbar}\right\rbrace_{|j}
\end{split}
\end{equation}
where the Ricci tensor, \(\mathcal{R}_{ij} dx^i \otimes dx^j\), of the metric \textit{g} has arisen from the commutation of covariant derivatives followed by contraction of the resultant curvature tensor. This formula is the `Bochner identity' referred to above and it naturally incorporates the Bakry-Emery Ricci tensor \(\mathcal{R}^{\mathcal{S}} = \mathcal{R}^{\mathcal{S}}_{ij} dx^i \otimes dx^j\) defined by
\begin{equation}\label{eq:216}
\mathcal{R}^{\mathcal{S}}_{ij} = \mathcal{R}_{ij} + \frac{2}{\hbar} \mathcal{S}_{|ij}.
\end{equation}

Taking, for the moment, \(\tilde{\varphi}\) to have compact support and integrating  (\ref{eq:215}) over \textit{M} one arrives at
\begin{equation}\label{eq:217}
\begin{split}
&2\left(\frac{2m}{\hbar^2}\right)^2 \int_M \mu_g \left[ (\hat{H} - \zero{E}) (\tilde{\varphi}^{\,\dagger} e^{-\mathcal{S}/\hbar})\right] \left[ (\hat{H} - \zero{E}) (\tilde{\varphi} e^{-\mathcal{S}/\hbar})\right] d^nx\\
 &= \int_M \mu_g \left\lbrace 2 \left(\mathcal{R}_{ij} + \frac{2}{\hbar} \mathcal{S}_{|ij}\right) \tilde{\varphi}^{\dagger|i} \tilde{\varphi}^{\,|j} e^{-2\mathcal{S}/\hbar} + 2\tilde{\varphi}^{\,\dagger}_{|ij} \tilde{\varphi}^{\,|ij} e^{-2\mathcal{S}/\hbar}\right\rbrace d^nx\\
 &= 2\left(\frac{2m}{\hbar^2}\right)^2 \int_M \mu_g \left\lbrace (\tilde{\varphi}^{\,\dagger} e^{-\mathcal{S}/\hbar}) (\hat{H} - \zero{E})^2 (\tilde{\varphi} e^{-\mathcal{S}/\hbar})\right\rbrace d^nx
\end{split}
\end{equation}
where the final equality results from the self-adjoincy of the operator \(\hat{H} - \zero{E}\).

If now \(\one{\Psi} = \one{\varphi} e^{-\mathcal{S}/\hbar}\) is an eigenstate of \(\hat{H}\) with eigenvalue \(\one{E}\) corresponding (in this case of a discrete spectrum) to a minimally excited state then one can approximate this state by a sequence of functions of compact support, \(\tilde{\Psi}_\ell = \tilde{\varphi}_\ell e^{-\mathcal{S}/\hbar} \xrightarrow[\ell \rightarrow \infty]{} \one{\Psi} = \one{\varphi} e^{-\mathcal{S}/\hbar}\), the space of which densely filling the relevant Hilbert space, and conclude from (\ref{eq:217}) that, in the limit, \(\one{\Psi} = \one{\varphi} e^{-\mathcal{S}/\hbar}\) satisfies
\begin{equation}\label{eq:218}
\begin{split}
2\left(\frac{2m}{\hbar^2}\right)^2 (\one{E} - \zero{E})^2 \int_M \mu_g \one{\Psi}^{\,\dagger} \one{\Psi} d^nx &= 2\left(\frac{2m}{\hbar^2}\right)^2 (\one{E} - \zero{E})^2 \int_M \mu_g \one{\varphi}^{\,\dagger} \one{\varphi} e^{-2\mathcal{S}/\hbar} d^nx\\
&= \int_M \mu_g \left\lbrace 2\left(\mathcal{R}^{ij} + \frac{2}{\hbar} \mathcal{S}^{\,|ij}\right) \one{\varphi}^{\,\dagger}_{|i} \one{\varphi}_{|j} e^{-2\mathcal{S}/\hbar} + 2\one{\varphi}^{\,\dagger}_{|ij} \one{\varphi}^{\,|ij} e^{-2\mathcal{S}/\hbar}\right\rbrace d^nx\\
&= 2\left(\frac{2m}{\hbar^2}\right) (\one{E} - \zero{E}) \int_M \mu_g \one{\varphi}^{\,\dagger}_{|k} \one{\varphi}^{\,|k} e^{-2\mathcal{S}/\hbar} d^nx
\end{split}
\end{equation}
where the last equality results from applying (\ref{eq:213}) to the case at hand.

Since \(\one{\Psi} = \one{\varphi} e^{-\mathcal{S}/\hbar}\) is orthogonal to the ground state \(\one{\varphi}\) cannot be constant and thus one gets from (\ref{eq:218}) that
\begin{equation}\label{eq:219}
\begin{split}
(\one{E} - \zero{E}) &= \left\lbrace\frac{\displaystyle \frac{\hbar^2}{2m} \int_M \mu_g \left\lbrace\left(\mathcal{R}^{ij} + \frac{2}{\hbar}\mathcal{S}^{\,|ij}\right) \one{\varphi}^{\,\dagger}_{|i} \one{\varphi}_{|j} e^{-2\mathcal{S}/\hbar} + \one{\varphi}^{\,\dagger}_{|ij} \one{\varphi}^{\,|ij} e^{-2\mathcal{S}/\hbar}\right\rbrace d^nx}{\displaystyle \int_M \mu_g \left\lbrace \one{\varphi}^{\,\dagger}_{|k} \one{\varphi}^{\,|k} e^{-2\mathcal{S}/\hbar}\right\rbrace d^nx}\right\rbrace\\
&\geq \frac{\displaystyle \frac{\hbar^2}{2m} \int_M \mu_g \left\lbrace e^{-2\mathcal{S}/\hbar} \left(\mathcal{R}^{ij} + \frac{2}{\hbar} \mathcal{S}^{\,|ij}\right) \one{\varphi}^{\,\dagger}_{|i} \one{\varphi}_{|j}\right\rbrace d^nx}{\displaystyle \int_M \mu_g \left\lbrace e^{-2\mathcal{S}/\hbar} \one{\varphi}^{\,\dagger}_{|k} \one{\varphi}^{\,|k}\right\rbrace d^nx}\\
&\geq \inf_{\tilde{\varphi} \in \mathcal{A}} \frac{\displaystyle \frac{\hbar^2}{2m} \int_M \mu_g \left\lbrace e^{-2\mathcal{S}/\hbar} \mathcal{R}^{\mathcal{S}}_{ij} \tilde{\varphi}^{\dagger|i} \tilde{\varphi}^{\,|j}\right\rbrace d^nx}{\displaystyle \int_M \mu_g \left\lbrace e^{-2\mathcal{S}/\hbar} \tilde{\varphi}^{\,\dagger}_{|k} \tilde{\varphi}^{\,|k}\right\rbrace d^nx}
\end{split}
\end{equation}
where \(\mathcal{A}\) is the space of smooth functions on \textit{M} satisfying
\begin{equation}\label{eq:220}
\int_M \mu_g e^{-2\mathcal{S}/\hbar} \tilde{\varphi}^{\,\dagger} \tilde{\varphi} d^nx = 1
\end{equation}
and
\begin{equation}\label{eq:221}
\int_M \mu_g e^{-2\mathcal{S}/\hbar} \tilde{\varphi}^{\,\dagger} \cdot 1 d^nx = 0.
\end{equation}

From the foregoing it follows that if the Bakry-Emery Ricci tensor, \(\mathcal{R}^{\mathcal{S}} = \mathcal{R}^{\mathcal{S}}_{ij} dx^i \otimes dx^j\) satisfies the global positivity condition,
\begin{equation}\label{eq:222}
\mathcal{R}^{\mathcal{S}}_{ij} v^i v^j \geq \frac{1}{\ell_o^2} g_{ij} v^i v^j,
\end{equation}
for an arbitrary vector field \(\mathbf{v} = v^i\partial_i\) on \textit{M}, for some constant \(\ell_o > 0\) (with the dimensions of length), then the spectral gap satisfies
\begin{equation}\label{eq:223}
\one{E} - \zero{E} \geq \frac{\hbar^2}{2m} \frac{1}{\ell_o^2}.
\end{equation}

As a special case of the above consider a (multi-dimensional) \textit{harmonic} oscillator on Euclidean \(\mathbb{R}^n\) with oscillation frequencies \(0 < \omega_1 \leq \omega_2 \leq \cdots \leq \omega_n\) along the various Cartesian coordinate axes. The function \(\mathcal{S}\) is then given by
\begin{equation}\label{eq:224}
\mathcal{S} = \frac{1}{2} \sum_{j=1}^n m\omega_j (x^j)^2
\end{equation}
so that
\begin{equation}\label{eq:225}
\frac{2}{\hbar} \frac{\partial^2\mathcal{S}}{\partial x^j\partial x^\ell} = \frac{2m}{\hbar} \omega_j \delta_{j\ell}\qquad \text{ (no sum on \textit{j})}
\end{equation}
and thus that
\begin{equation}\label{eq:226}
\mathcal{R}^{\mathcal{S}}_{j\ell} v^j v^\ell \geq \frac{2m\omega_1}{\hbar} \delta_{j\ell} v^j v^\ell
\end{equation}
It follows from (\ref{eq:223}), taking \(\frac{1}{\ell_o^2} = \frac{2m\omega_1}{\hbar}\), that
\begin{equation}\label{eq:227}
\one{E} - \zero{E} \geq \hbar\omega_1.
\end{equation}
That the gap estimate is sharp in this case results from the fact that \(\one{\varphi}\) is a first order Hermite polynomial in \(x^1\) which, being linear in \(x^1\), satisfies \(\one{\varphi}_{|ij} = 0\).

In the foregoing we assumed that the excited state spectrum was discrete. Suppose instead that it is continuous with \(\one{E} > \zero{E}\) designating the infimum of the (continuous) excited state spectrum. From the spectral decomposition theorem \cite{Newton} it follows that, for any \(\epsilon > 0\), there will exist normalizable states, \(\Psi_\epsilon\), orthogonal to the ground state, satisfying
\begin{align}
\int_M \mu_g \Psi^{\,\dagger}_\epsilon (\hat{H} - \one{E}) \Psi_\epsilon d^nx &\geq 0,\label{eq:228}\\
\int_M \mu_g \left\lbrace\left[ (\hat{H} - \one{E}) \Psi_\epsilon\right]^\dagger \left[ (\hat{H} - \one{E}) \Psi_\epsilon\right]\right\rbrace d^nx &\leq \epsilon^2 \int_M \mu_g \Psi^{\,\dagger}_\epsilon \Psi_\epsilon d^nx\label{eq:229}
\end{align}
and
\begin{equation}\label{eq:230}
\int_M \mu_g \Psi^{\,\dagger}_\epsilon \zero{\Psi} d^nx = 0
\end{equation}
Note that the imposition of (\ref{eq:230}) is essential for the validity of (\ref{eq:228}) since otherwise one could simply take \(\Psi_\epsilon \rightarrow \zero{\Psi}\) to get a counterexample. One can assume for convenience though that \(\Psi_\epsilon\) has compact support and is smooth since the space of such functions is dense in the Hilbert space of interest.

from the Schwarz inequality one has, upon appealing to (\ref{eq:229}), that
\begin{equation}\label{eq:231}
\begin{split}
0 &\leq \int_M \mu_g \Psi^{\,\dagger}_\epsilon (\hat{H} - \one{E}) \Psi_\epsilon d^nx\\
&\leq (\int_M \mu_g \Psi^{\,\dagger}_\epsilon \Psi_\epsilon d^nx)^{1/2} \left(\int_M \mu_g \left\lbrace\left[ (\hat{H} - \one{E}) \Psi_\epsilon\right]^\dagger \left[ (\hat{H} - \one{E})\Psi_\epsilon\right]\right\rbrace d^nx\right)^{1/2}\\
&\leq \epsilon \int_M \mu_g (\Psi^{\,\dagger}_\epsilon \Psi_\epsilon) d^nx
\end{split}
\end{equation}
Using the fact that \((\hat{H} - \one{E})\) is a real, self-adjoint operator it is easily verified that
\begin{equation}\label{eq:232}
\begin{split}
&\int_M \mu_g \left\lbrace\left[ (\hat{H} - \zero{E}) \Psi_\epsilon\right]^\dagger \left[ (\hat{H} - \zero{E}) \Psi_\epsilon\right]\right\rbrace d^nx = \int_M \mu_g \left\lbrace\left[ (\hat{H} - \one{E}) \Psi_\epsilon\right]^\dagger \left[ (\hat{H} - \one{E}) \Psi_\epsilon\right]\right.\\
 &\qquad\left. \vphantom{\left[ (\hat{H} - \one{E}) \Psi_\epsilon\right]^\dagger \left[ (\hat{H} - \one{E}) \Psi_\epsilon\right]} + (\one{E} - \zero{E})^2 \Psi^{\,\dagger}_\epsilon \Psi_\epsilon + 2(\one{E} - \zero{E}) \Psi^{\,\dagger}_\epsilon (\hat{H} - \one{E})\Psi_\epsilon\right\rbrace d^nx\\
&\qquad\leq \epsilon^2 \int_M \mu_g \Psi^{\,\dagger}_\epsilon \Psi_\epsilon d^nx + 2(\one{E} - \zero{E}) \epsilon \int_M \mu_g \Psi^{\,\dagger}_\epsilon \Psi_\epsilon d^nx + (\one{E} - \zero{E})^2 \int_M \mu_g \Psi^{\,\dagger}_\epsilon \Psi_\epsilon d^nx\\
&\qquad = (\one{E} - \zero{E} + \epsilon)^2 \int_M \mu_g \Psi^{\,\dagger}_\epsilon \Psi_\epsilon d^nx
\end{split}
\end{equation}
where, in the final step, we have applied (\ref{eq:229}) and (\ref{eq:231}).

Setting \(\Psi_\epsilon = \varphi_\epsilon e^{-\mathcal{S}/\hbar}\) and combining (\ref{eq:232}) with (\ref{eq:217}), with \(\tilde{\varphi} \rightarrow \varphi_\epsilon\), we get
\begin{equation}\label{eq:233}
\begin{split}
&2\left(\frac{2m}{\hbar^2}\right)^2 (\one{E} - \zero{E} + \epsilon)^2 \int_M \mu_g (\Psi^{\,\dagger}_\epsilon \Psi_\epsilon) d^nx \geq 2\left(\frac{2m}{\hbar^2}\right)^2 \int_M \mu_g \left[ (\hat{H} - \zero{E}) \Psi^{\,\dagger}_\epsilon\right] \left[ (\hat{H} - \zero{E}) \Psi_\epsilon\right] d^nx\\
&\qquad = 2\left(\frac{2m}{\hbar^2}\right)^2 \int_M \mu_g \left\lbrace\left[ (\hat{H} - \zero{E}) (\varphi^{\,\dagger}_\epsilon e^{-\mathcal{S}/\hbar})\right] \left[ (\hat{H} - \zero{E}) (\varphi_\epsilon e^{-\mathcal{S}/\hbar})\right]\right\rbrace d^nx\\
&\qquad = 2 \int_M \mu_g \left\lbrace \left(\mathcal{R}_{ij} + \frac{2}{\hbar} \mathcal{S}_{|ij}\right) \varphi^{\dagger|i}_\epsilon \varphi^{|j}_\epsilon e^{-2\mathcal{S}/\hbar} + \varphi^{\,\dagger}_{\epsilon|ij} \varphi^{|ij}_\epsilon e^{-2\mathcal{S}/\hbar}\right\rbrace d^nx\\
&\qquad\qquad \geq 2 \int_M \mu_g \left\lbrace\left(\mathcal{R}_{ij} + \frac{2}{\hbar} \mathcal{S}_{|ij}\right) \varphi^{\dagger|i}_\epsilon \varphi^{|j}_\epsilon e^{-2\mathcal{S}/\hbar}\right\rbrace d^nx
\end{split}
\end{equation}

Thus, assuming the Bakry-Emery bound (\ref{eq:222}), one arrives at
\begin{equation}\label{eq:234}
\begin{split}
&(\one{E} - \zero{E} + \epsilon)^2 \int_M \mu_g (\Psi^{\,\dagger}_\epsilon \Psi_\epsilon) d^nx \geq \left(\frac{\hbar^2}{2m}\right)^2 \frac{1}{\ell_o^2} \int_M \mu_g \varphi^{\dagger|i}_\epsilon \varphi^{|j}_\epsilon g_{ij} e^{-2\mathcal{S}/\hbar} d^nx\\
&= \frac{\hbar^2}{2m\ell_o^2} \int_M \mu_g \Psi^{\,\dagger}_\epsilon (\hat{H} - \zero{E}) \Psi_\epsilon d^nx \geq \frac{\hbar^2}{2m\ell_0^2} (\one{E} - \zero{E}) \int_M \mu_g \Psi^{\,\dagger}_\epsilon \Psi_\epsilon d^nx
\end{split}
\end{equation}
where, in the final steps, we have appealed to (\ref{eq:209}) and (\ref{eq:228}) together with an integration by parts. Setting \(\one{E} - \zero{E} := \Delta E > 0\) we thus get from (\ref{eq:234}) that
\begin{equation}\label{eq:235}
\Delta E + 2\epsilon + \frac{\epsilon^2}{\Delta E} \geq \frac{\hbar^2}{2m\ell_o^2}, \qquad \forall\; \epsilon > 0
\end{equation}
and thus that
\begin{equation}\label{eq:236}
\Delta E \geq \frac{\hbar^2}{2m\ell_0^2}
\end{equation}

One might still wonder whether \(\one{E} - \zero{E} = 0\), i.e., with the normalizable ground state embedded at the bottom of a continuous excited state spectrum, is a remaining possibility. To exclude this, at least heuristically, (under the Bakry-Emery assumption (\ref{eq:222})), note that (\ref{eq:233}) then gives
\begin{equation}\label{eq:237}
\int_M \mu_g \left\lbrace\varphi^{\,\dagger}_{\epsilon|ij} \varphi^{|ij}_\epsilon + \frac{1}{\ell_0^2} \varphi^{\,\dagger}_{\epsilon|j} \varphi^{|j}_\epsilon\right\rbrace e^{-2\mathcal{S}/\hbar} d^nx \leq \left(\frac{2m}{\hbar^2}\right)^2 \epsilon^2 \int_M \mu_g \varphi^{\,\dagger}_\epsilon \varphi_\epsilon e^{-2\mathcal{S}/\hbar} d^nx
\end{equation}
But a sequence, \(\varphi_{1/\ell}\), of normalizable functions whose gradients converge to zero in \(H^1 (M,\mu_g e^{-2\mathcal{S}/\hbar})\)-norm would have their gradients converging to zero almost everywhere and thus could not converge to a smooth limit orthogonal to the ground state.

Although global positivity of the Bakry-Emery Ricci tensor yields the quantitative lower bound (\ref{eq:236}) for the spectral gap it is almost surely not strictly needed for the existence of at least \textit{some} gap. Suppose for example that \(\mathcal{R}^{\mathcal{S}} = \mathcal{R}^{\mathcal{S}}_{ij}\; dx^i \otimes dx^j\) actually vanishes on some lower dimensional variety embedded in \textit{M} but is strictly positive on the complement. In view of the Hessian terms occurring in (\ref{eq:219}) and (\ref{eq:233}) one cannot simply arrive at a vanishing gap by assuming that the gradients of \(\overset{(1)}{\varphi}\) and \(\varphi_\epsilon\) respectively have their supports concentrated on the zero set of \(\mathcal{R}^{\mathcal{S}}\). To convert this intuition to a quantitative estimate however would require a more detailed analysis which we shall not pursue here. It is worth emphasizing though that (\ref{eq:222}) is almost certainly only a sufficient condition for the existence of a spectral gap.

The foregoing has primarily been a rather straightforward application of some familiar techniques of geometric analysis (e.g. Bochner identities, the Schwarz inequality, Rayleigh quotient variational arguments, spectral theory) to the specific context of Schr\"{o}dinger eigenvalue problems formulated on \textit{curved manifolds}. In the mathematical literature on metric measure spaces and Bakry-Emery curvature (c.f., \cite{Lott:2003,Wei:2009} and references cited therein) one often simply \textit{specifies} the metric measure factor (the analogue of our \(e^{-2\mathcal{S}/\hbar}\)) and requires it to have certain desirable  analytical properties (e.g., boundedness of \(\mathcal{S}\) or of its gradient) depending upon the theorem to be proven (e.g., a generalization of the Bonnet-Myers theorem implying manifold compactness). For us on the other hand \(\zero{\Psi} := \zero{N} e^{-\mathcal{S}/\hbar}\) is the ground state wave function for the Schr\"{o}dinger eigenvalue problem under study and the `background' Riemannian manifold \(\lbrace M,g\rbrace\) is non-compact for the cases of most interest. Thus, for us, \(\mathcal{S}\) is never freely specifiable but must satisfy the relevant differential equation and associated boundary conditions. In particular \(\mathcal{S}\) will not be bounded (since this is incompatible with a normalizable ground state on a non-compact manifold of infinite volume) nor will it (as already seen in elementary examples) have bounded gradient. Thus, unfortunately, many of the hypotheses imposed upon \(\mathcal{S}\) in the differential geometry literature are inappropriate for us and, of course, vice-versa.

Our ultimate aim, on the other hand, is to extend the ideas sketched above to the infinite dimensional `configuration' spaces (typically Riemannian Hilbert manifolds) arising in the functional analytic approach to certain quantum field theories. The first step in this direction is, of course, to make sense of the Schr\"{o}dinger operator itself. Whereas the covariant Hessian of a sufficiently smooth functional over such a space is still well-defined its corresponding metrical `trace', or `Laplacian', will not in general make sense without some suitable regularization since the Hessian under study will not, in general, be `trace class'. There have however been a number of proposals in the literature for how best to regularize the formal functional Laplacians that occur in the Schr\"{o}dinger operators for bosonic quantum field theories, in particular gauge theories. Singer, for example, proposed an elegant `zeta function' regularization scheme \cite{Singer:1981}. Later Hatfield \cite{Hatfield:1985} and quite recently Krug \cite{Krug:2014} have advanced alternative proposals, equally applicable to quantum gauge theories --- the latter, in particular, involving a gauge invariant `point splitting' technique.

If one tracks through the derivation above of the Bochner identity for the model, finite dimensional problem (\ref{eq:215}) and imagines extending this calculation to the field theoretic setting of primary interest herein, it becomes clear that the `same' regularized trace operation that arises in defining the functional Laplacian will act on the curvature tensor of the configuration space metric \textit{g} to yield its corresponding Ricci tensor. But the latter would also (as originally emphasized by Singer) not otherwise be well-defined since the curvature tensors of the relevant gauge theories are themselves \textit{not trace class}. On the other hand the needed regularization procedure also plays the vital role (uniquely in 3+1 spacetime dimensions) of allowing a \textit{length scale} to be introduced into the quantum formalism ---  a scale without which no hypothetical `mass gap' could even be expressed in terms of the naturally occurring constants of the theory (Planck's constant, the speed of light and the Yang-Mills coupling constant).

Another key element in the finite dimensional model problem sketched above is the occurrence of numerous integrals over the Riemannian configuration manifold \(\lbrace M,g\rbrace\). But thanks to the ubiquitous metric measure factor \(e^{-2\mathcal{S}/\hbar}\) these integrals are \textit{not} being taken with respect to the (Riemannian) Lebesgue measure \(\mu_g d^nx\) but instead with respect to the measure \(e^{-2\mathcal{S}/\hbar} \mu_g d^nx\)   which for a normalizable ground state, will give a finite total measure for the non-compact manifold \(\lbrace M,g\rbrace\). This distinction will prove to be crucial for our intended upgrade of the foregoing arguments to an infinite dimensional setting wherein Lebesgue measures no longer make sense but for which a normaliizable  ground state wave functional, together with its associated metric measure factor, is expected \textit{to exist}. Furthermore the integrals to be carried out have much in common with the (Euclidean-signature) functional integrals arising in the Feynman path integral formalism with the  important distinction that they only now involve the integrals over fields defined in \textit{one lower dimension} than for the Feynman formalism. More precisely the integrals envisioned here would  only be over `instantaneous' field configurations defined over say \(\mathbb{R}^3\) rather than over the (more technically problematic) spaces of field `paths' defined over \(\mathbb{R}^4\). This distinction is already dramatic in ordinary quantum mechanics wherein ordinary (finite dimensional) Lebesgue integrals must be upgraded to genuine functional integrals in passing to the Feynman path integral formalism.

The naturally occurring metric measure factor \(e^{-2\mathcal{S}/\hbar}\), which yields non-compact metric measure spaces \(\lbrace M, g, e^{-2\mathcal{S}/\hbar}\rbrace\) of finite total measure, is the principal feature in our setup that allows us to contemplate extending the foregoing arguments to interesting infinite dimensional  settings. Its absence was a key shortcoming in the original Singer proposal for exploiting Lichnerowicz type arguments for the existence of a spectral gap.\footnote{Singer, of course, was well aware of this limitation and does not explicitly mention the mass gap problem as motivation or the Lichnerowicz spectral gap estimate as a potentially useful tool in his original paper. He did however mention these both informally during a lecture at the Yale Mathematics Department in 1981 at which the senior author (V.M.) was present. Without this fortuitous clarification we would not have appreciated the potential for generalizing Singer's argument to allow for a normalizable ground state on a non-compact manifold.\label{note:04}} To carry out the needed extension (to field theoretic problems) in a technically precise way, on the other hand, would take us much further afield, analytically, than  we are currently prepared to wander. Our intuition though is that such developments should be mathematically possible if one could gain sufficient control over the fundamental, logarithm functional \(\mathcal{S}\). This latter step is, in large part, the aim of our Euclidean-signature semi-classical program which, for the convenience of the reader, we briefly review in the section to follow. 

\section{Euclidean Signature Semi-Classical Methods}
\label{sec:euclidean}
\subsection{Quantum Mechanical Systems}
\label{subsec:QMS}
Elegant `microlocal analysis' methods have long since been developed for the study of Schr{\"o}dinger operators of the form \eqref{eq:202} in the special cases for which \(M \approx \mathbb{R}^n\), the metric \textit{g} is flat and for which the potential energy function \(V:M \rightarrow \mathbb{R}\) is of a suitable `non-linear oscillatory' type \cite{Moncrief:2012,Dimassi:1999,Helfer:1988,Helfer:1984}. These methods\footnote{For reasons to be clarified below we here follow a recent reformulation of the traditional microlocal approach developed by the authors in \cite{Moncrief:2012}.} begin with an ansatz for the ground state wave function of the form
\begin{equation}\label{eq:401}
\overset{(0)}{\Psi}_{\!\hbar} (\mathbf{x}) = N_\hbar\; e^{-\mathcal{S}_\hbar (\mathbf{x})/\hbar}
\end{equation}
and proceed to derive asymptotic expansions for the logarithm, \(\mathcal{S}_\hbar:\mathbb{R}^n \rightarrow \mathbb{R}\), expressed formally as a power series in Planck's constant,
\begin{equation}\label{eq:402}
\begin{split}
\mathcal{S}_\hbar (\mathbf{x}) &\simeq \mathcal{S}_{(0)} (\mathbf{x}) + \hbar\mathcal{S}_{(1)} (\mathbf{x}) + \frac{\hbar^2}{2!} \mathcal{S}_{(2)} (\mathbf{x}) \\
 &+ \cdots + \frac{\hbar^n}{n!} \mathcal{S}_{(n)} (\mathbf{x}) + \cdots,
\end{split}
\end{equation}
together with the associated ground state energy eigenvalue \(\overset{(0)}{E}_{\!\hbar}\) expressed as
\begin{equation}\label{eq:403}
\overset{(0)}{E}_{\!\hbar} \simeq \hbar (\overset{(0)}{\mathcal{E}}_{\!(0)} + \hbar\overset{(0)}{\mathcal{E}}_{\!(1)} + \frac{\hbar^2}{2!}\overset{(0)}{\mathcal{E}}_{\!(2)} + \cdots
 + \frac{\hbar^n}{n!}\overset{(0)}{\mathcal{E}}_{\!(n)} + \cdots).
\end{equation}
\(N_\hbar\) is a corresponding (for us inessential) normalization constant which one could always evaluate at any (finite) level of the calculation.

When the above ans{\"a}tze are substituted into the time-independent Schr{\"o}dinger equation and the latter is required to hold order-by-order in powers of \(\hbar\) the leading order term in the expansion (\ref{eq:402}) is found to satisfy an \textit{inverted-potential-vanishing-energy} `Hamilton-Jacobi' equation given by
\begin{equation}\label{eq:404}
\frac{1}{2m} g^{ij} \mathcal{S}_{(0),i} \mathcal{S}_{(0),i} - V = 0.
\end{equation}
For a large class of (non-linear oscillatory) potential energy functions and when \textit{g} is flat (with \(g = \sum_{i=1}^n dx^i \otimes dx^i\)) this equation can be proven to have a globally-defined, smooth, positive `fundamental solution' that is unique up to a (trivial) additive constant. In particular this is true whenever
\begin{enumerate}
\item \textit{V} is smooth, non-negative and has a unique global minimum attained at the origin of \(\mathbb{R}^n\) where \textit{V} vanishes,
\item \textit{V} can be expressed as
\begin{equation}\label{eq:405}
V(x^1, \ldots ,x^n) = \frac{1}{2} \sum_{i=1}^n m\; \omega_i^2 (x^i)^2 + A(x^1, \ldots ,x^n)
\end{equation}
where each of the `frequencies' \(\omega_i > 0\) for \(i \in \lbrace 1, \ldots ,n\rbrace\) and wherein the smooth function \(A:\mathbb{R}^n \rightarrow \mathbb{R}\) satisfies
\begin{equation}\label{eq:406}
A(0, \ldots ,0) = \frac{\partial A(0, \ldots ,0)}{\partial x^i} = \frac{\partial^2 A(0, \ldots , 0)}{\partial x^i\partial x^j} = 0\quad \forall\; i,j\; \in \lbrace 1, \ldots ,n\rbrace
\end{equation}
and the \textit{coercivity} condition
\begin{equation}\label{eq:407}
A(x^1, \ldots ,x^n) \geq -\frac{1}{2} m \sum_{i=1}^n \lambda_i^2 (x^i)^2\quad \forall\; (x^1, \ldots ,x^n)\; \in \mathbb{R}^n
\end{equation}
and for some constants \(\lbrace\lambda_i\rbrace\) such that \(\lambda_i^2 < \omega_i^2\quad \forall\; i\; \in \lbrace 1, \ldots ,n\rbrace\), and
\item \textit{V} satisfies the \textit{convexity} condition
\begin{equation}\label{eq:408}
\begin{split}
&\sum_{i,j=1}^n \frac{\partial^2 V(x^1, \ldots ,x^n)}{\partial x^i\partial x^j} \xi^i\xi^j \geq 0\\
&\hphantom{\sum_{i,j=1}^n}\forall\; (x^1, \ldots ,x^n)\; \in \mathbb{R}^n\; \hbox{and all}\\
&\hphantom{\sum_{i,j=1}^n}(\xi^1, \ldots ,\xi^n)\; \in \mathbb{R}^n.
\end{split}
\end{equation}
\end{enumerate}
Since only the sufficiency of these conditions was actually established in \cite{Moncrief:2012} it is quite conceivable that a satisfactory \textit{fundamental solution} to Eq.~(\ref{eq:404}) exists under weaker hypotheses on the potential energy.

Our approach to proving the existence of a global, smooth fundamental solution to the (inverted-potential-vanishing-energy) Hamilton-Jacobi equation
\begin{equation}\label{eq:409}
\frac{1}{2m} \nabla\mathcal{S}_{(0)} \cdot \nabla\mathcal{S}_{(0)} - V = 0
\end{equation}
is quite different from that developed previously in the microlocal literature but has the advantage of being applicable to certain field theoretic problems whereas it seems the latter does not\footnote{The reasons for this apparent limitation are clarified in the discussion to follow.}.

To establish the existence of \(\mathcal{S}_{(0)}\) we began by proving that the (inverted potential) action functional
\begin{equation}\label{eq:410}
\begin{split}
\mathcal{I}_{ip}[\gamma] &:= \int_{-\infty}^0 \left\lbrace\frac{1}{2} m \sum_{i=1}^n \left\lbrack(\dot{x}^i(t))^2 + \omega_i^2 (x^i(t))^2\right\rbrack\right.\\
&{} + A\left.(x^i(t), \ldots, x^n(t))\vphantom{\sum_{i=1}^n}\right\rbrace\; dt,
\end{split}
\end{equation}
defined on an appropriate Sobolev space of curves \(\gamma :(-\infty,0] \rightarrow \mathbb{R}^n\), has a unique minimizer, \(\gamma_{\mathbf{x}}\), for any choice of boundary data
\begin{equation}\label{eq:411}
\mathbf{x} = (x^1, \ldots ,x^n) = \lim_{t\nearrow 0} \gamma_{\mathbf{x}}(t)\; \in\; \mathbb{R}^n
\end{equation}
and that this minimizer always obeys
\begin{equation}\label{eq:412}
\lim_{t\searrow -\infty} \gamma_{\mathbf{x}}(t) = (0, \ldots ,0).
\end{equation}
We then showed that every such minimizing curve is smooth and satisfies the (\textit{inverted potential}) Euler-Lagrange equation
\begin{equation}\label{eq:413}
m \frac{d^2}{dt^2} \gamma_{\mathbf{x}}^i(t) = \frac{\partial V}{\partial x^i} (\gamma_{\mathbf{x}}(t))
\end{equation}
with vanishing (inverted potential) energy
\begin{equation}\label{eq:414}
\begin{split}
E_{ip} (\gamma_{\mathbf{x}}(t),\dot{\gamma}_{\mathbf{x}}(t)) &:= \frac{1}{2}m \sum_{i=1}^n (\dot{\gamma}_{\mathbf{x}}^i(t))^2 - V(\gamma_{\mathbf{x}}(t))\\
&= 0\quad \forall\; t\; \in (-\infty,0] := I.
\end{split}
\end{equation}
Setting \(\mathcal{S}_{(0)}(\mathbf{x}) := \mathcal{I}_{ip} [\gamma_{\mathbf{x}}]\) for each \(\mathbf{x}\; \in \mathbb{R}^n\) we proceeded to prove, using the (Banach space) implicit function theorem, that the \(\mathcal{S}_{(0)}:\mathbb{R}^n \rightarrow \mathbb{R}\), so-defined, satisfies the Hamilton-Jacobi equation
\begin{equation}\label{eq:415}
\frac{1}{2m} |\nabla\mathcal{S}_{(0)}|^2 - V = 0
\end{equation}
globally on \(\mathbb{R}^n\) and regenerates the minimizers \(\gamma_{\mathbf{x}}\) as the integral curves of its gradient (semi-)flow in the sense that
\begin{equation}\label{eq:416}
\begin{split}
\frac{d}{dt} \gamma_{\mathbf{x}}(t) &= \frac{1}{m} \nabla\mathcal{S}_{(0)} (\gamma_{\mathbf{x}}(t))\\
 &\forall\; t\; \in I := (-\infty,0]\; \hbox{ and }\\
 &\forall\; \mathbf{x}\; \in \mathbb{R}^n
\end{split}
\end{equation}
Actually each such integral curves \(\gamma_{\mathbf{x}}:I \rightarrow \mathbb{R}^n\) extends to a larger interval, \((-\infty , t^\ast (\gamma_{\mathbf{x}}))\) with \(0 < t^\ast (\gamma_{\mathbf{x}}) \leq \infty\; \forall\; \mathbf{x}\; \in \mathbb{R}^n\) but since, in general, \(t^\ast (\gamma_{\mathbf{x}}) < \infty\) we only have a semi-flow rather than a complete flow generated by \(\frac{1}{m} \nabla\mathcal{S}_{(0)}\). Purely \textit{harmonic} oscillations on the other hand (for which \(A(x^1, \ldots ,x^n) = 0\)) are an exception, having \(t^\ast (\gamma_{\mathbf{x}}) = \infty\; \forall\; \mathbf{x}\; \in \mathbf{R}^n\).

Among the additional properties established for \(\mathcal{S}_{(0)}\) were the Taylor expansion formulas
\begin{align}
\mathcal{S}_{(0)}(\mathbf{x}) &= \frac{1}{2} m \sum_{i=1}^n \omega_i(x^i)^2 + O(|\mathbf{x}|^3),\label{eq:417}\\
\partial_j \mathcal{S}_{(0)}(\mathbf{x})  &= m\omega_jx^j + O(|\mathbf{x}|^2)\label{eq:418}\\
\intertext{and}
\partial_j\partial_k\mathcal{S}_{(0)}(\mathbf{x}) &= m\omega_k\delta_j^k + O(|\mathbf{x}|),\label{eq:419}
\end{align}
where here (exceptionally) no sum on the repeated index is to be taken, and the global lower bound
\begin{equation}\label{eq:420}
\mathcal{S}_{(0)}(\mathbf{x}) \geq \mathcal{S}_{(0)}^\ast := \frac{1}{2} m \sum_{i=1}^n \nu_i (x^i)^2
\end{equation}
where \(\nu_i := \sqrt{\omega_i^2 - \lambda_i^2} > 0\; \forall\; i\; \in \lbrace 1, \ldots ,n\rbrace\). Note especially that this last inequality guarantees that, in particular, \(e^{-\mathcal{S}_{(0)}/\hbar}\) will always be normalizable on \(\lbrace\mathbb{R}^n,g = \sum_{i=1}^n dx^i \otimes dx^i\rbrace\).

The higher order `quantum corrections' to \(\mathcal{S}^{(0)}\) (i.e., the functions \(\mathcal{S}_{(k)}\) for \(k = 1, 2, \ldots\)) can now be computed through the systematic integration of a sequence of (first order, linear) `transport equations', derived from Schr{\"o}dinger's equation, along the integral curves of the gradient (semi-)flow generated by \(\mathcal{S}_{(0)}\). The natural demand for global smoothness of these quantum `loop corrections' forces the (heretofore undetermined) energy coefficients \(\lbrace\overset{(0)}{\mathcal{E}}_{\!(0)}, \overset{(0)}{\mathcal{E}}_{\!(1)}, \overset{(0)}{\mathcal{E}}_{\!(2)}, \ldots\rbrace\) all to take on specific, computable values.

Excited states can now be analyzed by substituting the ansatz
\begin{equation}\label{eq:421}
\overset{(\ast)}{\Psi}_{\!\hbar} (\mathbf{x}) = \overset{(\ast)}{\phi}_{\!\hbar} (\mathbf{x}) e^{-\mathcal{S}_\hbar(\mathbf{x})/\hbar}
\end{equation}
into the time independent Schr{\"o}dinger equation and formally expanding the unknown wave functions \(\overset{(\ast)}{\phi}_{\!\hbar}\) and energy eigenvalues \(\overset{(\ast)}{E}_{\!\hbar}\) in powers of \(\hbar\) via
\begin{align}
\overset{(\ast)}{\phi}_{\!\hbar} &\simeq \overset{(\ast)}{\phi}_{\!(0)} + \hbar\overset{(\ast)}{\phi}_{\!(1)} + \frac{\hbar^2}{2!}\overset{(\ast)}{\phi}_{(2)} + \cdots\label{eq:422}\\
\overset{(\ast)}{E}_{\!\hbar} &\simeq \hbar\overset{(\ast)}{\mathcal{E}}_{\!\hbar} \simeq \hbar \left(\overset{(\ast)}{\mathcal{E}}_{\!(0)} + \hbar\overset{(\ast)}{\mathcal{E}}_{\!(1)} + \frac{\hbar^2}{2!}\overset{(\ast)}{\mathcal{E}}_{\!(2)} + \cdots\right) \label{eq:423}
\end{align}
while retaining the `universal' factor \(e^{-\mathcal{S}_\hbar (\mathbf{x})/\hbar}\) determined by the ground state calculations.

From the leading order analysis one finds that these excited state expansions naturally allow themselves to be labelled by an \textit{n}-tuple \(\mathbf{m} = (m_1, m_2, \ldots , m_n)\) of non-negative integer `quantum numbers', \(m_i\), so that the foregoing notation can be refined to
\begin{align}
\overset{(\mathbf{m})}{\Psi}_{\!\hbar} (\mathbf{x}) &= \overset{(\mathbf{m})}{\phi}_{\!\hbar} (\mathbf{x}) e^{-\mathcal{S}_\hbar (\mathbf{x})/\hbar}\label{eq:424}\\
\intertext{and}
\overset{(\mathbf{m})}{E}_{\!\hbar} &= \hbar\overset{(\mathbf{m})}{\mathcal{E}}_{\!\hbar}\label{eq:425}
\end{align}
with \(\overset{(\mathbf{m})}{\varphi}_{\!\hbar}\) and \(\overset{(\mathbf{m})}{\mathcal{E}}_{\!\hbar}\) expanded as before. Using methods that are already well-known from the microlocal literature \cite{Dimassi:1999} but slightly modified to accord with our setup \cite{Moncrief:2012} one can now compute all the coefficients \(\lbrace\overset{(\mathbf{m})}{\phi}_{\!(k)}, \overset{(\mathbf{m})}{\mathcal{E}}_{\!(k)}, k = 0, 1, 2 \ldots\rbrace\) through the solution of a sequence of linear, first order transport equation integrated along the semi-flow generated by \(\mathcal{S}_{(0)}\).

A key feature of this program, when applied to an n-dimensional \textit{harmonic} oscillator, is that it regenerates all the well-known, \textit{exact} results for both ground and excited states, correctly capturing not only the eigenvalues but the \textit{exact eigenfunctions} as well \cite{Moncrief:2012,Dimassi:1999,Helfer:1988}. One finds for example that the fundamental solution to the relevant (inverted-potential-vanishing-energy) Hamilton-Jacobi equation, for an n-dimensional oscillator (with mass \textit{m} and (strictly positive) oscillation frequencies \(\lbrace\omega_i\rbrace\)) is given by
\begin{equation}\label{eq:426}
\mathcal{S}_{(0)} (\mathbf{x}) = \frac{1}{2} m \sum_{i=1}^n \omega_i (x^i)^2
\end{equation}
and that all higher order corrections to the logarithm of the ground state wave function vanish identically leaving the familiar gaussian
\begin{equation}\label{eq:427}
\overset{(0)}{\Psi}_{\!\hbar} (\mathbf{x}) = \overset{(0)}{N}_{\!\hbar}\; e^{-\frac{m}{2\hbar} \sum_{i=1}^n \omega_i (x^i)^2}
\end{equation}
where \(\mathbf{x} = (x^1, \ldots ,x^n)\) and \(\overset{(0)}{N}_{\!\hbar}\) is a normalization constant.

The construction of excited states begins with the observation that the only globally regular solutions to the corresponding, leading order `transport equation' are composed of the monomials
\begin{equation}\label{eq:428}
\overset{(\mathbf{m})}{\phi}_{\!(0)} (\mathbf{x}) = (x^1)^{m_1} (x^2)^{m_2} \cdots (x^n)^{m_n},
\end{equation}
where \(\mathbf{m} = (m_1, m_2, \ldots , m_n)\) is an n-tuple of non-negative integers with \(|m| := \sum_{i=1}^n m_i > 0\), and proceeds after a finite number of unequivocal steps, to assemble the exact excited eigenstate prefactor
\begin{equation}\label{eq:429}
\begin{split}
\overset{(\mathbf{m})}{\phi}_{\!\hbar} (\mathbf{x}) &= \overset{(\mathbf{m})}{N}_{\!\hbar} H_{m_1} \left(\sqrt{\frac{m\omega_1}{\hbar}} x^1\right) H_{m_2} \left(\sqrt{\frac{m\omega_2}{\hbar}} x^2\right)\\
 &\cdots H_{m_n} \left(\sqrt{\frac{m\omega_n}{\hbar}} x^n\right)
\end{split}
\end{equation}
where \(H_k\) is the Hermite polynomial of order \textit{k} (and \(\overset{(\mathbf{m})}{N}_{\!k}\) is the corresponding normalization constant) \cite{Moncrief:2012,Dimassi:1999,Helfer:1988}.

While there is nothing especially astonishing about being able to rederive such well-known, exact results in a different way, we invite the reader to compare them with those obtainable via the textbook WKB methods of the physics literature \cite{Brack:2008,Ozorio:1988}. Even for purely \textit{harmonic} oscillators conventional WKB methods yield only rather rough approximations to the wave functions and are, in any case, practically limited to one-dimensional problems and to those reducible to such through a separation of variables. The lesser known Einstein Brillouin Keller (or EBK) extension of the traditional semi-classical methods does apply to higher (finite-)dimensional systems but only to those that are completely integrable at the classical level \cite{Stone:2005}. In sharp contrast to these well-established approximation methods the (Euclidean signature\footnote{The significance of this qualifying expression will become clear when we turn to field theoretic problems.}) semi-classical program that we are advocating here requires neither classical integrability nor (as we shall see) finite dimensionality for its implementation.

As was discussed in the concluding section of Ref.~\cite{Moncrief:2012} our fundamental solution, \(\mathcal{S}_{(0)} (\mathbf{x})\), to the (inverted-potential-vanishing-energy) Hamilton-Jacobi equation for a coupled system of nonlinear oscillators has a natural geometric interpretation. The graphs, in the associated phase space \(T^\ast\mathbb{R}^n\), of its positive and negative gradients correspond precisely to the stable \((W^{\mathrm{s}}(p) \subset T^\ast\mathbb{R}^n)\) and unstable \((W^{\mathrm{u}}(p) \subset T^\ast\mathbb{R}^n)\) Lagrangian submanifolds of the assumed, isolated equilibrium point \(p \in T^\ast\mathbb{R}^n\):
\begin{align}
W^{\mathrm{u}}(p) &= \left\lbrace (\mathbf{x},\mathbf{p}):\mathbf{x} \in \mathbb{R}^n, \mathbf{p} = \nabla\mathcal{S}_{(0)} (\mathbf{x})\right\rbrace\label{eq:430}\\
W^{\mathrm{s}}(p) &= \left\lbrace (\mathbf{x},\mathbf{p}):\mathbf{x} \in \mathbb{R}^n, \mathbf{p} = -\nabla\mathcal{S}_{(0)} (\mathbf{x})\right\rbrace\label{eq:431}
\end{align}

Another result established for the aforementioned nonlinear oscillators of Ref.~\cite{Moncrief:2012} is that the first quantum `loop correction', \(\mathcal{S}_{(1)} (x^1, \ldots , x^n)\), to the (`tree level') fundamental solution, \(\mathcal{S}_{(0)} (x^1, \ldots , x^n)\), also has a natural geometric interpretation in terms of `Sternberg coordinates' for the gradient (semi-)flow generated by this fundamental solution. Sternberg coordinates, by construction, linearize the Hamilton-Jacobi flow equation
\begin{align}
m \frac{dx^i (t)}{dt} &= \frac{\partial\mathcal{S}_{(0)}}{\partial x^i} (x^1(t), \ldots , x^n(t))\label{eq:432}\\
\intertext{to the form}
\frac{dy^i(t)}{dt} &= \omega_i y^i(t)\quad \hbox{(no sum on \textit{i})}\label{eq:433}
\end{align}
through, as was proven in Ref.~\cite{Moncrief:2012}, the application of a global diffeomorphism
\begin{align}
&\mu:\mathbb{R}^n \rightarrow \mu(\mathbb{R}^n) \subset \mathbb{R}^n = \left\lbrace (y^1, \ldots , y^n)\right\rbrace,\label{eq:434}\\
&\mathbf{x} \mapsto \mu(\mathbf{x}) = \left\lbrace y^1(\mathbf{x}), \ldots , y^n(\mathbf{x})\right\rbrace\label{eq:435}
\end{align}
that maps \(\mathbb{R}^n\) to a star-shaped domain \(K = \mu(\mathbb{R}^n) \subset \mathbb{R}^n\) with \(\mu^{-1}(K) \approx \mathbb{R}^n = \left\lbrace (x^1, \ldots , x^n)\right\rbrace\).

Though not strictly needed for the constructions of Ref.~\cite{Moncrief:2012}, Sternberg coordinates have the natural feature of generating a Jacobian determinant for the Hilbert-space integration measure that \textit{exactly cancels} the contribution of the first quantum `loop correction', \(\mathcal{S}_{(1)}(\mathbf{x})\), to inner product calculations, taking, for example,
\begin{align}
\begin{split}
\left\langle\overset{(\mathbf{m})}{\Psi},\overset{(\mathbf{m})}{\Psi}\right\rangle &:= \int_{\mathbb{R}^n} \left|\overset{(\mathbf{m})}{\Psi}(\mathbf{x})\right|^2\; d^n x\\
&= \int_{\mu(\mathbb{R}^n)} \left|\overset{(\mathbf{m})}{\Psi} \circ \mu^{-1}(\mathbf{y})\right|^2\; \sqrt{\det{g_{\ast\ast}}(\mathbf{y})}\; d^n y\label{eq:436}
\end{split}
\intertext{to the form}
\begin{split}
\left\langle\overset{(\mathbf{m})}{\Psi},\overset{(\mathbf{m})}{\Psi}\right\rangle &= \int_{\mu(\mathbb{R}^n)} \left|\left\lbrack\overset{(\mathbf{m})}{\varphi} e^{\frac{-\mathcal{S}_{(0)}}{\hbar}-\frac{\hbar}{2!}\mathcal{S}_{(2)}+\cdots}\right\rbrack \circ \mu^{-1}(\mathbf{y})\right|^2\\
&\hphantom{=} \sqrt{\det{g_{\ast\ast}}(\mathbf{0})}\; d^n y\label{eq:437}
\end{split}
\end{align}
where, in the last integral, the contribution of \(\mathcal{S}_{(1)} \circ \mu^{-1}(\mathbf{y})\) to the wave function
\begin{equation}\label{eq:438}
\overset{(\mathbf{m})}{\Psi} \circ \mu^{-1} (\mathbf{y}) = \overset{(\mathbf{m})}{\varphi} e^{\frac{-\mathcal{S}_{(0)}}{\hbar}-\mathcal{S}_{(1)} - \frac{\hbar}{2!}\mathcal{S}_{(2)}\cdots}\; \circ \mu^{-1} (\mathbf{y})
\end{equation}
has precisely cancelled the non-Cartesian measure factor \(\sqrt{\det{g_{\ast\ast}} (\mathbf{y})}\), leaving the constant (Euclidean) factor \(\sqrt{\det{g_{\ast\ast}}(\mathbf{0})}\) in its place. Roughly speaking therefore, this role of \(\mathcal{S}_{(1)}\) is to `flatten out' the Sternberg coordinate volume element, reducing it to ordinary Lebesgue measure (albeit only over the star-shaped domain \(\mu(\mathbb{R}^n)\)), by exactly cancelling the Jacobian determinant that arises from the coordinate transformation.

For the nonlinear oscillators discussed in Ref.~\cite{Moncrief:2012}, Sternberg coordinates also have the remarkable property of allowing the leading order transport equation for \textit{excited states} to be solved in closed form. Indeed, the regular solutions to this equation are comprised of the monomials
\begin{equation}\label{eq:439}
\overset{(\mathbf{m})}{\varphi}_{\!(0)} (\mathbf{y}) = (y^1)^{m_1} (y^2)^{m_2} \cdots (y^n)^{m_n}
\end{equation}
wherein, precisely as for the harmonic case, the \(m_i\) are non-negative integers with \(|\mathbf{m}| := \sum_{i=1}^n m_i > 0\). On the other hand the higher order corrections, \(\left\lbrace\overset{(\mathbf{m})}{\varphi}_{\!(k)}(\mathbf{y}); k = 1, 2, \ldots\right\rbrace\), to these excited state prefactors will not in general terminate at a finite order as they do for strictly \textit{harmonic} oscillators but they are nevertheless systematically computable through the sequential integration of a set of well-understood linear transport equations \cite{Moncrief:2012,Dimassi:1999}. Formal expansions (in powers of \(\hbar\)) for the corresponding (ground and excited state) energy \textit{eigenvalues} are uniquely determined by the demand for global regularity of the associated eigenfunction expansions. More precisely one finds, upon integrating the relevant transport equation at a given order, that the only potential breakdown of smoothness for the solution would necessarily occur at the `origin' \(\mathbf{x} = 0\) (chosen here to coincide with the global minimum of the potential energy) but that this loss of regularity can always be uniquely avoided by an appropriate choice of eigenvalue coefficient at the corresponding order.

A number of explicit calculations of the eigenfunctions and eigenvalues for a family of 1-dimensional \textit{anharmonic} oscillators of quartic, sectic, octic, and dectic types were carried out in Ref.~\cite{Moncrief:2012} and compared with the corresponding results from conventional Rayleigh/Schr{\"o}dinger perturbation theory. To the orders considered (and, conjecturally, to all orders) our eigenvalue expansions agreed with those of Rayleigh/Schr{\"o}dinger theory whereas our wave functions, even at leading order, more accurately captured the more-rapid-than-gaussian decay known rigorously to hold for the exact solutions to these problems. For the quartic oscillator in particular our results strongly suggested that both the ground state energy eigenvalue expansion and its associated wave function expansion are Borel summable to yield natural candidates for the actual exact ground state solution and its energy.

Remarkably all of the integrals involved in computing the quantum corrections \(\left\lbrace\mathcal{S}_{(1)}, \mathcal{S}_{(2)}, \mathcal{S}_{(3)}, \cdots\right\rbrace\) to \(\mathcal{S}_{(0)}\) (up to the highest order computed in \cite{Moncrief:2012}, namely \(\mathcal{S}_{(25)}\)) were expressible explicitly in terms of elementary functions for the \textit{quartic} and \textit{sectic} oscillators whereas for the octic and dectic cases some (but not all) of the quantum corrections required, in addition, hypergeometric functions for their evaluation. It seems plausible to conjecture that these patterns persist to all orders in \(\hbar\) and thus, for the quartic and sectic\footnote{These results were subsequently extended to significantly higher orders by P.~Tang~\cite{misc-orbit02}.} cases in particular, lead to formal expansions for \(\mathcal{S}_\hbar\) in terms of elementary functions. The evidence supporting the conjectured Borel summability of this formal expansion in the quartic case is discussed in detail in Section~V.A. of \cite{Moncrief:2012}.

For the Lagrangians normally considered in classical mechanics it would not be feasible to define their corresponding action functionals over (semi-) infinite domains, as we have done, since the integrals involved, when evaluated on solutions to the Euler-Lagrange equations, would almost never converge. It is only because of the special nature of our problem, with its inverted potential energy function and associated boundary conditions, that we could define a convergent action integral for the class of curves of interest and use this functional to determine corresponding minimizers.

A remarkable feature of our construction, given the hypotheses of convexity and coercivity imposed upon the potential energy \(V(\mathbf{x})\), is that it led to a \textit{globally smooth} solution to the corresponding Hamilton-Jacobi equation. Normally the solutions to a Hamilton-Jacobi equation in mechanics fail to exist globally, even for rather elementary problems, because of the occurrence of caustics in the associated families of solution curves. For our problem however caustics were non-existent for the (semi-)flow generated by the gradient of \(\mathcal{S}_{(0)}(\mathbf{x})\). The basic reason for this was the inverted potential character of the forces considered which led to the development of diverging (in the future time direction) solution curves having, in effect, uniformly positive Lyapunov exponents that served to prevent the occurrence of caustics altogether.

By contrast, the more conventional approach (in the physics literature) to semi-classical methods leads instead to a standard (non-inverted-potential-non-vanishing-energy) Hamilton-Jacobi equation for which, especially in higher dimensions, caustics are virtually unavoidable and for which, even in their absence, a nontrivial matching of solutions across the boundary separating classically allowed and classically forbidden regions must be performed. While Maslov and others have developed elegant methods for dealing with these complications \cite{Maslov:1981} their techniques are more appropriate in the short wavelength limit wherein wave packets of highly excited states are evolved for finite time intervals. On the other hand our approach is aimed at the ground and lower excited states though, in principle, it is not limited thereto.

As we have already mentioned though, our approach is a natural variation of one that has been extensively developed in the microlocal analysis literature but it also differs from this innovative work in fundamental ways that are crucial for our ultimate, intended application to field theoretic problems. In the microlocal approach \cite{Dimassi:1999,Helfer:1988,Helfer:1984} one begins by analyzing the (classical, inverted potential) dynamics locally, near an equilibrium, by appealing to the stable manifold theorem of mechanics \cite{Abraham:1978}. One then shows, by a separate argument, that, for an equilibrium \textit{p} (lying in some neighborhood \(U \subset \mathbb{R}^n\)) the corresponding stable (\(W^s(p) \subset T^\ast U\)) and unstable (\(W^u(p) \subset T^\ast U\)) submanifolds of the associated phase space \(T^\ast U\) are in fact Lagrangian submanifolds that can be characterized as graphs of the (positive and negative) gradients of a smooth function \(\phi:U \rightarrow \mathbb{R}\):
\begin{align}
W^s(p) &= \left\lbrace (\mathbf{x},\mathbf{p}) | \mathbf{x} \in U, \mathbf{p} = \nabla\phi (\mathbf{x})\right\rbrace\label{eq:440}\\
W^u(p) &= \left\lbrace (\mathbf{x},\mathbf{p}) | \mathbf{x} \in U, \mathbf{p} = -\nabla\phi (\mathbf{x})\right\rbrace .\label{eq:441}
\end{align}
This function is shown to satisfy a certain `eikonal' equation (equivalent to our inverted-potential-vanishing-energy Hamilton-Jacobi equation restricted to \textit{U}) and \(\phi (\mathbf{x})\) itself is, of course, nothing but the (locally defined) analogue of our action function \(\mathcal{S}_{(0)}(\mathbf{x})\). A further argument is then needed to extend \(\phi(x)\) to a solution globally defined on \(\mathbb{R}^n\).

The potential energies, \(V(\mathbf{x})\), dealt with in the microlocal literature often entail multiple local minima, or ``wells'', for which our global convexity and coercivity hypotheses are not appropriate. Much of the detailed analysis therein involves a careful matching of locally defined approximate solutions (constructed on suitable neighborhoods of each well) to yield global asymptotic approximations to the eigenvalues and eigenfunctions for such problems. Since, however, we are focussed primarily on potential energies having single wells (corresponding to unique classical ``vacuum states''), many of the technical features of this elegant analysis are not directly relevant to the issues of interest herein.

For the case of a single well, however, we have essentially unified and globalized several of the, aforementioned, local arguments, replacing them with the integrated study of the properties of the (inverted potential) action functional (\ref{eq:410}). When one turns from finite dimensional problems to field theoretic ones \cite{Marini:2016,Maitra:inprep} this change of analytical strategy will be seen to play an absolutely crucial role. For the typical (relativistic, bosonic) field theories of interest to us in this context, the Euler Lagrange equations for the corresponding, inverted potential action functionals that now arise are the \textit{Euclidean signature}, elliptic analogues of the Lorentzian signature, hyperbolic field equations that one is endeavoring to quantize. While generalizations of the aforementioned stable manifold theorem do exist for certain types of infinite dimensional dynamical systems, the elliptic field equations of interest to us do not correspond to well-defined dynamical systems \textit{at all}. In particular their associated Cauchy initial value problems are never well-posed. This is the main reason, in our opinion, why the traditional microlocal methods have not heretofore been applicable to quantum field theories.

On the other hand the direct method of the calculus of variations is applicable to the Euclidean signature action functionals of interest to us here and allows one to generalize the principle arguments discussed above to a natural infinite dimensional setting.

\subsection{Interacting Scalar Fields}
\label{subsec:interacting_scalar}
For a first glimpse at how these techniques can be applied to relativistic quantum field theories consider the formal Schr{\"o}dinger operator for the massive, quartically self-interacting scalar field on (3+1 dimensional) Minkowski spacetime given by
\begin{equation}\label{eq:442}
\begin{split}
\hat{H} &= \int_{\mathbb{R}^3} \left\lbrace-\frac{\hbar^2}{2}\; \frac{\delta^2}{\delta\phi^2(\mathbf{x})} + \frac{1}{2} \nabla\phi (\mathbf{x}) \cdot \nabla\phi (\mathbf{x}) \right.\\
 &{} \left.+ \frac{m^2}{2} \phi^2 (\mathbf{x}) + \lambda\phi^4 (\mathbf{x})\right\rbrace\; d^3x
\end{split}
\end{equation}
where \textit{m} and \(\lambda\) are constants \(> 0\). Though the functional Laplacian term, in particular, requires regularization to be well-defined, the influence of this regularization will only be felt at the level of quantum `loop' corrections and not for the `tree level' determination of a fundamental solution, \(\mathcal{S}_{(0)} [\phi(\cdot)]\), to the `vanishing-energy-Euclidean-signature' functional Hamilton-Jacobi equation given by
\begin{equation}\label{eq:443}
\begin{split}
\int_{\mathbb{R}^3} & \left\lbrace\frac{1}{2} \frac{\delta\mathcal{S}_{(0)}}{\delta\phi (\mathbf{x})} \frac{\delta\mathcal{S}_{(0)}}{\delta\phi (\mathbf{x})} - \frac{1}{2} \nabla\phi (\mathbf{x}) \cdot \nabla\phi (\mathbf{x})\right.\\
 &{} \left.- \frac{m^2}{2} \phi^2 (\mathbf{x}) - \lambda\phi^4 (\mathbf{x})\right\rbrace\; d^3 x = 0.
\end{split}
\end{equation}
As in the quantum mechanical examples discussed above this equation arises, at leading order, from substituting the ground state wave functional ansatz
\begin{equation}\label{eq:444}
\overset{(0)}{\Psi}_{\!\hbar} [\phi (\cdot)] = N_\hbar e^{-\mathcal{S}_\hbar[\phi(\cdot)]/\hbar}
\end{equation}
into the time-independent Schr{\"o}dinger equation
\begin{equation}\label{eq:445}
\hat{H}\overset{(0)}{\Psi}_{\!\hbar} = \overset{(0)}{E}_{\!\hbar}\overset{(0)}{\Psi}_{\!\hbar},
\end{equation}
and demanding satisfaction, order-by-order in powers of \(\hbar\), relative to the formal expansions
\begin{align}
\begin{split}
\mathcal{S}_\hbar [\phi(\cdot)] &\simeq \mathcal{S}_{(0)} [\phi(\cdot)] + \hbar\mathcal{S}_{(1)} [\phi(\cdot)]\\
 & {} + \frac{\hbar^2}{2!} \mathcal{S}_{(2)} [\phi(\cdot)] + \cdots\label{eq:446}
\end{split}
\intertext{and}
\overset{(0)}{E}_{\!\hbar} &\simeq \hbar\left\lbrace\overset{(0)}{\mathcal{E}}_{\!(0)} + \hbar\overset{(0)}{\mathcal{E}}_{\!(1)} + \frac{\hbar^2}{2!}\overset{(0)}{\mathcal{E}}_{\!(2)} + \cdots\right\rbrace.\label{eq:447}
\end{align}

In the foregoing formulas \(\phi(\cdot)\) symbolizes a real-valued distribution on \(\mathbb{R}^3\) belonging to a certain Sobolev `trace' space that we shall characterize more precisely below. In accordance with our strategy for solving the functional Hamilton-Jacobi equation (\ref{eq:443}) each such \(\phi(\cdot)\) will be taken to represent boundary data, induced on the \(t = 0\) hypersurface of (Euclidean)
\begin{equation}\label{eq:448}
\mathbb{R}^4 = \left\lbrace (t,\mathbf{x})| t \in \mathbb{R}, \mathbf{x} \in \mathbb{R}^3\right\rbrace,
\end{equation}
by a real (distributional) scalar field \(\Phi\) defined on the half-space \(\mathbb{R}^{4-} := (-\infty,0] \times \mathbb{R}^3\). Here \(\Phi\) plays the role of the curve \(\gamma:(-\infty,0] \rightarrow \mathbb{R}^n\) in the quantum mechanics problem and \(\phi(\cdot)\) the role of its right end point \((x^1, \ldots , x^n)\).

By generalizing the technique sketched above for the quantum mechanical problems the authors have proven the existence of a (globally-defined, Fr{\'e}chet smooth) `fundamental solution',  \(\mathcal{S}_{(0)} [\phi(\cdot)]\) to Eq.~(\ref{eq:443}) by first establishing the existence of unique minimizers, \(\Phi_\phi\), for the Euclidean-signature action functional
\begin{equation}\label{eq:449}
\begin{split}
\mathcal{I}_{es} [\Phi] &:= \int_{\mathbb{R}^3} \int_{-\infty}^0 \left\lbrace\frac{1}{2} \dot{\Phi}^2 + \frac{1}{2} \nabla\Phi \cdot \nabla\Phi\right.\\
& {} \left. + \frac{1}{2} m^2 \Phi^2 + \lambda\Phi^4\right\rbrace\; dt\; d^3 x
\end{split}
\end{equation}
for `arbitrary' boundary data \(\phi(\cdot)\), prescribed at \(t = 0\) and then setting
\begin{equation}\label{eq:450}
\mathcal{S}_{(0)} [\phi(\cdot)] = \mathcal{I}_{es} [\Phi_\phi].
\end{equation}
This was accomplished by defining the action functional \(\mathcal{I}_{es} [\Phi]\) on the Sobolev space \(H^1 (\mathbb{R}^{4-},\mathbb{R})\), with boundary data naturally induced on the corresponding trace space, and proving that this functional is coercive, weakly (sequentially) lower semi-continuous and convex \cite{Marini:2016}. Through an application of the (Banach space) implicit function theorem we then proved that the functional so-defined is Fr{\'e}chet smooth throughout its (Sobolev trace space) domain of definition and that it indeed satisfies the (Eucliedean-signature-vanishing-energy) functional Hamilton-Jacobi equation,
\begin{equation}\label{eq:451}
\begin{split}
\frac{1}{2} \int_{\mathbb{R}^3} & \left|\frac{\delta\mathcal{S}_{(0)} [\phi(\cdot)]}{\delta\phi (\mathbf{x})}\right|^2\; d^3 x\\
 &= \int_{\mathbb{R}^3} \left\lbrace\frac{1}{2} \nabla\phi (\mathbf{x}) \cdot \nabla\phi (\mathbf{x}) + \frac{1}{2} m^2\; \phi^2 (\mathbf{x})\right.\\
 & \qquad {} \left. \hphantom{\frac{1}{2}} + \lambda\phi^4 (\mathbf{x})\right\rbrace\; d^3 x,
\end{split}
\end{equation}
and thus provides the fundamental solution that one needs for the computation of all higher order quantum `loop' corrections. These analytical methods were shown to work equally well in lower spatial dimensions for certain higher-order nonlinearities, allowing, for example, \(\Phi^6\) in (Euclidean) \(\mathbb{R}^{3-}\) and \(\Phi^p\) for any even \(p > 2\) in \(\mathbb{R}^{2-}\), and also for more general convex polynomial interaction potentials \(\mathcal{P} (\Phi)\), allowing terms of intermediate degrees, replacing the \(\frac{1}{2} m^2\Phi^2 + \lambda\Phi^4\) of the example above. These correspond precisely to the usual `renormalizable' cases when treated by more conventional quantization methods. For us the restriction on the allowed polynomial degree in a given  spacetime dimension results from applying the Sobolev embedding theorem,
\begin{equation}\label{eq:452}
H^1 (\mathbb{R}^- \times \mathbb{R}^n) \hookrightarrow L^p (\mathbb{R}^- \times \mathbb{R}^n)
\end{equation}
for \(2 \leq p \leq 2 (n+1)/(n-1)\) if \(n > 1\) and for any \(p \geq 2\) if \(n = 1\) (noting here that the domain in question has dimension \(n+1\)), to the demand (needed in our analysis) that the higher order terms in the corresponding action functional be bounded by (some power of) the \(H^1 (\mathbb{R}^- \times \mathbb{R}^n)\) norm defined by the quadratic terms.

To compute higher order `loop' corrections in this field theoretic setting one will first need to regularize the formal functional Laplacian that arises in the Schr{\"o}dinger operator (\ref{eq:442}) and that will reoccur in each of the transport equations that result from substituting ans{\"a}tze such as (\ref{eq:444}), (\ref{eq:446}) and (\ref{eq:447}) into the time independent Schr{\"o}dinger equation (\ref{eq:445}) and requiring satisfaction order-by-order in powers of \(\hbar\). Solving these transport equations for the `loop corrections', \(\lbrace\mathcal{S}_{(1)} [\varphi(\cdot)], \mathcal{S}_{(2)} [\varphi(\cdot)], \ldots\rbrace\), to the ground state wave functional simply amounts to \textit{evaluating} sequentially computable, smooth functionals on the Euclidean signature action minimizers, \(\Phi_\phi\), for arbitrarily chosen boundary data \(\varphi(\cdot)\).

Solving the transport equations for excited states is somewhat more involved since these equations entail a lower order term in the unknown but the technology for handling this (at least in finite dimensions) is well-understood \cite{Moncrief:2012,Dimassi:1999,Helfer:1988}. If, in particular, a Sternberg diffeomorphism could be shown to exist for field theoretic problems of the type discussed herein then the leading order, excited state transport equation could be solved in closed form. Otherwise though one could simply fall back on the machinery developed in Refs.~\cite{Moncrief:2012,Dimassi:1999,Helfer:1988}, which does not assume the existence of Sternberg coordinates, and solve this and the corresponding higher order excited state equations in a less direct fashion since the aforementioned `machinery' apparently generalizes, in a straightforward way, to this infinite dimensional setting. In either case it is intriguing to note that the excited states for \textit{interacting} field theories would be naturally labeled by sequences of (integral) `particle excitation numbers' in much the same way that the Fock-space excited states of a free field are characterized.

Indeed, modulo some apparently quite modest technicalities, needed to  handle a continuous range of frequencies, it seems clear that when these same (Euclidean-signature-semi-classical)  methods are applied to \textit{free}, bosonic field theories they will simply regenerate the well-known (Fock-space) exact solutions for these systems. In particular the fundamental solutions to the relevant (Euclidean signature) Hamilton-Jacobi equations are explicitly known for the most interesting cases (\cite{Maitra:2007}, and from a different perspective \cite{Wheeler:1962}), the higher order `loop corrections' \(\lbrace\mathcal{S}_{(1)} [\varphi(\cdot)], \mathcal{S}_{(2)} [\varphi(\cdot)], \ldots\rbrace\) will be found all to vanish (as they do for finite dimensional, \textit{harmonic} oscillators) and the natural coordinates on the configuration manifold (i.e., the associated trace space described above) are already of Sternberg type.

One often hears that the fundamental particle interpretation of \textit{interacting} quantized fields hinges upon their approximation, asymptotically, by corresponding \textit{free} fields. This is somewhat unsatisfactory since, of course, an elementary particle cannot `turn off' its self-interactions to behave, even asymptotically, like a Fock-space, free field quantum. While we do not yet have a clear `physical interpretation' of the integral, `excitation numbers' that would label our excited states one of the natural features of this (Euclidean-signature-semi-classical) program is that it maintains the dynamical nonlinearities of an interacting quantum system intact at every level of the analysis rather than attempting to reinstate nonlinear effects gradually through a perturbative expansion. One of our main motivations for pursuing it is the expectation that it will ultimately provide much more accurate approximations for wave functionals and their associated, \textit{non-gaussian} integration measures than those generated by conventional (Rayleigh/Schr{\"o}dinger) perturbation theory.

\subsection{Yang-Mills Fields}
\label{subsec:yang-mills-fields}
In continuing research the authors are currently applying these (Euclidean-signature-semi-classical) techniques to the quantization of Yang-Mills fields \cite{Maitra:inprep}. While the methods in question apply equally well to both 3 and 4 dimensional gauge theories (i.e., to the renormalizable cases), we shall focus here on the physically most interesting case of Yang-Mills fields in 4 spacetime dimensions. The formal Schr{\"o}dinger operator for this system is expressible as
\begin{equation}\label{eq:453}
\begin{split}
\hat{H}_{YM} &:= \int_{\mathbb{R}^3} \Sigma_I \left\lbrace-\frac{\hbar^2}{2} \sum_{i=1}^3 \frac{\delta}{\delta A_i^I(\mathbf{x})} \frac{\delta}{\delta A_i^I(\mathbf{x})}\right.\\
 &{} + \left.  \frac{1}{4} \sum_{j,k=1}^3 F_{jk}^I F_{jk}^I (\mathbf{x})\right\rbrace\; d^3 x
\end{split}
\end{equation}
where the index \textit{I} labels a suitable basis for the Lie algebra of the gauge structure group \textit{G}, \(A_k^I\) is the spatial connection field with curvature
\begin{equation}\label{eq:454}
F_{jk}^I = \partial_j A_k^I - \partial_k A_j^I + q [A_j,A_k]^I,
\end{equation}
\textit{q} is the gauge coupling constant and \([\cdot,\cdot]\) the bracket in the Lie algebra of the structure group \textit{G} (under a matrix representation, the commutator).

As in the case of scalar field theory the functional Laplacian requires regularization to be well-defined even when acting on smooth functionals of the (spatial) connection but, since the influence of this regularization will not be felt until higher order quantum `loop' corrections are computed, we can temporarily ignore this refinement here and attempt first to construct a (gauge invariant) fundamental solution, \(\mathcal{S}_{(0)} [A(\cdot)]\), to the Euclidean-signature-vanishing-energy Hamilton-Jacobi equation
\begin{equation}\label{eq:455}
\begin{split}
\int_{\mathbb{R}^3} \Sigma_i & \left\lbrace\frac{1}{2} \sum_{i=1}^3 \frac{\delta\mathcal{S}_{(0)}}{\delta A_i^I (\mathbf{x})} \frac{\delta\mathcal{S}_{(0)}}{\delta A_i^I (\mathbf{x})}\right.\\
 & {} - \left. \frac{1}{4} \sum_{j,k=1}^3 F_{jk}^I (\mathbf{x}) F_{jk}^I (\mathbf{x})\right\rbrace\; d^3 x = 0
\end{split}
\end{equation}
by seeking minimizers of the corresponding Euclidean-signature action functional in the form of (spacetime) connections \(\lbrace\mathcal{A}_\mu^I\rbrace\) defined on \(\mathbb{R}^{4-} = (-\infty,0] \times \mathbb{R}^3\) with boundary data \(A_i^I\) prescribed at \(t = 0\).

As usual in our approach, Eq.~(\ref{eq:455}) results from substituting the ans{\"a}tze
\begin{align}
\overset{(0)}{\Psi}_{\!\hbar} [A(\cdot)] &= N_\hbar e^{-\mathcal{S}_\hbar [A(\cdot)]/\hbar},\label{eq:456}\\
\begin{split}
\mathcal{S}_\hbar [A(\cdot)] &\simeq \mathcal{S}_{(0)} [A(\cdot)] + \hbar\mathcal{S}_{(1)} [A(\cdot)] + \frac{\hbar^2}{2!} \mathcal{S}_{(2)} [A(\cdot)]\\
& {} + \cdots + \frac{\hbar^k}{k!} \mathcal{S}_{(k)} [A(\cdot)] + \cdots , \label{eq:457}
\end{split}\\
\begin{split}
\overset{(0)}{E}_{\!\hbar} &\simeq \hbar \left(\overset{(0)}{\mathcal{E}}_{\!(0)} + \hbar\overset{(0)}{\mathcal{E}}_{\!(1)} + \frac{\hbar^2}{2!}\overset{(0)}{\mathcal{E}}_{\!(2)} + \cdots\right.\\
 & {} \left. + \cdots + \frac{\hbar^k}{k!}\overset{(0)}{\mathcal{E}}_{\!(k)} + \cdots\right)\label{eq:458}
\end{split}\\
\intertext{into the Schr{\"o}dinger equation}
\hat{H}_{YM}\overset{(0)}{\Psi}_{\!\hbar} &= \overset{(0)}{E}_{\!\hbar}\overset{(0)}{\Psi}_{\!\hbar} \label{eq:459}
\end{align}
and demanding satisfaction order by order in \(\hbar\).

To construct the functional \(\mathcal{S}_{(0)} [A(\cdot)]\) we treat the (spatial) connection field \(A = \lbrace A_i^I\rbrace\) on \(\mathbb{R}^3\) as (tangential) boundary data for the Euclidean-signature Yang-Mills Dirichlet problem --- prescribing this data on the hypersurface \(\lbrace x^0 = ct =  0\rbrace\) of \(\mathbb{R}^4 = \lbrace (x^\mu) = (ct,\mathbf{x}): \mathbf{x} = (x^1, x^2, x^3)\rbrace\). Thus for `arbitrary' boundary data \textit{A} defined on \(\lbrace 0\rbrace \times \mathbb{R}^3\) (and lying in a suitable `trace space' for spacetime connection fields \(\mathcal{A} = \lbrace\mathcal{A}_\mu^I\rbrace\)), we seek an absolute minimizer, \(\mathcal{A}_A\), for the Euclidean-signature Yang-Mills action functional, \(\mathcal{I}_{es} [\mathcal{A}]\), defined on the half-space \(\mathbb{R}^- \times \mathbb{R}^3 := (-\infty,0] \times \mathbb{R}^3\) by
\begin{equation}\label{eq:460}
\begin{split}
\mathcal{I}_{es} [\mathcal{A}] &:= \frac{1}{4} \int_{\mathbb{R}^-\times\mathbb{R}^3} \left\lbrace\Sigma_I \sum_{\mu,\nu = 0}^3 \left\lbrack\mathcal{F}_{\mu\nu}^I\mathcal{F}_{\mu\nu}^I\right\rbrack\right\rbrace\; dt\; d^3 x\\
 &= \frac{1}{2} \int_{-\infty}^0 dt \int_{\mathbb{R}^3} d^3 x\; \left\lbrace\Sigma_I \left\lbrack\sum_{i=1}^3 (\partial_0\mathcal{A}_i^I - \partial_i\mathcal{A}_0^I)^2\right.\right.\\
 & {} \left.\left. + \frac{1}{2} \sum_{j,k=1}^3 \mathcal{F}_{jk}^I\mathcal{F}_{jk}^I\right\rbrack\right\rbrace
\end{split}
\end{equation}
where \(\mathcal{F} = \lbrace\mathcal{F}_{\mu\nu}^I\rbrace\), the curvature of the connection \(\mathcal{A}\), is given by
\begin{equation}\label{eq:461}
\mathcal{F}_{\mu\nu}^I := \partial_\mu\mathcal{A}_\nu^I - \partial_\nu\mathcal{A}_\mu^I + q [\mathcal{A}_\mu,\mathcal{A}_\nu]^I.
\end{equation}

The first question our construction must address is that of defining the function space from which Yang-Mills connections on $\{0\} \times \mathbb{R}^3$ (viewed as initial data for the Euclidean-signature Dirichlet problem on the half-space $\mathbb{R}^{4-}$) are to be drawn.  Modulo the action of gauge transformations, this function space of connections yields as its quotient the orbit space which is the true Yang-Mills configuration space.  

In particular, our construction proceeds differently depending on whether or not we require each connection to approach a coherent value at spatial infinity, as done for instance by Jackiw in \cite{Jackiw:1980}.  Under this requirement, the initial hypersurface $\{0\} \times \mathbb{R}^3$ effectively becomes a 3-sphere, introducing a distinction between `small' and `large' gauge transformations (homotopically trivial and nontrivial, respectively), and an attendant division of the Yang-Mills configuration space into distinct topological sectors.  The dichotomy between large and small gauge transformations is usually seen as the origin of the `vacuum angle' in quantum Yang-Mills theory, with wave functionals invariant only up to a phase under large gauge transformations \cite{Jackiw:1980}.  As in the treatment by Khoze \cite{Khoze:1994}, we allow connections to have no coherent limit at spatial infinity, and regard all gauge transformations on the same footing.  Nevertheless our approach, like others with the same definition of the configuration space, is not incompatible with the introduction of a vacuum angle, since such a feature (if present in nature) can be incorporated in the Lagrangian as observed in \cite{Jackiw:1980} and \cite{Khoze:1994}.

To prove existence of a minimizer for the Euclidean-signature Yang-Mills action with (tangential) initial data $A$ prescribed from our configuration space on $\{0\} \times \mathbb{R}^3$, we use the direct method in the calculus of variations to conclude that any action-minimizing sequence with given initial data has a convergent subsequence, on whose limit the Euclidean-signature Yang-Mills action is indeed minimized.  As in the physical models discussed in the preceding sections, we then define $\mathcal{S}_{(0)} [A(\cdot)]$ to assume the absolute minimizing value of the Euclidean-signature Yang-Mills action for initial data $A$.  

The existence of a convergent minimizing subsequence is essentially due to weak compactness of bounded sets in Sobolev spaces.  One is enabled to invoke Sobolev weak compactness by gauge transforming to a `Hodge' or 'Coulomb' gauge locally on neighborhoods of $\mathbb{R}^{4-}$ where the curvature of connections in the minimizing sequence has sufficiently small $L^2$ norm.  On such neighborhoods, a pivotal result of Uhlenbeck \cite{Uhlenbeck:1982a} states that one can transform to the Hodge gauge, and that the $L^2_1$ Sobolev norm of the transformed connection is bounded in terms of the $L^2$ norm of its curvature.  Additionally, use of the (local) Hodge gauge allows the top order term in the Yang-Mills equation to be viewed as a Laplace-de Rham operator, making available elliptic regularity results to establish smoothness of the solution.  For further details, the reader is referred to the work of Sedlacek \cite{Sedlacek:1982} for a compact manifold without boundary, Marini \cite{Marini:1992} for a compact manifold with boundary, and the present authors \cite{Maitra:2007,Maitra:inprep} for a possibly noncompact manifold with boundary.  

While the local Hodge gauge is key to achieving existence and regularity of a minimizer, it should be noted that this method is internal to the proof and thence the construction of $\mathcal{S}_{(0)}$.  Thus it does not introduce a Gribov ambiguity since it does not constitute a global gauge fixing within the Yang-Mills configuration space.  We treat the domain of $\mathcal{S}_{(0)}$ as a Sobolev space of connections, noting that gauge invariance of the Euclidean-signature Yang-Mills action immediately implies that $\mathcal{S}_{(0)} [A(\cdot)]$ is a (\textit{fully non-abelian}) gauge invariant solution to the Hamilton-Jacobi equation (\ref{eq:455}) and accordingly satisfies the corresponding Gauss-law constraint --- namely the vanishing of the gauge covariant divergence of its (electric field) functional gradient, \(\frac{\delta\mathcal{S}_{(0)}}{\delta A^I(\mathbf{x})} [A(\cdot)]\).  As such \(\mathcal{S}_{(0)} [A(\cdot)]\) will naturally pass to the quotient, orbit space whereon it will correspondingly satisfy the \textit{reduced} Hamilton-Jacobi equation for this (positively curved) infinite dimensional Riemannian manifold.  In establishing smoothness results for $\mathcal{S}_{(0)}$, we use the Sobolev topology on the space of connections which form its domain, employing the Banach space version of Rademacher's theorem to show that $\mathcal{S}_{(0)}$ is G\^ateaux differentiable almost everywhere in a suitable sense (for details, see \cite{Maitra:inprep}).  Application of the Banach space implicit function theorem to establish Fr\'echet differentiability of $\mathcal{S}_{(0)}$ to all orders is the topic of current investigations.



The self-interactions of `gluons' (the quanta of the Yang-Mills field) are closely connected to the non-abelian character of the associated gauge group. Thus a conventional perturbative approach to quantization, which disregards these interactions at the lowest order, necessarily `approximates' the gauge group as well, replacing it with the abelian structure group of the associated free field theory (i.e., several copies of the Maxwell field labelled by the index \textit{I}), and then attempts to reinstate both the interactions and the non-commutative character of the actual gauge group gradually, through the development of series expansions in the Yang-Mills coupling constant. By contrast the Euclidean-signature-semi-classical program that we are advocating for the Yang-Mills problem has the advantage of maintaining \textit{full, non-abelian gauge invariance} at every order of the calculation and of generating globally defined (approximate) wave functionals on the naturally associated Yang-Mills configuration manifold. 

\section{The Orbit Space Curvature for Scalar Electrodynamics}
\label{sec:orbit-space-curvature-scalar-electrodynamics}
The Lagrangian density for `scalar electrodynamics', as we shall use the term herein, is given by
\begin{equation}\label{eq:101}
\mathcal{L} = -\frac{1}{4} F_{\mu\nu}F^{\mu\nu} - \eta^{\mu\nu} (D_\mu\varphi)^\dagger (D_\nu\varphi) - \mathcal{U}(\varphi^\dagger\varphi)
\end{equation}
where \(\varphi := \varphi^1 + i\varphi^2\), with \(\varphi^a\) real, is a complex scalar field, \(\varphi^\dagger := \varphi^1 - i\varphi^2\) its complex conjugate and where \(F = F_{\mu\nu}\; dx^\mu \otimes dx^\nu\) is the electromagnetic field tensor expressible in terms of its associated connection or `vector potential' \(A = A_\mu dx^\mu\) as
\begin{equation}\label{eq:102}
F_{\mu\nu} = \partial_\mu A_\nu - \partial_\nu A_\mu.
\end{equation}
The gauge covariant derivatives \(D_\mu\varphi\), \((D_\mu\varphi)^\dagger\) are defined by
\begin{align}
D_\mu\varphi &= \partial_\mu\varphi - iqA_\mu\varphi\label{eq:103}\\
(D_\mu\varphi)^\dagger &= \partial_\mu\varphi^\dagger + iqA_\mu\varphi^\dagger\label{eq:104}
\end{align}
wherein \textit{q} is a gauge `coupling' constant having the dimensions
\begin{equation}\label{eq:105}
[q] = \left[\frac{e}{\hbar c}\right]
\end{equation}
with \textit{e} the fundamental constant of electric charge, \(\hbar\) the (reduced) Planck constant and \textit{c} the speed of light. The self-interaction potential \(\mathcal{U} : \mathbb{R} \rightarrow \mathbb{R}\) is assumed to be smooth and positive. In the standard (Lorentz frame) coordinates, \(\lbrace x^\mu; \mu = 0,1,2,3\rbrace = \lbrace ct, x^i; i = 1,2,3\rbrace\), that we shall use the Minbowski metric \(\eta = \eta_{\mu\nu} dx^\mu \otimes dx^\nu\) takes the form
\begin{equation}\label{eq:106}
\eta = -c^2 dt \otimes dt + \sum_{i=1}^3 dx^i \otimes dx^i
\end{equation}
with corresponding line element
\begin{equation}\label{eq:107}
ds^2 = -c^2 dt^2 + d\mathbf{x} \cdot d\mathbf{x}
\end{equation}
where \(\mathbf{x} := (x^1,x^2,x^3)\) and \(\cdot\) designates the Euclidean metric on \(\mathbb{R}^3\).

As is well-known \(\mathcal{L}\) is invariant with respect to the group \(\mathcal{G}\) of `gauge transformations' under which
\begin{equation}\label{eq:191}
A_\mu \rightarrow A_\mu + \partial_\mu\Lambda, \qquad \varphi \rightarrow \varphi e^{iq\Lambda}
\end{equation}
where \(\Lambda\) is an arbitrary, smooth function having the dimensions of `charge', \([e]\), and vanishing at infinity, \(|\mathbf{x}| := \sqrt{\mathbf{x} \cdot \mathbf{x}} \rightarrow \infty\). The action functional defined on any domain of the form \(\Omega = I \times \mathbb{R}^3\), with \(I = [t_0,t_1] \subset \mathbb{R}\), is given by
\begin{equation}\label{eq:108}
\mathcal{I}_\Omega [\varphi ,A] := \frac{1}{c} \int_\Omega d^4x\; \mathcal{L} = \int_I dt\; L
\end{equation}
where \textit{L} is the Lagrangian defined by
\begin{equation}\label{eq:109}
L := \int_{\mathbb{R}^3} d^3x\; \mathcal{L}.
\end{equation}
The Euler-Lagrange equations (for the domain \(\Omega\)) obtained by varying \(\mathcal{I}_\Omega [\varphi ,A]\) with respect to \(\varphi\) and \textit{A} are given (respectively) by
\begin{equation}\label{eq:110}
\eta^{\mu\nu} D_\mu D_\nu\varphi - \mathcal{U}' (\varphi^\dagger\varphi) \varphi = 0
\end{equation}
and
\begin{equation}\label{eq:111}
\partial_\nu F^{\mu\nu} = iq\eta^{\mu\nu} \left[(D_\nu \varphi)^\dagger\varphi - \varphi^\dagger (D_\nu \varphi)\right]
\end{equation}
wherein \(\mathcal{U}'(u) := \frac{d\mathcal{U}(u)}{du}\).

The time component, \(\mu \rightarrow 0\), of the Maxwell equation (\ref{eq:111}) gives, of course, the Gauss law `constraint'
\begin{equation}\label{eq:112}
\begin{split}
\partial_i F^{0i} &= -\partial_i F_{0i} \\
 &= -iq \left[(D_0\varphi)^\dagger\varphi - \varphi^\dagger (D_0\varphi)\right]
\end{split}
\end{equation}
which, expressed in terms of the vector potential \textit{A}, becomes
\begin{equation}\label{eq:113}
-\Delta A_0 + 2q^2\varphi^\dagger\varphi A_0 = -\partial_i (A_{i,0}) + iq \left[(\partial_0\varphi^\dagger)\varphi - \varphi^\dagger (\partial_0\varphi)\right]
\end{equation}
with \(\Delta\) the Laplacian on (Euclidean) \(\mathbb{R}^3\),
\begin{equation}\label{eq:114}
\Delta = \sum_{i=1}^3 \frac{\partial^2}{{\partial x^i}^2},
\end{equation}
and where, in the above, we have adopted the summation convention for sums over repeated \textit{spatial} indices (writing, e.g., \(\partial_iv^i\) for \(\sum_{i=1}^3 \partial_iv^i\)).

The operator \(\Delta_\varphi\) defined by
\begin{equation}\label{eq:115}
\Delta_\varphi := \Delta - 2q^2\varphi^\dagger\varphi
\end{equation}
will play a fundamental role in the following. In a suitable function space setting its inverse, \(\Delta_\varphi^{-1}\), will exist and allow one to solve the elliptic, Gauss law constraint for \(A_0\) by setting
\begin{equation}\label{eq:116}
A_0 = \Delta_\varphi^{-1} \left[(\partial_iA_{i,0}) - iq \left[(\partial_0\varphi^\dagger)\varphi - \varphi^\dagger(\partial_0\varphi)\right]\right].
\end{equation}

Reexpressed in this 3+1 dimensional notation the Lagrangian defined above now takes the form
\begin{equation}\label{eq:117}
L = \int_{\mathbb{R}^3} d^3x \left\lbrace\frac{1}{2} F_{0j}F_{0j} - \frac{1}{4} F_{jk}F_{jk} + (D_0\varphi)^\dagger (D_0\varphi) - (D_j\varphi)^\dagger (D_j\varphi) - U(\varphi^\dagger\varphi)\right\rbrace.
\end{equation}
Defining canonical momenta \(\pi_\varphi\) and \(\pi^j\) conjugate to \(\partial\) and \(A_j\) (respectively) by the Legendre transformation
\begin{align}
\pi_\varphi &:= \frac{\delta L}{\delta\varphi_{,t}} = \frac{1}{c} (D_0\varphi)^\dagger\label{eq:118}\\
\pi^j &:= \frac{\delta L}{\delta A_{j,t}} = \frac{1}{c} (A_{j,0} - A_{0,j}) = \frac{1}{c} F_{0j}\label{eq:119}
\end{align}
with, of course,
\begin{equation}\label{eq:120}
\pi_\varphi^\dagger := \frac{\delta L}{\delta\varphi^{\,\dagger}_{,t}} = \frac{1}{c} (D_0\varphi)
\end{equation}
and noting that
\begin{equation}\label{eq:121}
\pi^0 := \frac{\delta L}{\delta A_{0,t}} \equiv 0
\end{equation}
one arrives at the associated Hamiltonian density
\begin{equation}\label{eq:122}
\mathcal{H} := \pi_\varphi \varphi_{,t} + \pi_{\varphi^\dagger} \varphi^{\,\dagger}_{,t} + \pi^j A_{j,t} - \mathcal{L}.
\end{equation}
The corresponding Hamiltonian takes the explicit form
\begin{equation}\label{eq:123}
\begin{split}
H &:= \int_{\mathbb{R}^3} dx^3 \mathcal{H}\\
 &= \int_{\mathbb{R}^3} d^3x \left\lbrace\frac{1}{2} c^2\pi^j\pi^j + c^2\pi_\varphi^\dagger\pi_\varphi + \frac{1}{4} F_{jk}F_{jk} + (D_j\varphi)^\dagger (D_j\varphi) + U(\varphi^\dagger\varphi)\right.\\
 &\quad - \left.\vphantom{\frac{1}{4}} A_0 \left[\partial_j (c\pi^j) - iqc (\varphi\pi_\varphi - \varphi^\dagger\pi_\varphi^\dagger)\right]\right\rbrace\\
 &\quad + \int_{\mathbb{R}^3} d^3x (\partial_j \left(A_0 c\pi^j)\right)
\end{split}
\end{equation}
wherein \(A_0\) now plays the role of a `Lagrange multiplier' with respect to whose variation one recovers the Hamiltonian form of the Gauss constraint
\begin{equation}\label{eq:124}
\partial_j (c\pi^j) = iq \left(\varphi (c\pi_\varphi) - \varphi^\dagger (c\pi_\varphi^\dagger)\right).
\end{equation}
Noting that \(c\pi^j = F_{0j} = -E^j\), where \(\mathbf{E} = E^j \frac{\partial}{\partial x_j}\) is the electric field, one sees that the (gauge invariant) charge density \(\rho\) of the \(\varphi\) field is given by
\begin{equation}\label{eq:125}
\begin{split}
4\pi\rho &= -iq \left(\varphi (c\pi_\varphi) - \varphi^\dagger (c\pi_\varphi^\dagger)\right)\\
 &= -iq \left(\varphi (D_0\varphi)^\dagger - \varphi^\dagger (D_0\varphi)\right).
\end{split}
\end{equation}

Again in a suitable function space setting one can decompose \(\boldsymbol{\pi} = \pi^j \frac{\partial}{\partial x^j}\) into \(L^2\)-orthogonal `transverse' and `longitudinal' components,
\begin{equation}\label{eq:126}
\boldsymbol{\pi} = \boldsymbol{\pi}^T + \boldsymbol{\pi}^L,
\end{equation}
with
\begin{align}
\nabla \cdot \boldsymbol{\pi}^T &= \partial_j (\pi^T)^j = 0,\label{eq:127}\\
\boldsymbol{\pi}^L &= \nabla\lambda\label{eq:128}
\end{align}
so that
\begin{equation}\label{eq:129}
\nabla \cdot \boldsymbol{\pi} = \nabla \cdot \boldsymbol{\pi}^L = \Delta\lambda
\end{equation}
and thereby express the solution of the Gauss constraint in the (Hamiltonian) form
\begin{equation}\label{eq:130}
\begin{split}
-(c\boldsymbol{\pi}^L)^j &:= (\mathbf{E}^L)^j\\
 &= -\nabla^j \left(\Delta^{-1} \left[ iq\left(\varphi (c\pi_\varphi) - \varphi^\dagger (c\pi_\varphi^\dagger)\right)\right]\right)\\
 &= \nabla^j \left(\Delta^{-1} (4\pi\rho)\right)
\end{split}
\end{equation}
where, more explicitly,
\begin{equation}\label{eq:131}
\left(\Delta^{-1} (4\pi\rho)\right) (\mathbf{x}) = -\int_{\mathbb{R}^3} d^3x' \frac{\rho (\mathbf{x}')}{|\mathbf{x} - \mathbf{x}'|}
\end{equation}
with \(|\mathbf{x} - \mathbf{x}'|\) the Euclidean distance from \(\mathbf{x}\) to \(\mathbf{x}'\).

In parallel with the above decomposition of \(\boldsymbol{\pi}\), we can also express \(A_i\) in terms of \(L^2\)-orthogonal transverse and longitudinal summands via
\begin{equation}\label{eq:132}
A_i = A_i^T + A_i^L
\end{equation}
with
\begin{equation}\label{eq:133}
\nabla \cdot \mathbf{A}^T = \partial_jA_j^T = 0
\end{equation}
and
\begin{equation}\label{eq:134}
\nabla \times \mathbf{A}^L = 0
\end{equation}
with \(\mathbf{A}^L\) given explicitly by
\begin{equation}\label{eq:135}
A_j^L (\mathbf{x}) = -\partial_j \left[\int_{\mathbb{R}^3} d^3 x' \left(\frac{\left(\partial_k A_k(\mathbf{x}')\right)}{4\pi |\mathbf{x} - \mathbf{x}'|}\right)\right].
\end{equation}
Note accordingly that one can always achieve the `Coulomb gauge' condition \(\mathbf{A}^L = 0\) with the \(\mathcal{G}\) action generated by
\begin{equation}\label{eq:136}
\Lambda (\mathbf{x}) = \int_{\mathbb{R}^3} d^3x' \left(\frac{\partial_k A_k (\mathbf{x}')}{4\pi |\mathbf{x} - \mathbf{x}'|}\right)
\end{equation}
under which \(\varphi\) undergoes the corresponding change of `phase' \(\varphi \rightarrow \varphi e^{iq\Lambda}\). In an arbitrary gauge it is easily verified that \(\lbrace\mathbf{A}^T,\boldsymbol{\pi}^T\rbrace\) and \(\lbrace\mathbf{A}^L,\boldsymbol{\pi}^L\rbrace\) are canonically conjugate variables. Since \(\boldsymbol{\pi}^L\) is uniquely determined by the charge density however (c.f. \ref{eq:130}) and since its conjugate partner can be eliminated by the choice of Coulomb gauge it is natural to pass to a \textit{reduced Hamiltonian} framework.

We therefore define a `reduced' Hamiltonian by substituting the above expression (\ref{eq:130}) for \(\boldsymbol{\pi}^L\) into \textit{H}, dropping the boundary integral, \(\int_{\mathbb{R}^3} d^3x \left(\partial_j (A_0c\pi^j)\right)\), (which makes no contribution to the field equations) and imposing the Coulomb gauge condition under which \(A_j \rightarrow A_j^T\). The result is
\begin{equation}\label{eq:137}
\begin{split}
H_{\mathrm{reduced}} &:= \int_{\mathbb{R}^3} d^3x \left\lbrace\frac{1}{2} c^2\boldsymbol{\pi}^T \cdot \boldsymbol{\pi}^T + c^2\pi_\varphi^\dagger\pi_\varphi + \frac{1}{4} F_{jk}F_{jk}\right.\\
 &\quad\quad  + (D_j\varphi)^\dagger (D_j\varphi) + U(\varphi^\dagger\varphi)\\
 &\quad \left.\vphantom{\frac{1}{4}} + \frac{1}{2} q^2c^2 (\varphi\pi_\varphi - \varphi^\dagger\pi_\varphi^\dagger) \Delta^{-1} (\varphi\pi_\varphi - \varphi^\dagger\pi_\varphi^\dagger)\right\rbrace
\end{split}
\end{equation}
wherein
\begin{equation}\label{eq:138}
\left(\Delta^{-1} (\varphi\pi_\varphi - \varphi^\dagger\pi_\varphi^\dagger)\right) (\mathbf{x}) = -\frac{1}{4\pi} \int_{\mathbb{R}^3} d^3x' \left(\frac{(\varphi\pi_\varphi - \varphi^\dagger\pi_\varphi^\dagger)(\mathbf{x}')}{|\mathbf{x} - \mathbf{x}'|}\right).
\end{equation}
Note that, in this gauge, Eq.~(\ref{eq:116}) for \(A_0\) simplifies to
\begin{equation}\label{eq:139}
A_0 = \Delta_\varphi^{-1} \left[ -iq \left[ (\partial_0\varphi^\dagger)\varphi - \varphi^\dagger (\partial_0\varphi)\right]\right]
\end{equation}
or, since
\begin{equation}\label{eq:140}
\Delta_\varphi A_0 = -iq \left[ (\partial_0\varphi^\dagger)\varphi - \varphi^\dagger (\partial_0\varphi)\right]
\end{equation}
can be expressed as
\begin{equation}\label{eq:141}
\Delta A_0 = -iq \left[ (c\pi_\varphi)\varphi - (c\pi_\varphi^\dagger)\varphi^\dagger\right],
\end{equation}
also to
\begin{equation}\label{eq:142}
A_0 = \Delta^{-1} \left[ -iq \left[ (c\pi_\varphi)\varphi - (c\pi_\varphi^\dagger)\varphi^\dagger\right]\right].
\end{equation}
The reduced Lagrangian that corresponds to \(H_{\mathrm{reduced}}\) may be equivalently derived by substituting (\ref{eq:139}) and \(\mathbf{A}^L = 0\) into \textit{L} or by inverting the Legendre transformation determined by \(H_{\mathrm{reduced}}\). The result is:
\begin{equation}\label{eq:143}
\begin{split}
L_{\mathrm{reduced}} &:= \int_{\mathbb{R}^3} d^3x \left\lbrace\frac{1}{2c^2} \mathbf{A}_{,t}^T \cdot \mathbf{A}_{,t}^T + \frac{1}{c^2} (\varphi_{,t}^\dagger) (\varphi_{,t})\right.\\
 &\quad - \frac{1}{4} F_{jk}F_{jk} - (D_j\varphi)^\dagger (D_j\varphi) - U(\varphi^\dagger\varphi)\\
 &\quad \left.\vphantom{\frac{1}{2c^2}} -\frac{q^2}{2c^2} (\varphi_{,t}^\dagger\varphi - \varphi_{,t}\varphi^\dagger) \Delta_\varphi^{-1} (\varphi_{,t}^\dagger\varphi - \varphi_{,t}\varphi^\dagger)\right\rbrace
\end{split}
\end{equation}

Prior to reduction the configuration manifold \(\mathcal{Q}\) can be regarded as the product of the space of (spatial) connections \(\mathcal{A}\) with the space of complex scalar fields \(\mathcal{S}\), all defined over \(\mathbb{R}^3\):
\begin{equation}\label{eq:144}
\mathcal{Q} = \mathcal{A} \times \mathcal{S}
\end{equation}
The Hamiltonian \textit{H} is defined on its associated cotangent bundle
\begin{equation}\label{eq:145}
\mathcal{P} = T^*\mathcal{Q}
\end{equation}
but depends not only on the corresponding canonical variables but also on the (at this point still arbitrary) `Lagrange multiplier' field \(A_0\). The natural \textit{reduced} configuration manifold, \(\mathcal{Q}_{\mathrm{reduced}}\), can be viewed as the abstract quotient of \(\mathcal{Q}\) by the gauge group \(\mathcal{G}\)
\begin{equation}\label{eq:146}
\mathcal{Q}_{\mathrm{reduced}} := \mathcal{Q}/\mathcal{G}
\end{equation}
so that, in more geometric language, \(\mathcal{Q}\) is a \(\mathcal{G}\)-bundle over \(\mathcal{Q}_{\mathrm{reduced}}\). By the same token the reduced phase space (over which \(H_{\mathrm{reduced}}\) is defined) can be regarded as the cotangent bundle of \(\mathcal{Q}_{\mathrm{reduced}}\)
\begin{equation}\label{eq:147}
\mathcal{P}_{\mathrm{reduced}} := T^*\mathcal{Q}_{\mathrm{reduced}}.
\end{equation}

The Coulomb gauge condition defines a smooth, global cross-section of this (topologically trivial) bundle
\begin{equation}\label{eq:148}
\mathcal{Q} \rightarrow \mathcal{Q}_{\mathrm{reduced}} = \mathcal{Q}/\mathcal{G}
\end{equation}
and thus may be viewed as providing a concrete realization of this abstract quotient space in terms of an explicit submanifold of \(\mathcal{Q}\). In this setting the reduced-space canonical variables \(\lbrace\mathbf{A}^T,\varphi\rbrace\) effectively provide a global coordinate system for the quotient manifold, \(\mathcal{Q}_{\mathrm{reduced}}\), and, together with their conjugate momenta \(\lbrace\boldsymbol{\pi}^T,\pi_\varphi\rbrace\), define global canonical coordinates for \(\mathcal{P}_{\mathrm{reduced}}\). A different choice of gauge up in the bundle (other than the Coulomb one that we have made) would have induced a different coordinate system down in the base without, however, modifying the (gauge) invariant dynamics unfolding in the quotient, `orbit' space.

From the purely `kinetic energy' terms in \(L_{\mathrm{reduced}}\) (i.e., those bilinear in \(\varphi_{,t}\) and \(\varphi_{,t}^\dagger\)) and in \(H_{\mathrm{reduced}}\) (i.e., those bilinear in \(\pi_\varphi\) and \(\pi_\varphi^\dagger\)) one can read off coordinate expressions for the naturally induced (product) Riemannian metric, \({}^{\mathcal{Q}}\mathsf{g}\), defined on \(\mathcal{Q}_{\mathrm{reduced}}\) and its inverse, \({}^{\mathcal{Q}}\mathsf{g}^{-1}\). The metric in the \(\mathcal{A}^T\) factor is manifestly `Euclidean' whereas that on the \(\mathcal{S}\) factor takes (in a notation explicitly geared to the chosen coordinate system) the form:
\begin{equation}\label{eq:149}
\mathsf{g}_{\varphi^a(\mathbf{x})\varphi^b(\mathbf{x}')} := \frac{2}{c^2} \left\lbrace \delta_{ab}\delta (\mathbf{x},\mathbf{x}') + 2q^2 \epsilon_{ac} \varphi^c (\mathbf{x}) \Delta_\varphi^{-1} (\mathbf{x},\mathbf{x}') \epsilon_{bd} \varphi^d(\mathbf{x}')\right\rbrace
\end{equation}
where \(\Delta_\varphi^{-1}(\mathbf{x},\mathbf{x}')\) is the kernel function for the operator \(\Delta_\varphi^{-1}\) and where \(\epsilon^{ab} = -\epsilon^{ba}\) with \(\epsilon^{12} = 1\). The inverse (i.e., contra-variant) form of this metric is given by
\begin{equation}\label{eq:150}
\mathsf{g}^{\varphi^a(\mathbf{x})\varphi^b(\mathbf{x}')} := \frac{c^2}{2} \left\lbrace\delta^{ab} \delta (\mathbf{x},\mathbf{x}') + 2q^2 \frac{\epsilon^{ac}\varphi_c(\mathbf{x}) \epsilon^{bd}\varphi_d(\mathbf{x}')}{4\pi |\mathbf{x} - \mathbf{x}'|}\right\rbrace
\end{equation}
with \(\varphi_a = \delta_{ab}\varphi^b = \varphi^a\) and \(\epsilon^{ab} = \epsilon_{ab}\). With these definitions the kinetic energy term, \(\mathcal{K}_\varphi\), for the \(\mathcal{S}\) factor can be written as
\begin{equation}\label{eq:151}
\mathcal{K}_\varphi = \frac{1}{2} \int_{\mathbb{R}^3} d^3 x \int_{\mathbb{R}^3} d^3x' \left\lbrace\mathsf{g}_{\varphi^a(\mathbf{x})\varphi^b(\mathbf{x}')} \varphi_{,t}^a(\mathbf{x})\varphi_{,t}^b(\mathbf{x}')\right\rbrace
\end{equation}
or, equivalently, as
\begin{equation}\label{eq:152}
\mathcal{K}_\varphi = \frac{1}{2} \int_{\mathbb{R}^3} d^3x \int_{\mathbb{R}^3} d^3x' \left\lbrace\mathsf{g}^{\varphi^a(\mathbf{x}) \varphi^b(\mathbf{x}')} \pi_a(\mathbf{x}) \pi_b(\mathbf{x}')\right\rbrace
\end{equation}
where
\begin{equation}\label{eq:153}
\pi_1 := \pi_\varphi + \pi_\varphi^\dagger
\end{equation}
and
\begin{equation}\label{eq:154}
\pi_2 := i (\pi_\varphi - \pi_\varphi^\dagger)
\end{equation}
are the momenta conjugate to \(\varphi^1\) and \(\varphi^2\) (respectively) so that, in particular,
\begin{equation}\label{eq:155}
\pi_\varphi\varphi_{,t} + \pi_\varphi^\dagger\varphi^{\,\dagger}_{,t} = \pi_1\varphi^1_{,t} + \pi_2\varphi^2_{,t}.
\end{equation}

Recalling that the kernel function, \(\Delta^{-1}(\mathbf{x},\mathbf{x}')\), for the operator \(\Delta^{-1}\) is given by
\begin{equation}\label{eq:156}
\Delta^{-1}(\mathbf{x},\mathbf{x}') = \frac{-1}{4\pi |\mathbf{x} - \mathbf{x}'|}
\end{equation}
it is not difficult to verify directly that \(\mathsf{g}\) and \(\mathsf{g}^{-1}\) are indeed inverses of one another and hence satisfy
\begin{equation}\label{eq:157}
\int_{\mathbb{R}^3} d^3x' \left(\mathsf{g}_{\varphi^a(\mathbf{x})\varphi^b(\mathbf{x}')} \mathsf{g}^{\varphi^b(\mathbf{x}')\varphi^c(\mathbf{x}'')}\right) = \delta_a^c \delta (\mathbf{x},\mathbf{x}'').
\end{equation}
This identity plays a key role in the Legendre transformation relating \(L_{\mathrm{reduced}}\) to \(H_{\mathrm{reduced}}\).

While it would now be straightforward to compute the curvature of the manifold (\(\mathcal{S},\mathsf{g}\)) directly in the global chart defined above there is an alternative approach that allows for an easier comparison of the curvatures at different points of \(\mathcal{S}\) as well as for an illuminating comparison with the corresponding results for Yang-Mills fields derived in \cite{Singer:1981,Babelon:1981,Vergeles:1983}. This alternative involves solving the geodesic equations for the manifold  (\(\mathcal{S},\mathsf{g}\)), constructing the exponential map associated to an \textit{arbitrary point} of \(\mathcal{S}\) and thereby introducing an analogue of \textit{normal coordinates} centered at the chosen point. In normal coordinates the connection components vanish at the chosen point thereby dramatically simplifying the evaluation of the corresponding curvature at that point.

The reduced Hamilton equations for the \(\varphi\) field are readily found to be
\begin{equation}\label{eq:158}
\begin{split}
\varphi_{,t} &= \frac{\delta H_{\mathrm{reduced}}}{\delta\pi_\varphi}\\
 &= c^2\pi_\varphi^\dagger + iqc A_0\varphi
\end{split}
\end{equation}
and
\begin{equation}\label{eq:159}
\begin{split}
(\pi_\varphi^\dagger)_{,t} &= -\frac{\delta H_{\mathrm{reduced}}}{\delta\varphi^\dagger}\\
 &= iqc A_0 \pi_\varphi^\dagger - \frac{\delta}{\delta\varphi^\dagger} \int_{\mathbb{R}^3} d^3x \left\lbrace (D_j\varphi)^\dagger (D_j\varphi) + U(\varphi^\dagger\varphi)\right\rbrace
\end{split}
\end{equation}
in which
\begin{equation}\label{eq:160}
A_0 = \Delta^{-1} \left[ -iq (\varphi c\pi_\varphi - \varphi^\dagger c\pi_\varphi^\dagger)\right]
\end{equation}
as was shown (in Eq.~(\ref{eq:142})) above. The geodesic equations result from simply dropping the `forcing term' in the \((\pi_{\varphi}^\dagger)_{,t}\) equation and thus correspond to
\begin{equation}\label{eq:161}
\varphi_{,0} - iq A_0\varphi = c\pi_\varphi^\dagger = D_0\varphi
\end{equation}
and
\begin{equation}\label{eq:162}
(c\pi_\varphi^\dagger)_{,0} - iq A_0 (c\pi_\varphi^\dagger) = 0.
\end{equation}
It follows immediately from differentiating Eq.~(\ref{eq:160}) for \(A_0\) that, \textit{for the geodesics problem}
\begin{equation}\label{eq:163}
A_{0,0} = 0.\qquad \text{(for \textit{geodesics})}
\end{equation}
Combining Eqs.~(\ref{eq:161}), (\ref{eq:162}) and (\ref{eq:163}) one arrives at a second order form for the geodesic equations
\begin{equation}\label{eq:164}
\begin{split}
D_0D_0\varphi &= \varphi_{,00} - 2iq A_0\varphi_{,0} - q^2A^2_0\varphi\\
 &= 0.
\end{split}
\end{equation}
The general solution of this equation is expressible as
\begin{equation}\label{eq:165}
\varphi = (\alpha + \beta x^0) e^{iqA_0x^0}
\end{equation}
where \(\alpha\) and \(\beta\) are `arbitrary' complex fields independent of \(x^0\). One easily finds that
\begin{equation}\label{eq:166}
\varphi (D_0\varphi)^\dagger - \varphi^\dagger (D_0\varphi) = \beta^\dagger\alpha - \beta\alpha^\dagger
\end{equation}
so that \(A_0\) becomes expressible as
\begin{equation}\label{eq:167}
\begin{split}
iqA_0 &= q^2\Delta^{-1} \left[\varphi (D_0\varphi)^\dagger - \varphi^\dagger (D_0\varphi)\right]\\
 &= q^2\Delta^{-1} (\beta^\dagger\alpha - \beta\alpha^\dagger)
\end{split}
\end{equation}
which explicitly displays its time independence.

For the exponential map however we want the geodesic expressed in terms of tangent space initial data \(\left\lbrace\varphi\right|_{x^0=0},\varphi_{,0}\left|_{x^0=0}\right\rbrace\) but, whereas \(\left.\varphi\right|_{x^0=0} = \alpha\), one finds that
\begin{equation}\label{eq:168}
\beta = \left.\varphi_{,0}\right|_{x^0=0} - iqA_0\left.\varphi\right|_{x^0=0}
\end{equation}
which, in view of (\ref{eq:167}), is difficult to solve for \(\beta\). Using the alternative expression for \(A_0\) given by (\ref{eq:139}), however, one can write
\begin{equation}\label{eq:169}
\begin{split}
iqA_0 &= \left.\left\lbrace q^2\Delta^{-1}_\varphi \left[\varphi (\partial_0\varphi^\dagger) - \varphi^\dagger (\partial_0\varphi)\right]\right\rbrace\right|_{x^0=0}\\
 &= q^2\Delta^{-1}_\alpha [\alpha\zeta^\dagger - \alpha^\dagger\zeta]
\end{split}
\end{equation}
where \(\zeta := \left.\varphi_{,0}\right|_{x^0=0}\). Substituting these expressions into (\ref{eq:165}) yields the derived formula for geodesics expressed in terms of tangent space initial data \(\lbrace\alpha ,\zeta\rbrace\):
\begin{equation}\label{eq:170}
\varphi = \left(\alpha (1 - iqA_0x^0) + \zeta x^0\right) e^{iqA_0x^0}.
\end{equation}
Evaluating this at a fixed `unit of time' \(x^0 = \ell^0 = ct^0\) and defining the \textit{`normal' coordinate} \textit{h} by\footnote{More precisely, actual normal coordinates would be the components of an expansion of the coordinate vector \textit{h} in terms of an orthonormal basis for the tangent space, \(T_\alpha\mathcal{S}\), to \(\mathcal{S}\) at the point \(\varphi = \alpha\). Since there is no apparent `canonical' choice for such a basis we shall leave it unspecified in the discussion to follow. In terms of any such (herein suppressed) choice of actual normal coordinates, however, the metric at \(\varphi = \alpha\) would simplfy to an explicitly `Cartesian' form.}
\begin{equation}\label{eq:171}
h := \ell^0\zeta = \left.\ell^0 (\partial_0\varphi)\right|_{x^0=0}
\end{equation}
one arrives at our explicit formula for the exponential map
\begin{equation}\label{eq:172}
\varphi_h = \left\lbrace\alpha \left( 1 - q^2\Delta_\alpha^{-1} (\alpha h^\dagger - \alpha^\dagger h)\right) + h\right\rbrace e^{q^2\Delta_\alpha^{-1} [\alpha h^\dagger - \alpha^\dagger h]}
\end{equation}
which for arbitrary fixed \(\alpha\), will be smoothly invertible on a sufficiently small `normal' neighborhood of this chosen point which, of course, corresponds to the `origin' \(h = 0\).

To compute the metric \(\mathsf{g}\) in normal coordinates we need only evaluate the kinetic energy term \(\mathcal{K}_\varphi\) (c.f. Eq.~(\ref{eq:151})) along an arbitrary differentiable curve (in the chosen chart for \(\mathcal{S}\)) after substituting \(\varphi_h\) for \(\varphi\) everywhere. To calculate the curvature tensor at the (arbitrary) reference point \(\alpha\), however, one only needs the transformed expression for \(\mathsf{g}\) expanded out to second order in \textit{h}. To this end note that
\begin{equation}\label{eq:173}
\begin{split}
\Delta_{\varphi_h} &:= \Delta - 2q^2\varphi^{\,\dagger}_h\varphi_h\\
 &= \Delta_\alpha + \mathcal{F}
\end{split}
\end{equation}
where
\begin{equation}\label{eq:174}
\Delta_\alpha := \Delta - 2q^2\alpha^\dagger\alpha\
\end{equation}
and
\begin{equation}\label{eq:175}
\begin{split}
\mathcal{F} &= -2q^2 (\alpha^\dagger h + \alpha h^\dagger) - 2q^2 \left\lbrace h^\dagger h - \alpha^\dagger\alpha \left( q^2\Delta_\alpha^{-1} (\alpha h^\dagger - \alpha^\dagger h)\right) \left( q^2\Delta_\alpha^{-1} (\alpha h^\dagger - \alpha^\dagger h)\right)\right.\\
 &\quad \left. + (\alpha^\dagger h - h^\dagger\alpha) q^2\Delta_\alpha^{-1} (\alpha h^\dagger - \alpha^\dagger h)\right\rbrace.
\end{split}
\end{equation}
The latter expresses \(\mathcal{F}\) as an explicit sum of first and second order terms,
\begin{equation}\label{eq:176}
\mathcal{F} := \one{\mathcal{F}} + \two{\mathcal{F}}
\end{equation}
with
\begin{equation}\label{eq:177}
\one{\mathcal{F}} = -2q^2 (\alpha^\dagger h + \alpha h^\dagger).
\end{equation}
What we actually need however is the inverse operator \(\Delta_{\varphi_h}^{-1}\) expanded to second order in \textit{h}.

Note, however, that, for any field \(\mathcal{B}\) lying in the range of \(\Delta_{\varphi_h}^{-1}\), we have
\begin{equation}\label{eq:178}
\begin{split}
\mathcal{B} &= \Delta_{\varphi_h} (\Delta_{\varphi_h}^{-1}\mathcal{B})\\
 &= (\Delta_\alpha + \mathcal{F}) (\Delta_{\varphi_h}^{-1}\mathcal{B})
\end{split}
\end{equation}
so that
\begin{equation}\label{eq:179}
\begin{split}
\Delta_{\varphi_h}^{-1}\mathcal{B} &= \Delta_\alpha^{-1}\mathcal{B} - \Delta_\alpha^{-1} \left[\mathcal{F} (\Delta_{\varphi_h}^{-1}\mathcal{B})\right]\\
 &= \Delta_\alpha^{-1}\mathcal{B} - \Delta_\alpha^{-1} \left[\mathcal{F}\left(\Delta_\alpha^{-1}\mathcal{B} - \Delta_\alpha^{-1} \left[\mathcal{F} (\Delta_{\varphi_h}^{-1}\mathcal{B})\right]\right)\right]\\
 &= \Delta_\alpha^{-1} \left\lbrace\mathcal{B} - \mathcal{F} \left(\Delta_\alpha^{-1} \left[\mathcal{B} - \mathcal{F} (\Delta_\alpha^{-1}\mathcal{B})\right]\right)\right\rbrace\\
 &\quad + \mathcal{O} (|h|^3)
\end{split}
\end{equation}
One could have iterated the intermediate steps above to get the result expressed to an arbitrary high order in \textit{h} but, for the present purposes, the formula given here will suffice.

To evaluate the transformed kinetic energy we need to apply \(\Delta_{\varphi_h}^{-1}\) to the specific quantity
\begin{equation}\label{eq:180}
\mathcal{B} = (\varphi_h^\dagger)_{,0} \varphi_h - (\varphi_h)_{,0} \varphi_h^\dagger.
\end{equation}
Expanding this expression out through the use of (\ref{eq:172}) one arrives at
\begin{equation}\label{eq:181}
\mathcal{B} = \zero{\mathcal{B}} + \one{\mathcal{B}} + \two{\mathcal{B}}
\end{equation}
where
\begin{align}
\zero{\mathcal{B}} &:= (\alpha h^{\,\dagger}_{,0} - \alpha^\dagger h_{,0})\label{eq:182}\\
\begin{split}
\one{\mathcal{B}} &= \left[ hh^{\,\dagger}_{,0} - h^\dagger h_{,0} - (\alpha h^{\,\dagger}_{,0} + \alpha^\dagger h_{,0}) q^2\Delta_\alpha^{-1} (\alpha h^\dagger - a^\dagger h)\right.\\
 &\quad \left. + (\alpha^\dagger h + h^\dagger\alpha) q^2 \Delta_\alpha^{-1} (\alpha^\dagger h_{,0} - \alpha h^{\,\dagger}_{,0})\right]\label{eq:183}
\end{split}
\end{align}
and
\begin{equation}\label{eq:184}
\begin{split}
\two{\mathcal{B}} &:= \left[ 2 (\alpha^\dagger h - h^\dagger\alpha) q^2\Delta_\alpha^{-1} (\alpha h^\dagger - \alpha^\dagger h)\right.\\
 &\quad - 2\alpha^\dagger\alpha \left(q^2\Delta_\alpha^{-1} (\alpha h^\dagger - \alpha^\dagger h)\right) \left( q^2\Delta_\alpha^{-1} (\alpha h^\dagger - \alpha^\dagger h)\right)\\
 &\quad + \left. 2 h^\dagger h\right] q^2\Delta_\alpha^{-1} (\alpha^\dagger h_{,0} - \alpha h^{\,\dagger}_{,0}).
\end{split}
\end{equation}
A useful identity satisfied by the \(\eye{\mathcal{B}}\) and \(\eye{\mathcal{F}}\) is:
\begin{equation}\label{eq:185}
\two{\mathcal{B}} - \two{\mathcal{F}}\Delta_\alpha^{-1} \zero{\mathcal{B}} = 0.
\end{equation}

Assembling these various components for the kinetic energy \(\mathcal{K}_\varphi\) and retaining terms explicitly only through second order in \textit{h} one finally arrives at:
\begin{equation}\label{eq:186}
\begin{split}
\mathcal{K}_\varphi &= \int_{\mathbb{R}^3} d^3 x \left\lbrace h^{\,\dagger}_{,0} h_{,0} - \frac{q^2}{2} (\alpha h^{\,\dagger}_{,0} - \alpha^\dagger h_{,0}) \Delta_\alpha^{-1} (\alpha h^{\,\dagger}_{,0} - \alpha^\dagger h_{,0})\right\rbrace\\
 &\quad - \frac{q^2}{2} \int_{\mathbb{R}^3} d^3 x \left\lbrace(\one{\mathcal{B}} - \one{\mathcal{F}}\Delta_\alpha^{-1}\zero{\mathcal{B}}) \Delta_\alpha^{-1} (\one{\mathcal{B}} - \one{\mathcal{F}} \Delta_\alpha^{-1}\zero{\mathcal{B}})\right\rbrace\\
 &\quad + O (|h|^3)
\end{split}
\end{equation}
where
\begin{equation}\label{eq:187}
\begin{split}
\one{\mathcal{B}} - \one{\mathcal{F}} \Delta_\alpha^{-1}\zero{\mathcal{B}}
&= 2 i\epsilon_{ab} \left[ h_{,0}^a + \epsilon^{af} \alpha_f 2q^2 \Delta_\alpha^{-1} (\epsilon_{cd} h^c_{,0} \alpha^d)\right]\\
&\quad\times \left[ h^b + \epsilon^{bg} \alpha_g 2q^2 \Delta_\alpha^{-1} (\epsilon_{mn} h^m\alpha^n)\right]
\end{split}
\end{equation}
wherein
\begin{align}
\alpha_f = \delta_{fg}\alpha^g = \alpha^f,&\qquad \epsilon_{cd} = \delta_{cm}\delta_{dn}\epsilon^{mn} = \epsilon^{cd}\label{eq:188}\\
h = h^1 + ih^2,&\qquad \alpha = \alpha^1 + i\alpha^2\label{eq:189}\\
h^\dagger = h^1 - ih^2, &\qquad \alpha^\dagger = \alpha^1 - i\alpha^2\label{eq:190}
\end{align}
and \(\epsilon^{ab} = -\epsilon^{ba}\) with \(\epsilon^{12} = 1\) as before.

Noting that
\begin{equation}\label{eq:192}
\begin{split}
\alpha h^{\,\dagger}_{,0} - \alpha^{\dagger} h_{,0} &= -2i\epsilon_{cd} \alpha^c h^d_{,0}\\
 = -\frac{2i}{c} \epsilon_{cd} \alpha^c h^c_{,t}
 \end{split}
\end{equation}
and recalling Eq.~(\ref{eq:149}) it is straightforward to verify that the first integral on the right hand side of Eq.~(\ref{eq:186}) is simply 1/2 the squared norm of the velocity vectors \(h_{,t}\) evaluated in the metric at \(\varphi = \alpha\). As explained in the footnote for (\ref{eq:171}) this expression would simplify to purely `Cartesian' form if \(h_{,t}\) were expanded in actual normal coordinates there.

From the classical, Riemannian result for the expansion of a metric in normal coordinates it follows that the second integral on the right hand side of Eq.~(\ref{eq:186}) is \(-1/6\) of the curvature tensor of the metric (\ref{eq:149}) at \(\varphi = \alpha\) evaluated, on both its first and last pair of `slots', on the tangent plane spanned by the vectors \textit{h} and \(h_{,t}\). As such it corresponds (up to the usual normalization factor expressible in terms of the `dot' products of these vectors) to the sectional curvature of this metric at the point \(\alpha\). Again, as explained in the previous footnote, this expression would directly yield the normal coordinate components of the sectional curvature at \(\alpha\) if the vectors \textit{h} and \(h_{,t}\) were both expressed in a common orthonormal basis for the tangent space \(T_\alpha \mathcal{S}\).

Furthermore, in view of the factors of \textit{i} in the defining equation (\ref{eq:187}) of \(\one{\mathcal{B}} - \one{\mathcal{F}} \Delta^{-1}_\alpha \zero{\mathcal{B}}\) and of the negativity of the operator \(\Delta_\alpha\) defined by Eq.~(\ref{eq:115}), it is clear that the curvature defined via Eq.~(\ref{eq:186}) is everywhere non-negative (i.e., \(\forall\; \alpha\) and for any pair \(\lbrace h, h_{,t}\rbrace\) in \(T_\alpha\mathcal{S}\)) but also that it vanishes on those 2-planes in \(T_\alpha\mathcal{S}\) for which \(\one{\mathcal{B}} - \one{\mathcal{F}} \Delta^{-1}_\alpha \zero{\mathcal{B}}\) vanishes.

The authors have not, so far, decided which regularization scheme fits most naturally with their overall \textit{Euclidean-signature semi-classical} program. Such a decision is not needed until the higher order, quantum `loop corrections' to field theoretic problems are under construction. These latter however (as one can see from sections IIB and IVB of Ref.~\cite{Moncrief:2012} which treats the analogue quantum mechanical systems) will be governed entirely by the integration of first order, linear transport equations of a comparatively elementary type. By contrast we have instead focussed our efforts so far on solving the analytically more challenging, \textit{uniquely nonlinear} functional partial differential equations for the fields of interest --- namely the corresponding Euclidean-signature vanishing-energy functional Hamilton-Jacobi equations (c.f., sections 3.2 and 3.3 herein).

Until we do settle upon an appropriate regularization scheme we cannot, consistently,  carry out the regularized construction of the relevant orbit space Ricci tensors for our program (or, for that matter, their loop corrected Bakry-Emery `enhancements'). One hopes though, as is often the case in quantum field theory, that the result aimed for (e.g., positivity of the relevant Bakry-Emery Ricci tensor) will not crucially depend upon the method of regularization employed. 

\section{Euclidean-Signature Semi-Classical Methods for (Mini-Superspace) Quantum Cosmology}
\label{sec:euclidean-cosmology}
Though it is somewhat peripheral to the central issues discussed herein we have begun to explore the applicability of Euclidean-signature semi-classical methods to the problem of solving, at least asymptotically, the Wheeler-DeWitt equation of canonical quantum gravity. Since this (functional differential) equation has, at present however, only a formal significance we actually began by analyzing instead the mathematically well-defined model problem of constructing asymptotic solutions to the idealized Wheeler-DeWitt equation for spatially homogenous, Bianchi type IX (or `Mixmaster') universes. Though the (partial differential) Wheeler-DeWitt equation for the model problem was first formulated nearly a half century ago, techniques for solving it that bring to light the discrete, \textit{quantized} character naturally to be expected for its solutions have, only recently, been developed. In particular we shall sketch below how the \textit{microlocal} analytical methods, already well-established for the study of conventional Schr{\"o}dinger eigenvalue problems, can be modified in such a way as to apply to the (Mixmaster) Wheeler-DeWitt equation.

That some essential modification of the microlocal methods will be needed is evident from the fact that the Wheeler-DeWitt equation does not define an eigenvalue problem, in the conventional sense, \textit{at all}. For closed universe models, such as those of Mixmaster type, all of the would-be eigenvalues of the Wheeler-DeWitt operator, whether for `ground' or `excited' quantum states, are required to \textit{vanish identically}. But a crucial feature of standard microlocal methods, when applied to conventional Schr{\"o}dinger eigenvalue problems, exploits the flexibility to adjust the eigenvalues being generated, order-by-order in an expansion in Planck's constant, to ensure the global smoothness of the eigenfunctions, being constructed in parallel, at the corresponding order. But if, as in the Wheeler-DeWitt problem, there are no eigenvalues to adjust, wherein lies the flexibility needed to ensure the required smoothness of the hypothetical eigenfunctions? And, by the same token, where are the `quantum numbers' that one would normally expect to have at hand to label the distinct quantum states? Remarkably however, as was shown in detail in Ref.~\cite{Moncrief:2015}, the scope of microlocal methods can indeed, in spite of this apparent impasse, be broadened to provide creditable, aesthetically appealing answers to the questions raised above. We shall briefly review below the key steps in this analysis.

\subsection{Mixmaster Spacetimes}
\label{subsec:mixmaster_spacetimes}
The Bianchi IX, or `Mixmaster' cosmological models are spatially homogeneous spacetimes defined on the manifold \(\mathbb{S}^3 \times \mathbb{R}\). Their metrics can be conveniently expressed in terms of a basis, \(\lbrace\sigma^i\rbrace\), for the left-invariant one-forms of the Lie group \(SU(2)\) which of course is diffeomorphic to the `spatial' manifold under study. In a standard Euler angle coordinate system for \(\mathbb{S}^3\) these basis one-forms can be written as:
\begin{equation}\label{eq:501}
\begin{split}
\sigma^1 &= \cos{\psi}\; d\theta + \sin{\psi}\sin{\theta}\; d\varphi,\\
\sigma^2 &= \sin{\psi}\; d\theta - \cos{\psi}\sin{\theta}\; d\varphi,\\
\sigma^3 &= d\psi + \cos{\theta}\; d\varphi
\end{split}
\end{equation}
and satisfy
\begin{equation}\label{eq:502}
d\sigma^i = \frac{1}{2} \epsilon_{ijk}\; \sigma^j \wedge \sigma^k
\end{equation}
where \(\epsilon_{ijk}\) is completely antisymmetric with \(\epsilon_{123} = 1\).

In the absence of matter sources for the Einstein equations (ie., in the so-called `vacuum' case) it is well-known that the Mixmaster spacetime metric can always be put, after a suitable frame `rotation', into diagonal form. Thus, without essential loss of generality, one can write the line element for vacuum, Bianchi IX models in the form
\begin{equation}\label{eq:503}
\begin{split}
ds^2 &= {}^{(4)}g_{\mu\nu}\; dx^\mu\; dx^\nu\\
 &= -N^2 dt^2 + \frac{L^2}{6\pi} e^{2\alpha} (e^{2\beta})_{ij}\; \sigma^i\sigma^j
\end{split}
\end{equation}
where \(\lbrace x^\mu\rbrace = \lbrace t, \theta, \varphi, \psi, \rbrace\) with \(t \in \mathbb{R}, e^{2\beta}\) is a diagonal, positive definite matrix of unit determinant and \textit{L} is a positive constant with the dimensions of `length'.

In the notation introduced by Misner \cite{Misner:1973} one writes
\begin{equation}\label{eq:504}
(e^{2\beta}) = \hbox{diag}(e^{2\beta_+ + 2\sqrt{3}\beta_-},  e^{2\beta_+ - 2\sqrt{3}\beta_-}, e^{-4\beta_+})
\end{equation}
and thereby expresses \(e^{2\beta}\) in terms of his (arbitrary, real-valued) anisotropy parameters \(\lbrace\beta_+, \beta_-\rbrace\). These measure the departure from `roundness' of the homogeneous, Riemannian metric on \(\mathbb{S}^3\) given by
\begin{equation}\label{eq:505}
\gamma_{ij} dx^i \otimes dx^j := \frac{L^2}{6\pi} e^{2\alpha} (e^{2\beta})_{ij}\; \sigma^i \otimes \sigma^j
\end{equation}
whereas the remaining (arbitrary, real-valued) parameter \(\alpha\) determines the sphere's overall `size' (in units of \textit{L}).

To ensure spatial homogeneity the metric functions \(\lbrace N, \alpha, \beta_+, \beta_-\rbrace\) can only depend upon the time coordinate \textit{t} which, for convenience, we take to be dimensionless. To ensure the uniform Lorentzian signature of the metric \({}^{(4)}g\) the `lapse' function \textit{N} must be non-vanishing (and, with our conventions, have the dimension of length). Taken together the parameters \(\lbrace\alpha, \beta_+, \beta_-\rbrace\) coordinatize the associated `mini-superspace' of spatially homogeneous, diagonal Riemannian metrics on \(\mathbb{S}^3\). This mini-superspace is the natural configuration manifold for the Mixmaster dynamics.

The ADM (Arnowitt, Deser and Misner \cite{misc-misner}) action for these Bianchi IX models, which differ from the Hilbert action by an inessential boundary term, is given by
\begin{equation}\label{eq:506}
\begin{split}
I_{\mathrm{ADM}} &:= \frac{c^3L^3\pi}{G(6\pi)^{3/2}} \int_I dt\; \left\lbrace\frac{6e^{3\alpha}}{N} (-\dot{\alpha}^2 + \dot{\beta}_+^2 + \dot{\beta}_-^2)\right.\\
 &{} - \frac{(6\pi)Ne^\alpha}{2L^2} \left\lbrack e^{-8\beta_+} - 4e^{-2\beta_+} \cosh{(2\sqrt{3}\beta_-)}\right.\\
 &\quad {} \left.\vphantom{\frac{(6\pi)Ne^\alpha}{2L^2}}\left. + 2e^{4\beta_+} \left(\cosh{(4\sqrt{3}\beta_-)} - 1\right)\right\rbrack\right\rbrace\\
 &:= \int_I L_{\mathrm{ADM}} dt
\end{split}
\end{equation}
in which \(\dot{\alpha} = \frac{d\alpha}{dt}\) etc. and where \textit{I} is an arbitrary interval of the form \([t_0, t_1] \subset \mathbb{R}\) and \textit{G} is Newton's constant. The corresponding Hamiltonian formulation is arrived at via the Legendre transformation
\begin{align}
p_\alpha &:= \frac{\partial L_{\mathrm{ADM}}}{\partial\dot{\alpha}} = \frac{-c^3L^3\pi}{G(6\pi)^{3/2}} \frac{12e^{3\alpha}\dot{\alpha}}{N}\label{eq:507}\\
p_\pm &:= \frac{\partial L_{\mathrm{ADM}}}{\partial\dot{\beta}_\pm} = \frac{c^3L^3\pi}{G(6\pi)^{3/2}} \frac{12e^{3\alpha}\dot{\beta}_\pm}{N}.\label{eq:508}
\end{align}

In terms of the canonical variables \(\lbrace\alpha, \beta_+, \beta_-, p_\alpha, p_+, p_-\rbrace\) the ADM action takes the form
\begin{equation}\label{eq:509}
I_{\mathrm{ADM}} = \int_I dt\; \left\lbrace p_\alpha\dot{\alpha} + p_+\dot{\beta}_+ + p_-\dot{\beta}_- - N\mathcal{H}_\perp\right\rbrace
\end{equation}
where
\begin{equation}\label{eq:510}
\begin{split}
\mathcal{H}_\perp &:= \frac{(6\pi)^{1/2}G}{4c^3L^3e^{3\alpha}}\; \left\lbrace\vphantom{\frac{1}{2}} (-p_\alpha^2 + p_+^2 + p_-^2)\right.\\
 &{} + \left(\frac{c^3}{G}\right)^2 L^4 e^{4\alpha}\; \left\lbrack\frac{e^{-8\beta_+}}{3} - \frac{4e^{-2\beta_+}}{3}\; \cosh{(2\sqrt{3}\beta_-)}\right.\\
 &\quad {} + \left.\vphantom{\frac{1}{2}}\left.\frac{2}{3} e^{4\beta_+}\; \left(\cosh{(4\sqrt{3}\beta_-)} - 1\right)\right\rbrack\right\rbrace.
\end{split}
\end{equation}
Variation of the lapse function \textit{N}, which only appears now in Lagrange multiplier form, leads to that Einstein equation known as the `Hamiltonian constraint',
\begin{equation}\label{eq:511}
\mathcal{H}_\perp (\alpha, \beta_+, \beta_-, p_\alpha, p_+, p_-) = 0
\end{equation}
whereas variation of the canonical variables leads to the Hamiltonian evolution equations
\begin{align}
\dot{\alpha} &= \frac{\partial H_{\mathrm{ADM}}}{\partial p_\alpha}, \quad \dot{\beta}_\pm = \frac{\partial H_{\mathrm{ADM}}}{\partial p_\pm}\label{eq:512}\\
\dot{p}_\alpha &= -\frac{\partial H_{\mathrm{ADM}}}{\partial\alpha}, \quad \dot{p}_\pm = -\frac{\partial H_{\mathrm{ADM}}}{\partial\beta_\pm}\label{eq:513}
\end{align}
with so-called super-Hamiltonian given by
\begin{equation}\label{eq:514}
H_{\mathrm{ADM}} := N\mathcal{H}_\perp.
\end{equation}
The choice of lapse function \textit{N} is essentially arbitrary but determines the coordinate `gauge' by assigning a geometrical meaning to the time function \textit{t}. For example the choice \(N = L\) corresponds to taking \(t = \frac{c}{L}\tau\) where \(\tau\) is `proper time' normal to the hypersurfaces of homogeneity. The Hamiltonian constraint (\ref{eq:511}) is conserved in time by the evolution equations (\ref{eq:512}--\ref{eq:513})  independently of the choice of lapse. Equations (\ref{eq:511}) and (\ref{eq:512}--\ref{eq:513}) comprise the full set of Einstein equations for these models.

Though the general solution to the Mixmaster equations of motion is not known, much is known about the dynamical behavior and asymptotics of the resulting spacetimes. One can show for example that each such cosmological model expands from a `big bang' singularity of vanishing spatial volume, \(\alpha \rightarrow -\infty\), a finite proper time in the past, achieves a momentary maximal volume at some finite proper time from the big bang and then `recollapses' to another vanishing-volume, `big crunch' singularity a finite proper time in the future \cite{Lin:1989,Lin:1990,Rendall:1997,Ringstrom:2001}. For the generic solution spacetime curvature can be proven to blow up at these singular boundaries \cite{Ringstrom:2000} but some exceptional cases, so-called Taub universes \cite{Taub:1951,Misner:1969c}, develop (compact, null hypersurface) Cauchy horizons \(\approx \mathbb{S}^3\) instead of curvature singular boundaries and are analytically extendable through these horizons to certain acausal  NUT (Newman, Unti, Tamburino) spacetimes that admit closed timelike curves \cite{Newman:1963,Misner:1963}. The inextendability of the generic, vacuum Mixmaster spacetime is consistent with Penrose's (strong) \textit{cosmic censorship conjecture} according to which the maximal Cauchy developments of generic, globally hyperbolic solutions to the (vacuum) Einstein field equations should not allow such acausal extensions.

The dynamical behavior of the generic solution to equations (\ref{eq:511}--\ref{eq:513}), between its big bang and big crunch singular boundaries, entails an infinite sequence of intricate `bounces' of the evolving system point in mini-superspace, \((\alpha(t), \beta_+(t), \beta_-(t))\), off of the `walls' provided by the potential energy function
\begin{equation}\label{eq:515}
\begin{split}
\mathcal{U}(\alpha,\beta_+,\beta_-) &:= \frac{c^3(6\pi)^{1/2}Le^\alpha}{4G}\; \left\lbrack\frac{e^{-8\beta_+}}{3} - \frac{4}{3} e^{-2\beta_+} \cosh{(2\sqrt{3}\beta_-)}\right.\\
 &\vphantom{\frac{4}{3}}\left. + \frac{2}{3} e^{4\beta_+}\; \left(\cosh{(4\sqrt{3}\beta_-)} - 1\right)\right\rbrack
\end{split}
\end{equation}
appearing in the gravitational super-Hamiltonian \(H_{\mathrm{ADM}} = N\mathcal{H}_\perp.\) This sequence of bounces has been extensively analyzed with various analytical and numerical approximation methods beginning with the fundamental investigations of Belinski\v{\i}, Khalanikov and Lifshitz (BKL) \cite{Belinskii:1969,Belinskii:1970} and Misner \cite{Misner:1969a}. The insights gained therefrom led Belinski\v{\i}, et al to the bold conjecture that the Mixmaster dynamics provides a paradigm for the behavior of a generic, \textit{non-symmetric} cosmological model at a spacelike singular boundary \cite{Belinskii:1982,Uggla:2013}.
The study of such BKL oscillations within models of increasing generality and complexity is a continuing, significant research area within mathematical cosmology \cite{Berger:2001,Berger:1998,Berger:2000}. Though Newtonian definitions of `chaos' do not strictly apply to the Mixmaster dynamical system certain natural extensions of this concept have led to the conclusion that Mixmaster dynamics is indeed `chaotic' in a measurably meaningful sense \cite{Hobill:1994,Cornish:1997}.

At the same time it has long been suspected that quantum effects should dramatically modify the nature of the Mixmaster evolutions especially when the evolving universe models reach a size comparable to the so-called Planck length, i.e., when \(Le^\alpha\) becomes comparable to \(L_{\mathrm{Planck}} \simeq 1.616 \times 10^{-33}~\mathrm{cm}\). This suspicion led Misner to initiate the study of Mixmaster quantum cosmology \cite{Misner:1969b}, the subject to which we now turn.

\subsection{The Wheeler-DeWitt Equation for Mixmaster Universes}
\label{subsec:wheeler-dewitt}
One can formally quantize the Mixmaster dynamical system described above by working in the Schr\"{o}dinger representation wherein quantum states are expressed as `wave' functions of the canonical coordinates, \(\Psi(\alpha,\beta_+,\beta_-)\), and the conjugate momenta to these variables are replaced by differential operators:
\begin{equation}\label{eq:516}
\begin{split}
p_\alpha &\longrightarrow \hat{p}_\alpha := \frac{\hbar}{i}\; \frac{\partial}{\partial\alpha},\\
p_+ &\longrightarrow \hat{p}_+ := \frac{\hbar}{i}\; \frac{\partial}{\partial\beta_+},\\
p_- &\longrightarrow \hat{p}_- := \frac{\hbar}{i}\; \frac{\partial}{\partial\beta_-}.
\end{split}
\end{equation}
Here \(\hbar = \frac{h}{2\pi}\) where \textit{h} is Planck's constant given by \(h \simeq 6.62606957 \times 10^{-27}~\mathrm{erg \cdot sec}\).

In this picture one converts, after making a suitable choice of operator ordering, the classical Hamiltonian constraint function \(\mathcal{H}_\perp\) into a quantum operator \(\hat{\mathcal{H}}_\perp\) and imposes it, \`{a} la Dirac, as a fundamental constraint on the allowed quantum states by setting
\begin{equation}\label{eq:517}
\hat{\mathcal{H}}_\perp \Psi = 0.
\end{equation}
Since this equation is an idealized, finite dimensional model for the formal equation proposed by Wheeler and DeWitt for full, non-symmetric, canonical quantum gravity (formulated on the infinite dimensional `superspace' of Riemannian geometries \cite{Fischer:1970,Giulini:2009}) we shall refer to it as the Wheeler-DeWitt (WDW) equation for Mixmaster spacetimes.

For simplicity we shall limit our attention here to a particular one-parameter family of operator orderings for \(\hat{\mathcal{H}}_\perp\), first introduced by Hartle and Hawking \cite{Hartle:1983}, and characterized by the specific substitutions
\begin{align}
-e^{-3\alpha}\; p_\alpha^2 &\longrightarrow \frac{\hbar^2}{e^{(3-B)\alpha}}\; \frac{\partial}{\partial\alpha}\; \left(e^{-B\alpha}\frac{\partial}{\partial\alpha}\right),\label{eq:518}\\
e^{-3\alpha}\; p_+^2 &\longrightarrow \frac{-\hbar^2}{e^{3\alpha}}\; \frac{\partial^2}{\partial\beta^2_+},\label{eq:519}\\
e^{-3\alpha}\; p_-^2 &\longrightarrow \frac{-\hbar^2}{e^{3\alpha}}\; \frac{\partial^2}{\partial\beta^2_-},\label{eq:520}
\end{align}
for the `kinetic energy' terms appearing in \(\hat{\mathcal{H}}_\perp\). Here \textit{B} is an arbitrary real parameter whose specification determines a particular ordering of the family. For any such ordering the WDW equation can be written as
\begin{equation}\label{eq:521}
\begin{split}
& \left(\frac{L_{\mathrm{Planck}}}{L}\right)^3\; \left\lbrace e^{-(3-B)\alpha} \frac{\partial}{\partial\alpha}\; \left( e^{-B\alpha} \frac{\partial\Psi}{\partial\alpha}\right) - e^{-3\alpha} \left(\frac{\partial^2\Psi}{\partial\beta_+^2} + \frac{\partial^2\Psi}{\partial\beta_-^2}\right)\right\rbrace\\
& + \left(\frac{L}{L_{\mathrm{Planck}}}\right) e^\alpha \left\lbrack\frac{e^{-8\beta_+}}{3} - \frac{4}{3} e^{-2\beta_+} \cosh{(2\sqrt{3}\beta_-)} + \frac{2}{3} e^{4\beta_+} \left(\cosh{(4\sqrt{3}\beta_-)} - 1\right)\right\rbrack \Psi\\
&  = 0
\end{split}
\end{equation}
where \(L_{\mathrm{Planck}}\) is the Planck length defined by
\begin{equation}\label{eq:522}
L_{\mathrm{Planck}} = \left(\frac{G\hbar}{c^3}\right)^{1/2} \simeq 1.616199 \times 10^{-33}~\mathrm{cm}.
\end{equation}
Notice that the arbitrary `length' constant \textit{L} always occurs in the combination \(Le^\alpha\) so that a change of its value merely corresponds to a shift of \(\alpha\) by an additive constant.

Notice in addition that when the WDW equation, \(\hat{\mathcal{H}}_\perp\Psi = 0\), is imposed to constrain the allowed, so-called `physical', quantum states, then the conventional Schr\"{o}dinger equation, which would be expected to have the form
\begin{equation}\label{eq:523}
i\hbar \frac{\partial\Psi}{\partial t} = \hat{H}_{\mathrm{ADM}}\Psi = N\hat{\mathcal{H}}_\perp\Psi,
\end{equation}
reduces to the seemingly mysterious implication that physical states do not evolve in `time', i.e., to the conclusion that \(\frac{\partial\Psi}{\partial t} = 0\).

This result is a reflection of the conceptual `problem of time' in canonical quantum cosmology for the case of (spatially) closed universes. It leads one inexorably to the conclusion that actual temporal evolution must be measured not with respect to some external, `absolute' time, as in Newtonian or even special relativistic physics, but rather with respect to some internal `clock' contained within the system itself. The most obvious such clock variable for the Mixmaster models is the logarithmic scale parameter \(\alpha\) whose value, classically, determines the instantaneous spatial `size' of the model universe and which, again classically, evolves in an \textit{almost monotonic} fashion. More precisely \(\alpha\) increases monotonically during the epoch of cosmological expansion, stops for an instant at the moment of maximal volume and then decreases monotonically during the followup epoch of cosmological collapse until the final `big crunch'.

But, as Misner was the first to realize, the Wheeler-DeWitt equation for Mixmaster models does not have Schr\"{o}dinger form and so many of the usual constructions, familiar from ordinary quantum mechanics, such as the eigenfunctions and eigenvalues of a self-adjoint Hamiltonian operator acting on a naturally associated Hilbert space of quantum states and the conservation, in `time', of the Hilbert space norm of such evolving states, no longer seem to apply. The Wheeler-DeWitt equation is indeed a wave equation (though not one of Schr\"{o}dinger type), but where is the discreteness, expected of a normal \textit{quantum} system, to be found among its solutions?

In the section below we shall bring certain microlocal analysis techniques, already well-developed for the study of conventional Schr\"{o}dinger eigenvalue problems \cite{Dimassi:1999,Helfer:1988,Helfer:1984,Moncrief:2012}, to bear on such questions and sketch how these techniques can indeed be extended to apply to the Mixmaster Wheeler-DeWitt equation.

At first sight though it is not apparent that such microlocal methods can be applied at all. In particular, for a conventional Schr\"{o}dinger eigenvalue problem, they make crucial use of the freedom to adjust the eigenvalues under construction, order-by-order in an expansion in Planck's constant, to ensure the global smoothness of the eigenfunctions being generated at  the corresponding order. But for the Wheeler-DeWitt problem all eigenvalues of \(\hat{\mathcal{H}}_\perp\), whether for `ground' or `excited' states (whatever those terms might ultimately be taken to mean) are required to vanish to all orders with \textit{no flexibility whatsoever}. And if no meaningful eigenvalues can be defined wherein are the `quanta' naturally demanded of a \textit{quantized} system?

As we shall see however the special structure of the Wheeler-DeWitt operator, \(\hat{\mathcal{H}}_{\!\perp}\), and the fact that it is \textit{not} of Schr\"{o}dinger type, comes to the rescue and allows one to generate smooth, globally defined expansions (to all orders in Planck's constant) for both ground and excited states. These states are labeled by a pair of non-negative integers that can be naturally interpreted as \textit{graviton excitation numbers} for the ultra-long-wavelength gravitational waves modes represented by the quantum dynamics of the anisotropy degrees of freedom, \(\beta_+\) and \(\beta_-\).

\subsection{Microlocal Techniques for the Mixmaster Wheeler-DeWitt Equation}
\label{subsec:microlocal-wheeler-dewitt}
In view of the resemblance of \(\hat{\mathcal{H}}_\perp\) to a conventional Schr\"{o}dinger operator one is motivated to propose a `ground state' wave function of \textit{real, nodeless} type and thus to introduce an ansatz of the form
\begin{equation}\label{eq:524}
\overset{(0)}{\Psi}_{\!\hbar} = e^{-S_\hbar/\hbar},
\end{equation}
where \(S_\hbar = S_\hbar (\alpha,\beta_+,\beta_-)\) is a real-valued function on the Mixmaster mini-superspace having the dimensions of `action'. It will be convenient to define a dimensionless stand-in for \(S_\hbar\) by setting
\begin{equation}\label{eq:525}
\mathcal{S}_\hbar := \frac{G}{c^3L^2} S_\hbar
\end{equation}
and to assume that \(\mathcal{S}_\hbar\) admits a formal expansion in powers of the dimensionless ratio
\begin{equation}\label{eq:526}
X := \frac{L_{\mathrm{Planck}}^2}{L^2} = \frac{G\hbar}{c^3L^2}
\end{equation}
given by
\begin{equation}\label{eq:527}
\mathcal{S}_\hbar = \mathcal{S}_{(0)} + X\mathcal{S}_{(1)} + \frac{X^2}{2!}\mathcal{S}_{(2)} + \cdots + \frac{X^k}{k!}\mathcal{S}_{(k)} + \cdots
\end{equation}
so that \(\overset{(0)}{\Psi}_{\!\hbar}\) now becomes
\begin{equation}\label{eq:528}
\overset{(0)}{\Psi}_{\!\hbar} = e^{-\frac{1}{X}\mathcal{S}_{(0)} - \mathcal{S}_{(1)} - \frac{X}{2!}\mathcal{S}_{(2)} - \cdots}.
\end{equation}

Substituting this ansatz into the Wheeler-DeWitt equation, \(\hat{\mathcal{H}}_\perp\overset{(0)}{\Psi}_{\!\hbar} = 0\), and requiring satisfaction, order-by-order in powers of \textit{X} leads immediately to the sequence of equations:
\begin{align}
\begin{split}
&\left(\frac{\partial\mathcal{S}_{(0)}}{\partial\alpha}\right)^2 - \left(\frac{\partial\mathcal{S}_{(0)}}{\partial\beta_+}\right)^2 - \left(\frac{\partial\mathcal{S}_{(0)}}{\partial\beta_-}\right)^2 \\
&\quad + e^{4\alpha} \left\lbrack\frac{e^{-8\beta_+}}{3} - \frac{4}{3} e^{-2\beta_+} \cosh{(2\sqrt{3}\beta_-)} + \frac{2}{3} e^{4\beta_+} \left(\cosh{(4\sqrt{3}\beta_-)} - 1\right) \right\rbrack = 0,\\
\end{split}\label{eq:529}\\
\begin{split}
&2\left\lbrack\frac{\partial\mathcal{S}_{(0)}}{\partial\alpha} \frac{\partial\mathcal{S}_{(1)}}{\partial\alpha} - \frac{\partial\mathcal{S}_{(0)}}{\partial\beta_+} \frac{\partial\mathcal{S}_{(1)}}{\partial\beta_+} - \frac{\partial\mathcal{S}_{(0)}}{\partial\beta_-} \frac{\partial\mathcal{S}_{(1)}}{\partial\beta_-}\right\rbrack\\
&\quad + B\frac{\partial\mathcal{S}_{(0)}}{\partial\alpha} - \frac{\partial^2\mathcal{S}_{(0)}}{\partial\alpha^2} + \frac{\partial^2\mathcal{S}_{(0)}}{\partial\beta_+^2} + \frac{\partial^2\mathcal{S}_{(0)}}{\partial\beta_-^2} = 0,
\end{split}\label{eq:530}
\end{align}
and, for \(k \geq 2\),
\begin{equation}\label{eq:531}
\begin{split}
&2\left\lbrack\frac{\partial\mathcal{S}_{(0)}}{\partial\alpha} \frac{\partial\mathcal{S}_{(k)}}{\partial\alpha} - \frac{\partial\mathcal{S}_{(0)}}{\partial\beta_+} \frac{\partial\mathcal{S}_{(k)}}{\partial\beta_+} - \frac{\partial\mathcal{S}_{(0)}}{\partial\beta_-} \frac{\partial\mathcal{S}_{(k)}}{\partial\beta_-}\right\rbrack\\
&\quad +k\left\lbrack B\frac{\partial\mathcal{S}_{(k-1)}}{\partial\alpha} - \frac{\partial^2\mathcal{S}_{(k-1)}}{\partial\alpha^2} + \frac{\partial^2\mathcal{S}_{(k-1)}}{\partial\beta_+^2} + \frac{\partial^2\mathcal{S}_{(k-1)}}{\partial\beta_-^2}\right\rbrack\\
&\quad + \sum_{\ell = 1}^{k-1} \frac{k!}{\ell!(k-\ell)!} \left(\frac{\partial\mathcal{S}_{(\ell)}}{\partial\alpha} \frac{\partial\mathcal{S}_{(k-\ell)}}{\partial\alpha} - \frac{\partial\mathcal{S}_{(\ell)}}{\partial\beta_+} \frac{\partial\mathcal{S}_{(k-\ell)}}{\partial\beta_+} - \frac{\partial\mathcal{S}_{(\ell)}}{\partial\beta_-} \frac{\partial\mathcal{S}_{(k-\ell)}}{\partial\beta_-}\right) = 0.
\end{split}
\end{equation}

One recognizes Eq.~(\ref{eq:529}) as the \textit{Euclidean signature} analogue of the Hamilton-Jacobi equation for Mixmaster spacetimes that results from making the canonical substitutions
\begin{equation}\label{eq:532}
\begin{split}
p_\alpha &\longrightarrow \frac{\partial S}{\partial\alpha} = \frac{c^3L^2}{G} \frac{\partial\mathcal{S}}{\partial\alpha},\\
p_+ &\longrightarrow \frac{\partial S}{\partial\beta_+} = \frac{c^3L^2}{G} \frac{\partial\mathcal{S}}{\partial\beta_+},\\
p_- &\longrightarrow \frac{\partial S}{\partial\beta_-} = \frac{c^3L^2}{G} \frac{\partial\mathcal{S}}{\partial\beta_-}
\end{split}
\end{equation}
for the momenta in the Euclidean signature Hamiltonian constant, \(\mathcal{H}_{\perp\;\mathrm{Eucl}} = 0\), where
\begin{equation}\label{eq:533}
\begin{split}
\mathcal{H}_{\perp\;\mathrm{Eucl}} &:= \frac{(6\pi)^{1/2}G}{4c^3L^3e^{3\alpha}} \left\lbrace (p_\alpha^2 - p_+^2 - p_-^2)\vphantom{\left(\frac{c^3}{G}\right)^2}\right.\\
 &\left. + \left(\frac{c^3}{G}\right)^2 L^4e^{4\alpha} \left\lbrack\frac{e^{-8\beta_+}}{3} - \frac{4}{3}e^{-2\beta_+} \cosh{(2\sqrt{3}\beta_-)}\right.\right.\\
 &\left.\left.\vphantom{\left(\frac{c^3}{G}\right)^2} + \frac{2}{3} e^{4\beta_+} \left(\cosh{(4\sqrt{3}\beta_-)} - 1\right)\right\rbrack\right\rbrace .
\end{split}
\end{equation}
This expression results from repeating the derivation of \(I_{\mathrm{ADM}}\) given in Sect.~\ref{subsec:mixmaster_spacetimes}, but now for a \textit{Euclidean signature} Bianchi IX metric,
\begin{equation}\label{eq:534}
{}^{(4)}g_{\!\mu\nu}|_{\mathrm{Eucl}}\; dx^\mu \otimes dx^\nu = N|_{\mathrm{Eucl}}^2\; dt \otimes dt + \frac{L^2}{6\pi} e^{2\alpha} (e^{2\beta})_{ij} \sigma^i \otimes \sigma^j,
\end{equation}
and differs from Eq.~(\ref{eq:510}) only in the sign of the kinetic energy term.

The remaining equations (\ref{eq:530}, \ref{eq:531}) are \textit{linear} `transport' equations to be integrated along the flow generated by a solution for \(\mathcal{S}_{(0)}\) to sequentially determine the quantum corrections \(\left\lbrace\mathcal{S}_{(k)},\; k = 1, 2, \ldots\right\rbrace\) in the formal expansion (\ref{eq:527}) for \(\mathcal{S}_\hbar\).

There are two known, globally defined, smooth solutions to Eq.~(\ref{eq:529}) that share the rotational symmetry of the Wheeler-DeWitt operator under rotations by \(\pm \frac{2\pi}{3}\) in the \(\beta\)-plane. By virtue of the geometrical characters of the Euclidean signature `spacetimes' they respectively generate they are sometimes referred to as the `wormhole' solution,
\begin{equation}\label{eq:535}
\mathcal{S}^{\mathrm{wh}}_{(0)} := \frac{1}{6} e^{2\alpha} \left(e^{-4\beta_+} + 2e^{2\beta_+} \cosh{(2\sqrt{3}\beta_-)}\right),
\end{equation}
and the `no boundary' solution
\begin{equation}\label{eq:536}
\mathcal{S}^{\mathrm{nb}}_{(0)} := \frac{1}{6} e^{2\alpha} \left\lbrack\left( e^{-4\beta_+} + 2e^{2\beta_+} \cosh{(2\sqrt{3}\beta_-)}\right) - 2 \left( e^{2\beta_+} + 2e^{-\beta_+} \cosh{(\sqrt{3}\beta_-)}\right)\right\rbrack .
\end{equation}
The first of these was discovered in the present context by Ryan and the author in \cite{Moncrief:1991} and independently, in a somewhat related, but supersymmetric setting by Graham in \cite{Graham:1991} who then, together with Bene, proceeded to construct the second solution \cite{Bene:1993,Bene:1994}. An additional, non-symmetric solution, together with its (geometrically equivalent) images under \(\pm \frac{2\pi}{3}\) rotations in the \(\beta\)-plane, was later uncovered by Barbero and Ryan in a systematic, further search \cite{Barbero:1996}.

On the other hand the Euclidean signature Mixmaster `spacetimes' generated by these various solutions, together with a characterization of their global geometric properties, were actually known much earlier, having been discovered through extensive searches for self-dual-curvature solutions to the field equations by Gibbons and Pope in \cite{Gibbons:1979} and by Belinski\v{\i} et al. in \cite{Belinskii:1978}.
With respect to a certain time function \(\eta\), which corresponds to our choice
\begin{equation}\label{eq:536a}
N|_{\mathrm{Eucl}} = \frac{Le^{3\alpha}}{(6\pi)^{1/2}}
\end{equation}
for the Euclidean signature lapse, these authors found that the metric functions
\begin{equation}\label{eq:537}
\begin{split}
\omega_1 &:= e^{2\alpha - \beta_+ - \sqrt{3}\beta_-}\\
\omega_2 &:= e^{2\alpha - \beta_+ + \sqrt{3}\beta_-}\\
\omega_3 &:= e^{2\alpha + 2\beta_+}
\end{split}
\end{equation}
satisfied the evolution equations
\begin{equation}\label{eq:538}
\begin{split}
\frac{d\omega_1}{d\eta} &= \omega_2\omega_3,\\
\frac{d\omega_2}{d\eta} &= \omega_1\omega_3,\\
\frac{d\omega_3}{d\eta} &= \omega_1\omega_2
\end{split}
\end{equation}
for the `wormhole' family and
\begin{equation}\label{eq:539}
\begin{split}
\frac{d\omega_1}{d\eta} &= \omega_2\omega_3 - \omega_1(\omega_2 + \omega_3),\\
\frac{d\omega_2}{d\eta} &= \omega_1\omega_3 - \omega_2(\omega_1 + \omega_3),\\
\frac{d\omega_3}{d\eta} &= \omega_1\omega_2 - \omega_3(\omega_1 + \omega_2)
\end{split}
\end{equation}
for the `no boundary' family. One can easily recover these flow equations from our Hamilton-Jacobi formalism by making the substitutions (\ref{eq:532}) and (\ref{eq:536a}) for \(\left\lbrace p_\alpha,p_+,p_-\right\rbrace\) and \(N|_{\mathrm{Eucl}}\) in the Euclidean signature Hamilton equations
\begin{align}
\dot{\alpha} &= \frac{(6\pi)^{1/2}G}{2c^3L^3e^{3\alpha}} N|_{\mathrm{Eucl}}\; p_\alpha\label{eq:540}\\
\dot{\beta}_+ &= \frac{-(6\pi)^{1/2}G}{2c^3L^3e^{3\alpha}} N|_{\mathrm{Eucl}}\; p_+\label{eq:541}\\
\dot{\beta}_- &= \frac{-(6\pi)^{1/2}G}{2c^3L^3e^{3\alpha}} N|_{\mathrm{Eucl}}\; p_-\label{eq:542}
\end{align}
and choosing \(\mathcal{S} = \mathcal{S}^{\mathrm{wh}}_{(0)}\) or \(\mathcal{S} = \mathcal{S}^{\mathrm{nb}}_{(0)}\) accordingly.

Because of its remarkable correspondence to the Euler equations for an asymmetric top \cite{misc:03}
the `Euler' system (\ref{eq:538}) was integrated long ago by Abel and Jacobi in terms of elliptic functions \cite{Gibbons:1979,Latifi:1994,Takhtajan:1992}.
But system (\ref{eq:539}) also long predated general relativity having been discovered by Darboux in connection with a pure geometry problem \cite{Darboux:1878}.
This `Darboux' system was subsequently integrated by Halphen \cite{Halphen:1918}
and later Bureau \cite{Bureau:1987}
in terms of Hermite modular elliptic functions. Both systems also occur as reductions of the self-dual Yang-Mills equations \cite{Latifi:1994,Takhtajan:1992}.

Since the asymptotically Euclidean behavior of the wormhole `spacetimes', as elucidated by Belinski\v{\i}, et al. in \cite{Belinskii:1978}
and by Gibbons and Pope in \cite{Gibbons:1979},
fits most naturally with our current perspective on appropriate boundary conditions for a ground state wave function \(\overset{(0)}{\Psi}_{\!\hbar}\) we shall focus exclusively on the `wormhole' solution, \(\mathcal{S}^{\mathrm{wh}}_{(0)}\), and its associated `flow', in the analysis to follow. It is worth remarking however that the same (microlocal) methods could also be brought to bear on the `no boundary' solution, \(\mathcal{S}^{\mathrm{nb}}_{(0)}\), and its `flow'.

Though the classical solution to the Euler system (\ref{eq:538}) entails elliptic functions \cite{Gibbons:1979,Belinskii:1978},
J. Bae was recently able, using a choice for the Euclidean signature lapse proposed by one of us, to reintegrate this system purely in terms of elementary functions and thus to simplify some of the subsequent analysis \cite{Bae:2015}.
With the lapse function taken to be
\begin{equation}\label{eq:543}
N|_{\mathrm{Eucl}} = \frac{-Le^{\alpha - 2\beta_+}}{(2\pi)^{1/2}}
\end{equation}
the wormhole flow equations become
\begin{align}
\frac{d\beta_-}{dt} &= \sinh{(2\sqrt{3}\beta_-)},\label{eq:544}\\
\frac{d\beta_+}{dt} &= -\frac{1}{\sqrt{3}} \left(e^{-6\beta_+} - \cosh{(2\sqrt{3}\beta_-)}\right)\label{eq:545}\\
\frac{d\alpha}{dt} &= -\frac{1}{2\sqrt{3}} \left(e^{-6\beta_+} + 2\cosh{(2\sqrt{3}\beta_-)}\right)\label{eq:546}
\end{align}
and can be readily integrated in the order given.\footnote{Since the chosen lapse (\ref{eq:543}) does not share the triangular symmetry of \(S_{(0)}^{\mathrm{wh}}\) in the \(\beta\)-plane, geometrically equivalent solutions to the flow equations (\ref{eq:544}--\ref{eq:546}) will often be parametrized differently.}

In terms of initial values \(\left\lbrace\alpha_0,\beta_{+0},\beta_{-0}\right\rbrace\) prescribed at \(t = 0\) Bae's solution is expressible as
\begin{align}
e^{12\alpha (t)} &= e^{12\alpha_0 - 6\beta_{+0}} H_+ (h_+h_-)^2,\label{eq:547}\\
e^{6\beta_+(t)} &= \frac{H_+}{h_+h_-},\label{eq:548}\\
e^{2\sqrt{3}\beta_- (t)} &= \frac{h_+}{h_-}\label{eq:549}
\end{align}
where
\begin{align}
H_+ &= e^{6\beta_{+0}} - \cosh{(2\sqrt{3}\beta_{-0})} + \frac{1}{2} (h_+^2 + h_-^2) \label{eq:550}\\
&= e^{6\beta_{+0}} + (h_\pm)^2 - (h_{\pm 0})^2,\nonumber\\
h_+ &= e^{-\sqrt{3}t} \cosh{(\sqrt{3}\beta_{-0})} + e ^{\sqrt{3}t} \sinh{(\sqrt{3}\beta_{-0})},\label{eq:551}\\
h_- &= e^{-\sqrt{3}t} \cosh{(\sqrt{3}\beta_{-0})} - e^{\sqrt{3}t} \sinh{(\sqrt{3}\beta_{-0})}.\label{eq:552}
\end{align}
Several useful identities that follow from these formulas are given by
\begin{align}
\cosh{(2\sqrt{3}\beta_-(t))} &= \frac{h_+^2 + h_-^2}{2h_+h_-},\label{eq:553}\\
e^{2\alpha(t)+2\beta_+(t)} &= e^{2\alpha_0-\beta_{+0}} \sqrt{H_+},\label{eq:554}\\
e^{4\alpha(t)-2\beta_+(t)} &= e^{4\alpha_0-2\beta_{+0}} h_+h_-.\label{eq:555}
\end{align}

It is not difficult to verify that every solution is globally, smoothly defined on a maximal interval of the form \((-\infty,t_*)\) where \(t_* > 0\) so that, in particular, every solution curve is well-defined on the sub-interval \((-\infty,0]\). Furthermore \(\beta_+(t)\) and \(\beta_-(t)\) each decay exponentially rapidly to zero as \(t \rightarrow -\infty\) with
\begin{equation}\label{eq:556}
\beta_\pm(t) \sim \mathrm{const}_{\pm} e^{2\sqrt{3}t}
\end{equation}
while \(\alpha\) diverges, asymptotically linearly,
\begin{equation}\label{eq:557}
\alpha (t) \sim -\frac{\sqrt{3}}{2}t + \mathrm{const}
\end{equation}
in this limit. This behavior of the solution curves will play a crucial role in the integration of the transport equations (\ref{eq:530}, \ref{eq:531}).

It is worth noting that one can \textit{linearize} the \(\beta\)-plane flow equations (\ref{eq:544}--\ref{eq:545}) through an explicit transformation to `Sternberg coordinates' \(\lbrace y_+,y_-\rbrace\) in terms of which these equations reduce to
\begin{equation}\label{eq:558}
\frac{dy_+}{dt} = 2\sqrt{3}y_+,\quad \frac{dy_-}{dt} = 2\sqrt{3}y_-.
\end{equation}
These Sternberg coordinates are defined by
\begin{align}
y_+ &= \frac{1}{6} \left(\frac{e^{6\beta_+} - \cosh{(2\sqrt{3}\beta_-)}}{\cosh^2{(\sqrt{3}\beta_-)}}\right),\label{eq:559}\\
y_- &= \frac{1}{\sqrt{3}}\; \frac{\sinh{(\sqrt{3}\beta_-)}}{\cosh{(\sqrt{3}\beta_-)}}\label{eq:560}
\end{align}
which has the explicit inverse
\begin{align}
e^{6\beta_+} &= 3y_+ + (3y_+ +1) \left(\frac{1 + 3y_-^2}{1 - 3y_-^2}\right),\label{eq:561}\\
e^{2\sqrt{3}\beta_-} &= \frac{1 + \sqrt{3}y_-}{1 - \sqrt{3}y_-}\label{eq:562}
\end{align}
and maps the \(\beta\)-plane diffeomorphically onto the `strip' given by
\begin{align}
-\frac{1}{\sqrt{3}} < y_- < \frac{1}{\sqrt{3}},\label{eq:563}\\
y_+ > -\frac{1}{6} (1 + y_-^2).\label{eq:564}
\end{align}

Taking \(\mathcal{S}_{(0)} = \mathcal{S}^{\mathrm{wh}}_{(0)}\) Bae found a particular solution to the first transport equation (\ref{eq:530}) given by
\begin{equation}\label{eq:566}
\mathcal{S}_{(1)} = -\frac{1}{2} (B + 6)\alpha .
\end{equation}
Though one would be free to add an arbitrary solution to the corresponding homogeneous equation we shall reserve such flexibility for the subsequent construction of \textit{excited states}, retaining Bae's particular solution as the natural choice to make for a \textit{ground state}.

The ensuing transport equations (\ref{eq:531}) can now be solved inductively by making the ansatz
\begin{equation}\label{eq:567}
\mathcal{S}^{\mathrm{wh}}_{(k)} = 6e^{-2(k-1)\alpha} \Sigma_{(k)}^{\mathrm{wh}} (\beta_+,\beta_-)
\end{equation}
for \(k = 2, 3, \ldots\) and, for convenience, defining
\begin{equation}\label{eq:568}
\Sigma_{(0)}^{\mathrm{wh}} = e^{-4\beta_+} + 2e^{2\beta_+} \cosh{(2\sqrt{3}\beta_-)}
\end{equation}
so that
\begin{equation}\label{eq:569}
\mathcal{S}^{\mathrm{wh}}_{(0)} = \frac{e^{2\alpha}}{6} \Sigma_{(0)}^{\mathrm{wh}} (\beta_+,\beta_-).
\end{equation}

The corresponding transport equations for the coefficients \(\left\lbrace\sum_{(k)}^{wh} (\beta_+,\beta_-)\right\rbrace\) were integrated explicitly in Sect.~4 of Ref.~\cite{Moncrief:2015} and the solutions were shown, by an inductive argument given therein, to be globally smooth on the \(\beta\)-plane to all orders in Planck's constant.

This construction began to resolve the `paradox' alluded to above concerning how microlocal methods could possibly be used to generate smooth quantum corrections to candidate `eigenfunctions' when there are no corresponding `eigenvalues' available to adjust. In a conventional Schr\"{o}dinger eigenvalue problem \cite{Moncrief:2012} the values, \(\lbrace\mathcal{S}_{(k)} (0, \ldots ,0)\rbrace\), of the functions under construction \(\lbrace\mathcal{S}_{(k)} (x^1, \ldots , x^n)\rbrace\) are, at the minimum of the potential energy (taken here to be the origin), arbitrary constants of integration that can be lumped into an overall normalization constant for the ground state wave function. Thus these adjustable constants play \textit{no role} in guaranteeing the smoothness of the \(\lbrace\mathcal{S}_{(k)}\rbrace\). Here however the functions being computed by the analogous `transport' analysis are the \(\lbrace\Sigma_{(k)}^{\mathrm{wh}} (\beta_+,\beta_-)\rbrace\). But, because they multiply correspondingly different powers of \(e^\alpha\) in the ansatz (\ref{eq:567}) for \(\mathcal{S}_{(k)}^{\mathrm{wh}}\), their values at the classical equilibrium (i.e., at the origin in \((\beta_+,\beta_-)\)-space) are \textit{not} arbitrary but instead provide precisely the flexibility needed, in the absence of eigenvalue coefficients, to ensure the smoothness of the functions \(\lbrace\Sigma_{(k)}^{\mathrm{wh}} (\beta_+,\beta_-)\rbrace\) and hence also that of the \(\lbrace\mathcal{S}_{(k)}^{\mathrm{wh}} (\alpha,\beta_+,\beta_-)\rbrace\). In the section below we shall encounter an analogous phenomenon occurring in the construction of \textit{excited states}.

\subsection{Conserved Quantities and Excited States}
\label{subsec:conserved}
To generate `excited state' solutions to the Wheeler-DeWitt equation we begin by making the ansatz
\begin{equation}\label{eq:570}
\overset{(*)}{\Psi}_{\!\hbar} = \overset{(*)}{\phi}_{\!\hbar} e^{-S_\hbar/\hbar}
\end{equation}
where \(S_\hbar = \frac{c^3L^2}{G} \mathcal{S}_\hbar = \frac{c^3L^2}{G} \left(\mathcal{S}_{(0)} + X\mathcal{S}_{(1)} + \frac{X^2}{2!} \mathcal{S}_{(2)} + \cdots\right)\) is the \textit{same} formal expansion derived in the preceding section for the \textit{ground state} solution and where the new factor \(\overset{(*)}{\phi}_{\!\hbar}\) is assumed to admit an expansion of similar type,
\begin{equation}\label{eq:571}
\overset{(*)}{\phi}_{\!\hbar} = \overset{(*)}{\phi}_{\!(0)} + X\overset{(*)}{\phi}_{\!(1)} + \frac{X^2}{2!}\overset{(*)}{\phi}_{\!(2)} + \cdots + \frac{X^k}{k!}\overset{(*)}{\phi}_{\!(k)} + \cdots,
\end{equation}
with \(X = \frac{L_{\mathrm{Planck}}^2}{L^2} = \frac{G\hbar}{c^3L^2}\) as before. Substituting this ansatz into the Mixmaster Wheeler-DeWitt equation and demanding satisfaction, order-by-order in \textit{X}, one arrives at the sequence of equations
\begin{align}
&-\frac{\partial\overset{(*)}{\phi}_{\!(0)}}{\partial\alpha} \frac{\partial\mathcal{S}_{(0)}}{\partial\alpha} + \frac{\partial\overset{(*)}{\phi}_{\!(0)}}{\partial\beta_+} \frac{\partial\mathcal{S}_{(0)}}{\partial\beta_+} + \frac{\partial\overset{(*)}{\phi}_{\!(0)}}{\partial\beta_-} \frac{\partial\mathcal{S}_{(0)}}{\partial\beta_-} = 0,\label{eq:572}\\
\begin{split}
&-\frac{\partial\overset{(*)}{\phi}_{\!(1)}}{\partial\alpha} \frac{\partial\mathcal{S}_{(0)}}{\partial\alpha} + \frac{\partial\overset{(*)}{\phi}_{\!(1)}}{\partial\beta_+} \frac{\partial\mathcal{S}_{(0)}}{\partial\beta_+} + \frac{\partial\overset{(*)}{\phi}_{\!(1)}}{\partial\beta_-} \frac{\partial\mathcal{S}_{(0)}}{\partial\beta_-}\\
 &\quad + \left(-\frac{\partial\overset{(*)}{\phi}_{\!(0)}}{\partial\alpha} \frac{\partial\mathcal{S}_{(1)}}{\partial\alpha} + \frac{\partial\overset{(*)}{\phi}_{\!(0)}}{\partial\beta_+} \frac{\partial\mathcal{S}_{(1)}}{\partial\beta_+} + \frac{\partial\overset{(*)}{\phi}_{\!(0)}}{\partial\beta_-} \frac{\partial\mathcal{S}_{(1)}}{\partial\beta_-}\right)\\
  &\quad + \frac{1}{2} \left(-B \frac{\partial\overset{(*)}{\phi}_{\!(0)}}{\partial\alpha} + \frac{\partial^2\overset{(*)}{\phi}_{\!(0)}}{\partial\alpha^2} - \frac{\partial^2\overset{(*)}{\phi}_{\!(0)}}{\partial\beta_+^2} - \frac{\partial^2\overset{(*)}{\phi}_{\!(0)}}{\partial\beta_-^2}\right) = 0,
\end{split}\label{eq:573}
\end{align}
and, for \(k \geq 2\)
\begin{equation}\label{eq:574}
\begin{split}
&-\frac{\partial\overset{(*)}{\phi}_{\!(k)}}{\partial\alpha} \frac{\partial\mathcal{S}_{(0)}}{\partial\alpha} + \frac{\partial\overset{(*)}{\phi}_{\!(k)}}{\partial\beta_+} \frac{\mathcal{S}_{(0)}}{\partial\beta_+} + \frac{\partial\overset{(*)}{\phi}_{\!(k)}}{\partial\beta_-} \frac{\partial\mathcal{S}_{(0)}}{\partial\beta_-}\\
&\quad +k \left(-\frac{\partial\overset{(*)}{\phi}_{\!(k-1)}}{\partial\alpha} \frac{\partial\mathcal{S}_{(1)}}{\partial\alpha} + \frac{\overset{(*)}{\phi}_{\!(k-1)}}{\partial\beta_+} \frac{\partial\mathcal{S}_{(1)}}{\partial\beta_+} + \frac{\partial\overset{(*)}{\phi}_{\!(k-1)}}{\partial\beta_-} \frac{\partial\mathcal{S}_{(1)}}{\partial\beta_-}\right)\\
&\quad + \frac{k}{2} \left(-B \frac{\partial\overset{(*)}{\phi}_{\!(k-1)}}{\partial\alpha} + \frac{\partial^2\overset{(*)}{\phi}_{\!(k-1)}}{\partial\alpha^2} - \frac{\partial^2\overset{(*)}{\phi}_{\!(k-1)}}{\partial\beta_+^2} - \frac{\partial^2\overset{(*)}{\phi}_{\!(k-1)}}{\partial\beta_-^2}\right)\\
&\quad + \sum_{\ell=2}^{k} \frac{k!}{\ell!(k-\ell)!} \left(-\frac{\partial\overset{(*)}{\phi}_{\!(k-\ell)}}{\partial\alpha} \frac{\partial\mathcal{S}_{(\ell)}}{\partial\alpha} + \frac{\partial\overset{(*)}{\phi}_{\!(k-\ell)}}{\partial\beta_+} \frac{\partial\mathcal{S}_{(\ell)}}{\partial\beta_+} + \frac{\partial\overset{(*)}{\phi}_{\!(k-\ell)}}{\partial\beta_-} \frac{\partial\mathcal{S}_{(\ell)}}{\partial\beta_-}\right) = 0.
\end{split}
\end{equation}
The first of these is easily seen to be the requirement that \(\overset{(*)}{\phi}_{\!(0)}\) be constant along the flow in mini-superspace generated by \(\mathcal{S}_{(0)}\), the chosen solution to the Euclidean-signature Hamilton-Jacobi equation (\ref{eq:529}). For the case of most interest here, \(\mathcal{S}_{(0)} \longrightarrow \mathcal{S}_{(0)}^{\mathrm{wh}}\), Bae discovered two such conserved quantities through direct inspection of his solution (\ref{eq:547}--\ref{eq:552}) of the corresponding flow equations, namely
\begin{equation}\label{eq:575}
C_{(0)} := \frac{1}{6} e^{4\alpha - 2\beta_+} \left(e^{6\beta_+} - \cosh{(2\sqrt{3}\beta_-)}\right)
\end{equation}
and
\begin{equation}\label{eq:576}
S_{(0)} := \frac{1}{2\sqrt{3}} e^{4\alpha - 2\beta_+} \sinh{(2\sqrt{3}\beta_-)}
\end{equation}
\cite{Bae:2015}. By reexpressing these in terms of the functions \(\lbrace\omega_1,\omega_2,\omega_3\rbrace\) defined previously, one arrives at the alternative forms
\begin{align}
C_{(0)} &= \frac{1}{12} (2\omega_3^2 - \omega_1^2 - \omega_2^2)\label{eq:577}\\
S_{(0)} &= \frac{1}{4\sqrt{3}} (\omega_2^2 - \omega_1^2)\label{eq:578}
\end{align}
and can recognize them in terms of the well-known, conserved kinetic energy and squared angular momentum of the asymmetric top \cite{misc:03,Takhtajan:1992}.

Of course any differentiable function of \(C_{(0)}\) and \(S_{(0)}\) would be equally conserved but the Taylor expansions of these in particular,
\begin{align}
C_{(0)} &\simeq e^{4\alpha} \left(\beta_+ + \beta_+^2 - \beta_-^2 + O(\beta^3)\right),\label{eq:579}\\
S_{(0)} &\simeq e^{4\alpha} \left(\beta_- - 2\beta_+\beta_- + O(\beta^3)\right),\label{eq:580}
\end{align}
reveal their preferred features of behaving linearly in \(\beta_+\) and \(\beta_-\) (respectively) near the origin in \(\beta\)-space. It therefore seems natural to seek to construct a `basis' of excited states by taking
\begin{equation}\label{eq:581}
\begin{split}
\overset{(*)}{\phi}_{\!(0)} \longrightarrow \overset{(\mathbf{m})}{\phi}_{\!(0)} &:= C_{(0)}^{m_1} S_{(0)}^{m_2}\\
 &\quad \simeq e^{4(m_1 + m_2)\alpha} (\beta_+^{m_1} \beta_-^{m_2} + \cdots)
\end{split}
\end{equation}
as seeds for the computation of higher order quantum corrections. Here \(\mathbf{m} = (m_1,m_2)\) is a pair of non-negative integers that can be plausibly interpreted as \textit{graviton excitation numbers} for the ultralong wavelength gravitational wave modes embodied in the \(\beta_+\) and \(\beta_-\) degrees of freedom.

To see this more concretely note that, to leading order in \textit{X} and near the origin in \(\beta\)-space, one then gets
\begin{equation}\label{eq:582}
\overset{(\mathbf{m})}{\Psi}_{\!\hbar} \simeq e^{4(m_1 + m_2)\alpha} \beta_+^{m_1} \beta_-^{m_2} e^{-\frac{e^{2\alpha}}{X}\left(\frac{1}{2} + 2(\beta_+^2 + \beta_-^2)+\cdots\right)}
\end{equation}
which, for any fixed \(\alpha\), has the form of the top order term in the product of Hermite polynomials multiplied by a gaussian that one would expect to see for an actual, \textit{harmonic oscillator} wave function.

One wishes, however, to construct wave functions that share the invariance of the Wheeler-DeWitt operator under rotations by \(\pm \frac{2\pi}{3}\) in the \(\beta\)-plane since these correspond to residual gauge transformations. The functions \(\lbrace\mathcal{S}_{(k)}\rbrace\) constructed in the preceding section have this property automatically by virtue of the rotational invariance of the flow generated by the chosen \(\mathcal{S}_{(0)} = \mathcal{S}_{(0)}^{\mathrm{wh}}\) and the corresponding invariance of the technique employed for generating \textit{initial conditions} for the \(\lbrace\mathcal{S}_{(k)}, k = 1,2,\cdots\rbrace\). On the other hand the functions \(\overset{(\mathbf{m})}{\phi}_{\!(0)} := C_{(0)}^{m_1} S_{(0)}^{m_2}\) are not, in general, invariant but can be modified to become so by the straightforward technique of averaging over the group of rotations in question: \(\lbrace I,\pm \frac{2\pi}{3}\rbrace\). Some elegant graphical depictions of the lowest few such invariant states (to leading order in \textit{X}) have been given by Bae in \cite{Bae:2015}. The linearity of equations (\ref{eq:572}--\ref{eq:574}) in the \(\lbrace\overset{(\mathbf{m})}{\phi}_{\!(k)}\rbrace\) and the rotational invariance of the operators therein acting upon these functions will allow one to construct rotationally invariant quantum corrections to all orders, either by starting with an invariant `seed' of the type described above or, alternatively, carrying out the group averaging at the end of the sequence of calculations. We shall follow the latter approach here.

We begin by setting
\begin{equation}\label{eq:583}
\overset{(\mathbf{m})}{\phi}_{\!(0)} \longrightarrow C_{(0)}^{m_1} S_{(0)}^{m_2} := e^{4|m|\alpha} \overset{(\mathbf{m})}{\chi}_{\!(0)} (\beta_+,\beta_-)
\end{equation}
where \(|m| := m_1 + m_2\) and proceed by making the ansatz
\begin{equation}\label{eq:584}
\overset{(\mathbf{m})}{\phi}_{\!(k)} = e^{(4|m|-2k)\alpha} \overset{(\mathbf{m})}{\chi}_{\!(k)} (\beta_+,\beta_-)
\end{equation}
\(\forall\; k \geq 1\). Recalling the definitions of the functions \(\lbrace\Sigma_{(k)}^{\mathrm{wh}}(\beta_+,\beta_-)\rbrace\) given by (\ref{eq:567}--\ref{eq:569}) we now find that equations (\ref{eq:573}--\ref{eq:574}) can be reexpressed as flow equations in the \(\beta\)-plane for the unknowns \(\lbrace\overset{(\mathbf{m})}{\chi}_{\!(k)}(\beta_+,\beta_-);\; k = 1,2 \cdots\rbrace\):
\begin{equation}\label{eq:585}
\begin{split}
&\frac{\partial\overset{(\mathbf{m})}{\chi}_{\!(1)}}{\partial\beta_+} \frac{\partial\Sigma_{(0)}^{\mathrm{wh}}}{\partial\beta_+} + \frac{\partial\overset{(\mathbf{m})}{\chi}_{\!(1)}}{\partial\beta_-} \frac{\partial\Sigma_{(0)}^{\mathrm{wh}}}{\partial\beta_-} - 2\overset{(\mathbf{m})}{\chi}_{\!(1)} \left(4|m|-2\right) \Sigma_{(0)}^{\mathrm{wh}}\\
 &\quad + 3 \left\lbrack (16|m|^2 + 24|m|) \overset{(\mathbf{m})}{\chi}_{\!(0)} - \frac{\partial^2\overset{(\mathbf{m})}{\chi}_{\!(0)}}{\partial\beta_+^2} - \frac{\partial^2\overset{(\mathbf{m})}{\chi}_{\!(0)}}{\partial\beta_-^2}\right\rbrack = 0,
 \end{split}
\end{equation}
and, for \(k \geq 2\),
\begin{equation}\label{eq:586}
\begin{split}
&\frac{\partial\overset{(\mathbf{m})}{\chi}_{\!(k)}}{\partial\beta_+} \frac{\partial\Sigma_{(0)}^{\mathrm{wh}}}{\partial\beta_+} + \frac{\partial\overset{(\mathbf{m})}{\chi}_{\!(k)}}{\partial\beta_-} \frac{\partial\Sigma_{(0)}^{\mathrm{wh}}}{\partial\beta_-} - 2\overset{(\mathbf{m})}{\chi}_{\!(k)} (4|m| - 2k) \Sigma_{(0)}^{\mathrm{wh}}\\
 &\quad + 3k \left\lbrace\left\lbrack\left(4|m| - 2(k-1)\right)^2 + 6\left(4|m| - 2(k-1)\right)\right\rbrack \overset{(\mathbf{m})}{\chi}_{\!(k-1)} - \frac{\partial^2\overset{(\mathbf{m})}{\chi}_{\!(k-1)}}{\partial\beta_+^2}\right.\\
 &\quad\quad\left. - \frac{\partial^2\overset{(\mathbf{m})}{\chi}_{\!(k-1)}}{\partial\beta_-^2}\right\rbrace\\
 &\quad +36 \sum_{\ell = 2}^{k} \frac{k!}{\ell!(k-\ell)!} \left\lbrace 2(\ell - 1) \left(4|m| - 2 (k - \ell)\right) \overset{(\mathbf{m})}{\chi}_{\!(k-\ell)} \Sigma_{(\ell)}^{\mathrm{wh}}\vphantom{\frac{\partial\overset{(\mathbf{m})}{\chi}_{\!(k-\ell)}}{\partial\beta_+}}\right.\\
  &\quad\quad\left. + \frac{\partial\overset{(\mathbf{m})}{\chi}_{\!(k-\ell)}}{\partial\beta_+} \frac{\partial\Sigma_{(\ell)}^{\mathrm{wh}}}{\partial\beta_+} + \frac{\partial\overset{(\mathbf{m})}{\chi}_{\!(k-\ell)}}{\partial\beta_-} \frac{\partial\Sigma_{(\ell)}^{\mathrm{wh}}}{\partial\beta_-}\right\rbrace =0.
\end{split}
\end{equation}

As for the ground state problem our aim is to solve these transport equations sequentially and thereby to establish, for any given \(\mathbf{m} = (m_1,m_2)\), the existence of smooth, globally defined functions \(\lbrace\overset{(\mathbf{m})}{\chi}_{\!(k)} (\beta_+,\beta_-); k = 1, 2, \dots\rbrace\) on the \(\beta\)-plane. When \(k > 2|m|\) the relevant transport operator is of the same type dealt with in the previous section and the corresponding equation can be solved, for an arbitrary smooth `source' inhomogeneity, by the same methods exploited therein. When \(k \leq 2|m|\) however the associated integrating factor,
\begin{equation}\label{eq:587}
\frac{\mu_{(k)}(t)}{\mu_{(k)}(0)} = \frac{e^{(4|m| - 2k)\alpha (t)}}{e^{(4|m| - 2k)\alpha (0)}}
\end{equation}
is either constant or blows up at \(t \searrow -\infty\) and a different approach is needed. Fortunately there is a well-developed microlocal technique for handling such problems  \cite{Moncrief:2012,Dimassi:1999,Helfer:1988,Helfer:1984}. Details of the application of this method to the problem at hand are presented near the end of Section~5 of Ref.~\cite{Moncrief:2015}. 

\section{Euclidean-Signature Asymptotic Methods and the Wheeler-DeWitt Equation}
\label{sec:euclidean-signature-wheeler-dewitt}
Globally hyperbolic spacetimes, \(\lbrace{}^{(4)}V,{}^{(4)}g\rbrace\), are definable over manifolds with the product structure, \({}^{(4)}V \approx M \times \mathbb{R}\). We shall focus here on the `cosmological' case for which the spatial factor \textit{M} is a compact, connected, orientable 3-manifold without boundary. The Lorentzian metric, \({}^{(4)}g\), of such a spacetime is expressible, relative to a time function \(x^0 = t\), in the 3+1-dimensional form
\begin{equation}\label{eq:2155}
\begin{split}
{}^{(4)}g &= {}^{(4)}g_{\mu\nu}\; dx^\mu \otimes dx^\nu\\
 &= -N^2 dt \otimes dt + \gamma_{ij} (dx^i + Y^idt) \otimes (dx^j + Y^jdt)
\end{split}
\end{equation}
wherein, for each fixed \textit{t}, the Riemannian metric
\begin{equation}\label{eq:2156}
\gamma = \gamma_{ij} dx^i \otimes dx^j
\end{equation}
is the first fundamental form induced by \({}^{(4)}g\) on the corresponding \(t = \mathrm{constant}\), spacelike hypersurface. The unit, future pointing, timelike normal field to the chosen slicing (defined by the level surfaces of \textit{t}) is expressible in terms of the (strictly positive) `lapse' function \textit{N} and `shift vector' field \(Y^i \frac{\partial}{\partial x^i}\) as
\begin{equation}\label{eq:2157}
{}^{(4)}n = {}^{(4)}n^\alpha \frac{\partial}{\partial x^\alpha} = \frac{1}{N} \frac{\partial}{\partial t} - \frac{Y^i}{N} \frac{\partial}{\partial x^i}
\end{equation}
or, in covariant form, as
\begin{equation}\label{eq:2158}
{}^{(4)}n = {}^{(4)}n_\alpha dx^\alpha = -N\; dt.
\end{equation}
The canonical spacetime volume element of \({}^{(4)}g,\; \mu_{{}^{(4)}g} := \sqrt{-\det{{}^{(4)}g}}\), takes the 3+1-dimensional form
\begin{equation}\label{eq:2159}
\mu_{{}^{(4)}g} = N\mu_\gamma
\end{equation}
where \(\mu_\gamma := \sqrt{\det{\gamma}}\) is the volume element of \(\gamma\).

In view of the compactness of \textit{M} the Hilbert and ADM action functionals, evaluated on domains of the product form, \(\Omega = M \times I\), with \(I = \lbrack t_0,t_1\rbrack \subset \mathbb{R}\), simplify somewhat to
\begin{equation}\label{eq:2160}
\begin{split}
I_{\mathrm{Hilbert}} &:= \frac{c^3}{16\pi G}\; \int_\Omega \sqrt{-\det{{}^{(4)}g}}\; {}^{(4)}R({}^{(4)}g)\; d^4x\\
 &\hphantom{:}= \frac{c^3}{16\pi G}\; \int_\Omega \left\lbrace N\mu_\gamma \left( K^{ij} K_{ij} - (tr_\gamma K)^2\right) + N\mu_\gamma {}^{(3)}R(\gamma)\right\rbrace d^4x\\
 &\hphantom{:=} + \frac{c^3}{16\pi G}\; \int_M \left(-2\mu_\gamma tr_\gamma K\right) d^3x\Big|_{t_0}^{t_1}\\
 &:= I_{\mathrm{ADM}} + \frac{c^3}{16\pi G}\; \int_M \left(-2\mu_\gamma tr_\gamma K\right) d^3x\Big|_{t_0}^{t_1}
\end{split}
\end{equation}
wherein \({}^{(4)}R ({}^{(4)}g)\) and \({}^{(3)}R(\gamma)\) are the scalar curvatures of \({}^{(4)}g\) and \(\gamma\) and where
\begin{equation}\label{eq:2161}
K_{ij} := \frac{1}{2N} \left(-\gamma_{ij,t} + Y_{i|j} + Y_{j|i}\right)
\end{equation}
and
\begin{equation}\label{eq:2162}
tr_\gamma K := \gamma^{ij}K_{ij}
\end{equation}
designate the second fundamental form and mean curvature induced by \({}^{(4)}g\) on the constant \textit{t} slices. In these formulas spatial coordinate indices, \(i,j,k, \ldots,\) are raised and lowered with \(\gamma\) and the vertical bar, `\(|\)', signifies covariant differentiation with respect to this metric so that, for example, \(Y_{i|j} = \nabla_j (\gamma) \gamma_{i\ell} Y^\ell\). When the variations of \({}^{(4)}g\) are appropriately restricted, the boundary term distinguishing \(I_{\mathrm{Hilbert}}\) from \(I_{\mathrm{ADM}}\) makes no contribution to the field equations and so can be discarded.

Writing
\begin{equation}\label{eq:2163}
I_{\mathrm{ADM}} := \int_\Omega\; \mathcal{L}_{\mathrm{ADM}} d^4x,
\end{equation}
with Lagrangian density
\begin{equation}\label{eq:2164}
\mathcal{L}_{\mathrm{ADM}} := \frac{c^3}{16\pi G}\; \left\lbrace N\mu_\gamma \left( K^{ij} K_{ij} - (tr_\gamma K)^2\right) + N\mu_\gamma {}^{(3)}R(\gamma)\right\rbrace ,
\end{equation}
one defines the \textit{momentum} conjugate to \(\gamma\) via the Legendre transformation
\begin{equation}\label{eq:2165}
p^{ij} := \frac{\partial\mathcal{L}_{\mathrm{ADM}}}{\partial\gamma_{ij,t}} = \frac{c^3}{16\pi G}\; \mu_\gamma \left(-K^{ij} + \gamma^{ij} tr_\gamma K\right)
\end{equation}
so that \(p = p^{ij} \frac{\partial}{\partial x^i} \otimes \frac{\partial}{\partial x^j}\) is a symmetric tensor density induced on each \(t = \mathrm{constant}\) slice.

In terms of the variables \(\lbrace\gamma_{ij}, p^{ij}, N, Y^{i}\rbrace\) the ADM action takes the Hamiltonian form
\begin{equation}\label{eq:2166}
I_{\mathrm{ADM}} = \int_\Omega \left\lbrace p^{ij}\gamma_{ij,t} - N \mathcal{H}_\perp (\gamma,p) - Y^i\mathcal{J}_i (\gamma,p)\right\rbrace d^4x
\end{equation}
where
\begin{equation}\label{eq:2167}
\mathcal{H}_\perp (\gamma,p) := \left(\frac{16\pi G}{c^3}\right) \frac{\left(p^{ij}p_{ij} - \frac{1}{2}(p_m^m)^2\right)}{\mu_\gamma} - \left(\frac{c^3}{16\pi G}\right) \mu_\gamma\; {}^{(3)}R (\gamma)
\end{equation}
and
\begin{equation}\label{eq:2168}
\mathcal{J}_i(\gamma,p) := -2\; p_{i\hphantom{j}|j}^{\hphantom{i}j}.
\end{equation}
Variation of \(I_{\mathrm{ADM}}\) with respect to \textit{N} and \(Y^i\) leads to the Einstein (`Hamiltonian' and `momentum') \textit{constraint equations}
\begin{equation}\label{eq:2169}
\mathcal{H}_\perp (\gamma,p) = 0,\quad \mathcal{J}_i(\gamma,p) = 0,
\end{equation}
whereas variation with respect to the \textit{canonical variables}, \(\lbrace\gamma_{ij},p^{ij}\rbrace\), gives rise to the complementary Einstein \textit{evolution equations} in Hamiltonian form,
\begin{equation}\label{eq:2170}
\gamma_{ij,t} = \frac{\delta H_{\mathrm{ADM}}}{\delta p^{ij}},\quad p^{ij}_{\hphantom{ij},t} = -\frac{\delta H_{\mathrm{ADM}}}{\delta\gamma_{ij}}
\end{equation}
where \(H_{\mathrm{ADM}}\) is the `super' Hamiltonian defined by
\begin{equation}\label{eq:2171}
H_{\mathrm{ADM}} := \int_M \left( N\mathcal{H}_\perp (\gamma,p) + Y^i\mathcal{J}_i(\gamma,p)\right) d^3x.
\end{equation}
The first of equations (\ref{eq:2170}) regenerates (\ref{eq:2161}) when the latter is reexpressed in terms of \textit{p} via (\ref{eq:2165}). Note that, as a linear form in the constraints, the super Hamiltonian vanishes when evaluated on any solution to the field equations. There are neither constraints nor evolution equations for the lapse and shift fields which are only determined upon making, either explicitly or implicitly, a choice of spacetime coordinate \textit{gauge}. Bianchi identities function to ensure that the constraints are preserved by the evolution equations and thus need only be imposed `initially' on an arbitrary Cauchy hypersurface. Well-posedness theorems for the corresponding Cauchy problem exist for a variety of spacetime gauge conditions \cite{Choquet-Bruhat:2009,Andersson:2003}.

A formal `canonical' quantization of this system begins with the substitutions
\begin{equation}\label{eq:2172}
p^{ij} \longrightarrow \frac{\hbar}{i} \frac{\delta}{\delta\gamma_{ij}},
\end{equation}
together with a choice of operator ordering, to define quantum analogues \(\hat{\mathcal{H}}_\perp (\gamma,\frac{\hbar}{i} \frac{\delta}{\delta\gamma})\) and \(\hat{\mathcal{J}}_i(\gamma,\frac{\hbar}{i}\frac{\delta}{\delta\gamma})\) of the Hamiltonian and momentum constraints. These are then to be imposed, \`{a} la Dirac, as restrictions upon the allowed quantum states, regarded as functionals, \(\Psi [\gamma]\), of the spatial metric, by setting
\begin{equation}\label{eq:2173}
\hat{\mathcal{H}}_\perp \left(\gamma,\frac{\hbar}{i}\frac{\delta}{\delta\gamma}\right) \Psi [\gamma] = 0,
\end{equation}
and
\begin{equation}\label{eq:2174}
\hat{\mathcal{J}}_i \left(\gamma,\frac{\hbar}{i}\frac{\delta}{\delta\gamma}\right) \Psi [\gamma] = 0.
\end{equation}
The choice of ordering in the definition of the quantum constraints \(\lbrace\hat{\mathcal{H}}_\perp,\hat{\mathcal{J}}_i\rbrace\) is highly restricted by the demand that the \textit{commutators} of these operators should `close' in a natural way without generating `anomalous' new constraints upon the quantum states.

While a complete solution to this \textit{ordering problem} does not currently seem to be known it has long been realized that the operator, \(\hat{\mathcal{J}}_i (\gamma,\frac{\hbar}{i}\frac{\delta}{\delta\gamma})\), can be consistently defined so that the quantum constraint equation (\ref{eq:2174}), has the natural geometric interpretation of demanding that the wave functional, \(\Psi [\gamma]\), be invariant with respect to the action (by pullback of metrics on \textit{M}) of \(\mathcal{D}\mathit{iff}^0(M)\), the connected component of the identity of the group, \(\mathcal{D}\mathit{iff}^+(M)\), of orientation preserving diffeomorphisms of \textit{M}, on the space, \(\mathcal{M}(M)\), of Riemannian metrics on \textit{M}. In other words the quantized momentum constraint (\ref{eq:2174}) implies, precisely, that
\begin{equation}\label{eq:2176}
\Psi [\varphi^*\gamma] = \Psi [\gamma]
\end{equation}
\(\forall\; \varphi \in \mathcal{D}\mathit{iff}^0(M)\) and \(\forall\; \gamma \in \mathcal{M}(M)\). In terminology due to Wheeler wave functionals can thus be regarded as passing naturally to the quotient `superspace' of Riemannian \textit{3-geometries} \cite{Fischer:1970,Giulini:2009,misc:08} on \textit{M},
\begin{equation}\label{eq:2177}
\boldsymbol{\mathbb{S}}(M) := \frac{\mathcal{M}(M)}{\mathcal{D}\mathit{iff}^0(M)}.
\end{equation}

Insofar as a consistent factor ordering for the Hamiltonian constraint operator, \(\hat{\mathcal{H}}_\perp (\gamma,\frac{\hbar}{i}\frac{\delta}{\delta\gamma})\), also exists, one will be motivated to propose the (Euclidean-signature, semi-classical) ansatz
\begin{equation}\label{eq:2178}
\overset{(0)}{\Psi}_{\!\hbar} [\gamma] = e^{-S_\hbar [\gamma]/\hbar}
\end{equation}
for a `ground state' wave functional \(\overset{(0)}{\Psi}_{\!\hbar} [\gamma]\). In parallel with our earlier examples, the functional \(S_\hbar [\gamma]\) is assumed to admit a formal expansion in powers of \(\hbar\) so that one has
\begin{equation}\label{eq:2179}
S_\hbar [\gamma] = S_{(0)} [\gamma] + \hbar S_1 [\gamma] + \frac{\hbar^2}{2!} S_{(2)} [\gamma] + \cdots + \frac{\hbar^k}{k!} S_{(k)} [\gamma] + \cdots .
\end{equation}
Imposing the momentum constraint (\ref{eq:2174}) to all orders in \(\hbar\) leads to the conclusion that each of the functionals, \(\lbrace S_{(k)} [\gamma]; \; k = 0, 1, 2, \ldots\rbrace\), should be invariant with respect to the aforementioned action of \(\mathcal{D}\mathit{iff}^0(M)\) on \(\mathcal{M}(M)\), ie, that
\begin{equation}\label{eq:2180}
S_{(k)} [\varphi^*\gamma] = S_{(k)} [\gamma],\; k = 0, 1, 2, \ldots
\end{equation}
\(\forall\; \varphi \in \mathcal{D}\mathit{iff}^0(M)\) and \(\forall\; \gamma \in \mathcal{M}(M)\).

Independently of the precise form finally chosen for \(\hat{\mathcal{H}}_\perp (\gamma,\frac{\hbar}{i}\frac{\delta}{\delta\gamma})\), the leading order approximation to the \textit{Wheeler-DeWitt equation},
\begin{equation}\label{eq:2181}
\hat{\mathcal{H}}_\perp \left(\gamma,\frac{\hbar}{i}\frac{\delta}{\delta\gamma}\right) e^{-S_{(0)} [\gamma]/\hbar - S_{(1)}[\gamma] - \cdots} = 0,
\end{equation}
for the ground state wave functional will, inevitably reduce to the Euclidean-signature Hamilton-Jacobi equation
\begin{equation}\label{eq:2182}
\left(\frac{16\pi G}{c^3}\right)^2 \frac{\left(\gamma_{ik}\gamma_{j\ell} - \frac{1}{2} \gamma_{ij} \gamma_{k\ell}\right)}{\mu_\gamma} \frac{\delta S_{(0)}}{\delta\gamma_{ij}} \frac{\delta S_{(0)}}{\delta\gamma_{k\ell}} + \mu_\gamma {}^{(3)}R(\gamma) = 0.
\end{equation}
This equation coincides with that obtained from making the canonical substitution,
\begin{equation}\label{eq:2183}
p^{ij} \longrightarrow \frac{\delta S_{(0)} [\gamma]}{\delta\gamma_{ij}},
\end{equation}
in the Euclidean-signature version of the Hamiltonian constraint,
\begin{equation}\label{eq:2184}
\mathcal{H}_{\perp\mathrm{Eucl}} := -\left(\frac{16\pi G}{c^3}\right)\; \frac{\left(p^{ij} p_{ij} - \frac{1}{2} (p_m^m)^2\right)}{\mu_\gamma} - \left(\frac{c^3}{16\pi G}\right)\; \mu_\gamma\; {}^{(3)}R(\gamma) = 0,
\end{equation}
that, in turn, results from repeating the derivation sketched above for \(I_{\mathrm{ADM}}\) but now for the Riemannian metric form
\begin{equation}\label{eq:2185}
{}^{(4)}g\Big|_{\mathrm{Eucl}} = {}^{(4)}g_{\mu\nu}\Big|_{\mathrm{Eucl}} dx^\mu \otimes dx^\nu = N\Big|_{\mathrm{Eucl}}^2 dt \otimes dt + \gamma_{ij} (dx^i + Y^i dt) \otimes (dx^j + Y^j dt)
\end{equation}
in place of (\ref{eq:2155}). The resulting functional \(I_{\mathrm{ADM}\; \mathrm{Eucl}}\) differs from \(I_{\mathrm{ADM}}\) only in the replacements \(\mathcal{H}_\perp (\gamma,p) \longrightarrow \mathcal{H}_{\perp\mathrm{Eucl}} (\gamma,p)\) and \(N \longrightarrow N\Big|_{\mathrm{Eucl}}\).

The essential question that now comes to light is thus the following:
\medskip

\begin{quote}
\emph{Is there a well-defined mathematical method for establishing the existence of a \(\mathcal{D}\mathit{iff}^0(M)\)-invariant, fundamental solution to the Euclidean-signature functional differential Hamilton-Jacobi equation (\ref{eq:2182})?}
\end{quote}
\medskip

\noindent In view of the field theoretic examples discussed in Section~\ref{sec:euclidean} one's first thought might be to seek to minimize an appropriate Euclidean-signature action functional subject to suitable boundary and asymptotic conditions. But, as is well-known from the Euclidean-signature path integral program \cite{Gibbons:1993}, the natural functional to use for this purpose is \textit{unbounded from below} within any given conformal class --- one can make the functional arbitrarily large and negative by deforming any metric \({}^{(4)}g\Big|_{\mathrm{Eucl}}\) with a suitable conformal factor \cite{Gibbons:1979,Gibbons:1993}.

But the real point of the constructions of Section~\ref{sec:euclidean} was \textit{not} to minimize action functionals but rather to generate certain `fundamental sets' of solutions to the associated Euler-Lagrange equations upon which the relevant action functionals could then be evaluated. But the Einstein equations, in vacuum or even allowing for the coupling to conformally invariant matter sources, encompass, as a special case, the \textit{vanishing} of the 4-dimensional scalar curvature, \({}^{(4)}R({}^{(4)}g\Big|_{\mathrm{Eucl}})\). Thus there is no essential loss in generality, and indeed a partial simplification of the task at hand to be gained, by first restricting the relevant, Euclidean-signature action functional to the `manifold' of Riemannian metrics satisfying (in the vacuum case) \({}^{(4)}R({}^{(4)}g\Big|_{\mathrm{Eucl}}) = 0\) and then seeking to carry out a \textit{constrained minimization} of this functional.

Setting \({}^{(4)}R({}^{(4)}g\Big|_{\mathrm{Eucl}}) = 0\) freezes out the conformal degree of freedom that caused such consternation for the Euclidean path integral program \cite{Gibbons:1979,Gibbons:1993}, wherein one felt obligated to integrate over \textit{all possible} Riemannian metrics having the prescribed boundary behavior, but is perfectly natural in the present context and opens the door to appealing to the \textit{positive action theorem} which asserts that the relevant functional is indeed positive when evaluated on arbitrary, asymptotically Euclidean metrics that satisfy \({}^{(4)}R({}^{(4)}g\Big|_{\mathrm{Eucl}}) \geq 0\) \cite{Schoen:1979,Schoen:1979b,Zhang:1999,Dahl:1997}.

Another complication of the Euclidean path integral program was the apparent necessity to invert, by some still obscure means, something in the nature of a `Wick rotation' that had presumably been exploited to justify integrating over Riemannian, as opposed to Lorentzian-signature, metrics. Without this last step the formal `propagator' being constructed would presumably be that for the Euclidean-signature variant of the Wheeler-DeWitt equation and not the actual Lorentzian-signature version that one wishes to solve. In ordinary quantum mechanics the corresponding, well-understood step is needed to convert the Feynman-Kac propagator, derivable by rigorous path-integral methods, back to one for the actual Schr\"{o}dinger equation.

But in the present setting no such hypothetical `Wick rotation' would ever have been performed in the first place so there is none to invert. Our focus throughout is on constructing asymptotic solutions to the original, Lorentz-signature Wheeler-DeWitt equation and not to its Euclidean-signature counterpart. That a Euclidean-signature Einstein-Hamilton-Jacobi equation emerges in this approach has the very distinct advantage of leading one to specific problems in Riemannian geometry that may well be resolvable by established mathematical methods. By contrast, path integral methods, even for the significantly more accessible gauge theories discussed in Section~\ref{sec:euclidean}, would seem to require innovative new advances in measure theory for their rigorous implementation. Even the simpler scalar field theories, when formulated in the most interesting case of four spacetime dimensions, seem still to defy realization by path integral means. It is conceivable, as was suggested in the concluding section of \cite{Moncrief:2012}, that focusing predominantly on path integral methods to provide a `royal road' to quantization  may, inadvertently, render some problems more difficult to solve rather than actually facilitating their resolution.

The well-known `instanton' solutions to the Euclidean-signature Yang-Mills equations present a certain complication for the semi-classical program that we are advocating in that they allow one to establish the existence of \textit{non-unique minimizers} for the Yang-Mills action functional for certain special choices of boundary data \cite{Maitra:inprep}. This in turn can obstruct the global smoothness of the corresponding solution to the Euclidean-signature Hamilton-Jacobi equation. While it is conceivable that the resulting, apparent need to repair the associated `scars' in the semi-classical wave functionals may have non-perturbative implications for the Yang-Mills energy spectrum --- of potential relevance to the `mass-gap' problem --- no such corrections to the spectrum are expected or desired for the gravitational case. Thus it is reassuring to note that analogous `gravitational instanton' solutions to the Euclidean-signature Einstein equations have been proven \textit{not to exist} \cite{Gibbons:1979}.

We conclude by noting that other interesting, generally covariant systems of field equations exist to which our (`Euclidean-signature semi-classical') quantization methods could also be applied. Classical relativistic `membranes', for example, can be viewed as the evolutions of certain embedded submanifolds in an ambient spacetime --- their field equations determined by variation of the volume functional of the timelike `worldsheets' being thereby swept out. The corresponding Hamiltonian configuration space for such a system is comprised of the set of spacelike embeddings of a fixed \(n - 1\) dimensional manifold \textit{M} into the ambient \(n + k\) dimensional spacetime, each embedding representing a possible spacelike slice through some \textit{n}-dimensional membrane worldsheet. Upon canonical quantization wave functionals are constrained (by the associated, quantized momentum constraint equation) to be invariant with respect to the induced action of \(\mathcal{D}\mathit{iff}^0(M)\) on this configuration space of embeddings. The corresponding quantized Hamiltonian constraint, imposed \`{a} la Dirac, provides the natural analogue of the Wheeler-DeWitt equation for this problem.

A solution to the operator ordering problem for these quantized constraints, when the ambient spacetime is Minkowskian, was proposed by one of us in \cite{Moncrief:2006}. For the compact, codimension one case (i.e., when \textit{M} is compact and \(k = 1\)) it is not difficult to show that the relevant \textit{Euclidean-signature} Hamilton-Jacobi equation has a fundamental solution given by the volume functional of the maximal, spacelike hypersurface that uniquely spans, \`{a} la Plateau, the arbitrarily chosen embedding \cite{Moncrief:unpub}. It would be especially interesting to see whether higher-order quantum corrections and excited state wave functionals can be computed for this system in a way that realizes a quantum analogue of general covariance.

\section*{Acknowledgements}

Moncrief is grateful to the Swedish Royal Institute of Technology (KTH) for hospitality and support during a visit in March 2015 and especially to Lars Andersson for pointing out the relevance of mathematical literature on Bakry-Emery Ricci tensors to the research discussed herein. The authors are also grateful to the Albert Einstein Institute in Golm, Germany, the Institut des Hautes {\'E}tudes Scientifiques in Bures-sur-Yvette, France, the Erwin Schr{\"o}dinger Institute and the University of Vienna in Vienna, Austria for the hospitality and support extended to several of us during the course of this research.

\bibliographystyle{unsrt} 
\bibliography{BER_ArXiV}

\begin{thebibliography}{10}
\bibitem{orbit-misc01}
Note in particular that geodesic curves of infinite proper length of purely
  `polarized' connections (i.e., ones for which the {Lie} brackets all vanish)
  exist along which the magnetic energy monotonically increases without bound.
\end{thebibliography}

\end{document}